# Designing for Rich Collocated Social Interactions in the Age of Smartphones

by

Hüseyin Uğur Genç

A Dissertation Submitted to the Graduate School of Social Sciences and Humanities in Partial Fulfillment of the Requirements for the Degree of Doctor of Philosophy

in

Design, Technology, and Society
in the specification of Interaction Design

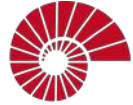

KOÇ ÜNİVERSİTESİ

December 20, 2022

# Designing for Rich Collocated Social Interactions in the Age of Smartphones

Koç University

Graduate School of Social Sciences and Humanities

This is to certify that I have examined this copy of a doctoral dissertation by

**Hüseyin Uğur Genç**

and have found that it is complete and satisfactory in all respects and that any and all revisions required by the final examining committee have been made.

Committee Members:

___________________________________________
Assoc. Prof. Aykut Coşkun

___________________________________________
Prof. Dr. Fatoş Gökşen

___________________________________________
Prof. Dr. Oğuzhan Özcan

___________________________________________
Prof. Dr. Jonna Häkkilä

___________________________________________
Assoc. Dr. Markus Löchtefeld

Date: 20/12/2022

*Dedicated to my family.*

# ABSTRACT

**Designing for Rich Collocated Social Interactions  
in the Age of Smartphones**

Hüseyin Uğur Genç

Doctor of Philosophy in Design, Technology, and Society

The quality of social interaction has great importance for psychological and physiological health. Previous research indicates that smartphones have adverse effects on collocated social interactions. Many HCI studies addressed this issue by restricting smartphone use during social interactions. Although the results of these studies indicated a decrease in smartphone use, restrictive and limiting approaches have limitations. Users should have high levels of self-regulation to comply with them, and they may lead to unintended outcomes like withdrawal symptoms. Considering the influence of smartphones on people's social relations and interactions, either positive or negative, there is a need for new solutions to mitigate the adverse effects of excessive smartphone use alternatives to restrictive approaches. To this end, this thesis aims to explore individuals' smartphone use behavior from the standpoint of social interactions and relations by employing diverse data collection techniques, i.e., how this behavior hinders and supports social interactions. We started investigating this question through in-situ observations and focus group sessions. With the information learned from this step, we developed two research prototypes to enrich social interactions without restricting smartphone use. We then collected users' thoughts, reactions towards, and concerns about these prototypes through user studies. Finally, we examined how these prototypes influenced conversation quality in social interactions through an experimental user study.

Along with the motivation and aim, this thesis makes knowledge and artifact contributions to the growing field of digital well-being. We identified 21 user insights, nine design implications, and four design approaches that will guide the design of innovations and solutions to enrich social interactions in the presence of smartphones. We developed two design concepts, which also served as the validation of these knowledge contributions.

# ÖZETÇE
## Akıllı Telefon Çağında Yüz Yüze Sosyal Etkileşimlerin Zenginleştirmesinin Tasarım Odaklı Araştırılması

Hüseyin Uğur Genç

Tasarım, Teknoloji ve Toplum, Doktora

Sosyal etkileşimin kalitesi, psikolojik ve fizyolojik sağlık için büyük öneme sahiptir. Araştırmalar, akıllı telefonların yüz yüze sosyal etkileşimler üzerinde olumsuz etkileri olduğunu göstermektedir. HCI alanında yapılan birçok çalışma, sosyal etkileşimler sırasında akıllı telefon kullanımını kısıtlayarak bu sorunu ele almıştır. Bu çalışmaların sonuçları akıllı telefon kullanımında bir azalma olduğunu gösterse de, sınırlayan ve yasaklayan yaklaşımların belirli kısıtları bulunmaktadır. Kullanıcılar, bu sınırlamalara uymak için yüksek düzeyde öz-düzenlemeye sahip olmalıdır ve bu sınırlamalar kullanıcılarda yoksunluk belirtileri gibi istenmeyen sonuçlara yol açabilirler. Akıllı telefonların insanların sosyal ilişkileri ve etkileşimleri üzerindeki olumlu ve olumsuz etkileri göz önüne alındığında, bu yaklaşımlara alternatif olarak olumsuz etkileri azaltacak yeni çözümlere ihtiyaç duyulmaktadır. Bu amaçla, bu tez, çeşitli veri toplama teknikleri kullanarak, bireylerin akıllı telefon kullanım davranışlarını sosyal etkileşimler ve ilişkiler açısından bir başka deyişle bu davranışın sosyal etkileşimleri nasıl engellediğini ve desteklediğini keşfetmeyi amaçlamaktadır. Bu soruyu gözlem ve odak grup oturumları aracılığıyla araştırıp, bu adımdan öğrenilen bilgilerle, akıllı telefon kullanımını kısıtlamadan sosyal etkileşimleri zenginleştirmek için iki araştırma prototipi geliştirdik. Daha sonra kullanıcı araştırmaları ile kullanıcıların bu prototipler hakkındaki düşüncelerini, tepkilerini ve endişelerini topladık. Son olarak, deneysel bir kullanıcı çalışması aracılığıyla bu prototiplerin sosyal etkileşimlerdeki konuşma kalitesini nasıl etkilediğini inceledik. Bu motivasyon ve amaç ile birlikte, bu tez önemi gitgide artan dijital iyi oluş alanına bilgi ve tasarım yolu ile katkılar sağlamaktadır. Akıllı telefonların varlığında sosyal etkileşimleri zenginleştirmek için yenilikçi fikirlerin ve çözümlerin tasarımına rehberlik edecek 21 kullanıcı içgörüsü, dokuz tasarım yönergesi, dört tasarım yaklaşımı belirledik ve bu bilgi katkılarının doğrulanması olarak da hizmet eden iki tasarım konsepti geliştirdik.

# ACKNOWLEDGMENTS

First and foremost, I would like to thank my father, mom, and sisters, Rükan, Rüyan, and Ezgi, for their endless love and support. In every obstacle and challenge I faced, I felt you by my side, and I found strength and motivation in every step of my life thanks to you. Without you, this day would not have been possible.

An immense thank you to my Ph.D. supervisor: Aykut Coşkun. Your support and guidance throughout this journey have been invaluable. Hocam, you've been a fantastic supervisor, and I feel very lucky for working with you.

Dear Burak, we had many different adventures together. We worked together, we learned, we produced, we won, and sometimes we lost because of fraudsters. But most importantly, we laughed and had fun all the time. Thank you for being there.

My beloved Kanki, Ceylan, with our gales of laughter and mind-opening discussions, this journey was the best years of my life. Dear Yağmur, I am very happy to meet you even though I met you in the middle of this story; your passion is inspiring. Dear Hakan, İpek, and Berre, my teamies, anytime I needed help, you were there with me, made me smile with your positivity. Your friendships mean a lot to me.

Thank you to all my professors who helped me get to this stage. Prof. Dr. Fatoş Gökşen, Prof. Dr. Oğuzhan Özcan, Assoc. Dr. Markus Löchtefeld and Assoc. Dr. Tilbe Göksun, you were a huge influence, without which I would have never followed this path.

I would also like to thank all my committee members, including Prof. Dr. Jonna Häkkilä, for making my defense enjoyable and for your brilliant comments and suggestions; thanks to you.

Other than all these great people who contributed to my dissertation, I also thank all my colleagues at Koç University – Arçelik Research Center for Creative Industries for making my last five years memorable and motivating me to finish this work.

Last but not least, my dear Nordichi *the cat* and Appa *the dog*. You bring joy into my life. Your contributions to this work are invaluable.

# TABLE OF CONTENTS

# LIST OF TABLES

# LIST OF FIGURES

# ABBREVIATIONS

HCI     Human-Computer Interaction

RtD     Research Through Design

BFI     Big Five Inventory

**How to read this dissertation**

This thesis is structured in a Collection of Papers, which investigates the research questions through four publications (three published and one in submission). These publications are referred to in the relevant chapters and are added to the manuscript.

# Chapter 1

# INTRODUCTION

## *1.1. Motivation*

*We expect more from technology and less from each other. We create technology to provide the illusion of companionship without the demands of friendship*

*– Sherry Turkle*

Smartphones accompany us in every aspect of our lives. We have become more dependent on them for almost every daily task, from making a payment at the coffee shop to connecting to our homes at work. Smartphone applications offer promising ways to ease our lives (e.g., navigation, contactless payment, etc.) and even to prevent and help treat chronic diseases such as diabetes (Årsand et al., 2015) or alcoholism (Gustafson et al., 2015). However, their overuse may lead to physical and mental health problems (e.g., sight problems, joint pain, neck pain, sleep disturbances, depression, and smartphone addiction) (H. Lee et al., 2014; S. Lee et al., 2017; Lemola et al., 2014; Lin et al., 2014). Furthermore, excessive smartphone use hinders our physiological and psychological well-being. It adversely affects our social interactions and relations, which covers our ability to communicate and develop meaningful relationships with others, such as family, friends, neighbors, and colleagues. In this regard, smartphone use may damage the level of intimacy and connection between friends, reduce conversation quality (Misra et al., 2016; Przybylski & Weinstein, 2013), and make companions feel awkward and excluded in social settings (Humphreys, 2005). This being the case, it should be noted that it is very crucial to examine users' excessive smartphone use behaviors from the perspective of their social interactions.

Smartphones and mobile apps have been intentionally designed to maximize user engagement. Apps with constant notifications invite users to check their smartphones regularly. Contrarily, there is an increasing trend towards creating solutions for supporting users' digital well-being, which concerns the responsible use (e.g., being aware of the impacts of smartphones on individuals' well-being) of technology in daily life. Some technology companies like Google introduced digital well-being features to support responsible smartphone use practices, such as monitoring and managing the time spent using social media apps, encouraging breaks in use, and promoting digital detoxes (Google, n.d.).

Studies in Human-Computer Interaction (HCI) literature have been exploring solutions to support responsible smartphone use. There are studies that deal with this use habit and prevent the side effects. However, most of those studies are based on the argument that using smartphones is a negative habit and commonly benefits from strategies helping users restrict and limit their smartphone use. For example, providing users with an opportunity to set use limits for themselves (Ko, Choi, et al., 2015; H. Lee et al., 2014; Löchtefeld et al., 2013) or setting use limits for each other via a mobile app (Ko, Chung, et al., 2015; Ko et al., 2016), and reminding excessive use through visual and haptic feedback (Choi & Lee, 2016; Okeke et al., 2016). Although these studies claimed they decreased excessive smartphone use, their strategies to influence user behavior have two limitations. First, behavior change strategies allowing users to restrict their smartphone use can only work when the users have i) an awareness of the adverse effects of their use-behavior, ii) an intention to change this behavior (Ajzen, 1991), and iii) high levels of self-regulation to maintain this change (B. J. Zimmerman, 2000). Thus, such strategies may fail to mitigate excessive smartphone use by individuals with low self-regulation levels. These individuals are more likely to be addicted to smartphones than individuals with high self-regulation levels (Gökçearslan et al., 2016; Jeong et al., 2016). Second, research showed that restriction-based approaches, described above, may lead to unintended outcomes (e.g., limiting the device use can often backfire, creating anxiety and withdrawal symptoms (Stibe & Cugelman, 2016) and increasing the time spent on social media (AP/NORC, n.d.)).

Considering the influence of smartphones on people's social relations and interactions, either positive or negative, there is a need for alternative solutions to mitigate the adverse effects of their excessive use, especially for individuals' social well-being, without following restrictive approaches.

## *1.2. Scope, Aim & Research Questions*

It is not easy for users to unplug from their smartphones. These devices have become increasingly integrated with their lives, positively and negatively affecting daily interactions. This challenges designers to develop solutions for mitigating excessive smartphone use. Our previous work examining smartphone usage during social interactions (Genç et al., 2018) showed that smartphone-checking behavior results from poor interaction between individuals but not the cause. Considering this, we should consider that smartphones do not just have negative effects. This implies that a deeper understanding of smartphone use during social interactions is required before developing solutions to mitigate its use.

This Ph.D. study focuses on the impact of technology use on our social interactions and relations and aims to explore it in the context of smartphone use during these interactions. Within this study's scope, we focus on the practices that support richer social interactions between individuals via technology while mitigating its negative effects. This study started with the question of whether we can reduce smartphone use, which has serious side effects as well as benefits, during collocated social interactions without using restrictive approaches. And instead of banning smartphones, we focused on creating positive and meaningful interactions (Desmet & Pohlmeyer, 2013) in a social context (for example, sharing a memory with a friend rather than focusing on the phone and ignoring the other person). With this perspective, this study aims to explore individuals' smartphone use from the standpoint of social interactions and relations (i.e., how this behavior hinders and supports social interaction). We intend to generate solutions for supporting social interactions without restricting smartphone use and for identifying design implications for creating such solutions. Aligned with this, our research questions are listed below.

RQ1. How do people conceptualize the responsible use of smartphones in collocated social interactions?

- How do these conceptualizations differ across different user groups?

- What factors influence an individual's smartphone use in collocated social interactions?

- How do users maintain their behavior for social interactions (i.e., user-driven strategies)?

RQ2. How do people use their smartphones during social interaction?

- What are the motivations/attitudes towards excessive phone use during social interactions?

- What are the situations that trigger smartphone use during social interactions?

- What are their feelings and reactions towards people who constantly interact with their phones instead of engaging in conversation?

RQ3. How can we design to support social interactions (without necessarily restricting smartphone use)?

- What would be the design strategies for supporting social interactions?

- What would be the dimensions for designing and assessing concepts that support social interactions?

- What would be the implications for designing to support social interactions?

RQ4. How would the design concepts aimed at supporting rich social interactions influence the interactions between individuals in the presence of smartphones?

Our aim was to explore how richer social interactions, in which people interact with people in social environments instead of interacting with their phones, can be supported through design. For this reason, we have developed prototypes to examine user behavior and enrich users' social interactions. We collected users' thoughts, reactions, and concerns about these prototypes in our user studies. Finally, we examined how conversion quality changed in social interactions during the existence of such a prototype. Therefore, it is beyond the scope of this study to develop an intervention that will reduce people's smartphone use and measure the effectiveness of this intervention.

## 1.3. *Research Approach*

Our methodological approach revolves around Research Through Design (RtD) (J. Zimmerman et al., 2007), which focuses on the role of design as an instrument of design knowledge inquiry. RtD involves an iterative research process, including the cycles between framing the problem, understanding the context, developing design concepts, and user tests.

We have also adopted the User-Centered Design (D. Norman, 2016) approach, which relates to RtD, as each of the above-mentioned iterative steps requires users' insights. This approach brings the involvement of users and stakeholders and enables meeting their needs and requirements in the design and its process.

This iterative research process is the backbone of this thesis, and we followed a path as follows (Figure 1). First, we provided an understanding of the problem and the context from the observations and focus groups we conducted with the users. With the information learned from this step, we aimed to uncover a design space and produce concepts from this design space. By using these concepts, our aim was to gain more profound information about the problem and context with user studies. Rather than aiming for the concepts we created to be a final product, we thought of these as auxiliary tools in parallel with the RtD approach in our information production process. The two conceptual products we created allowed us to get more comprehensive insights from users by putting these individuals directly in the context of the research questions we investigated and discussed.

## 1.4. *Contributions*

Along with the motivation and aim mentioned above, this thesis has two main contributions: Knowledge and Artifact. We have provided 21 insights, nine implications, and four approaches to the Design and HCI research field with the studies conducted with users and the knowledge we gained from them (Figure 1.2). While collecting this knowledge, we designed two artifacts contributing to every research stage.

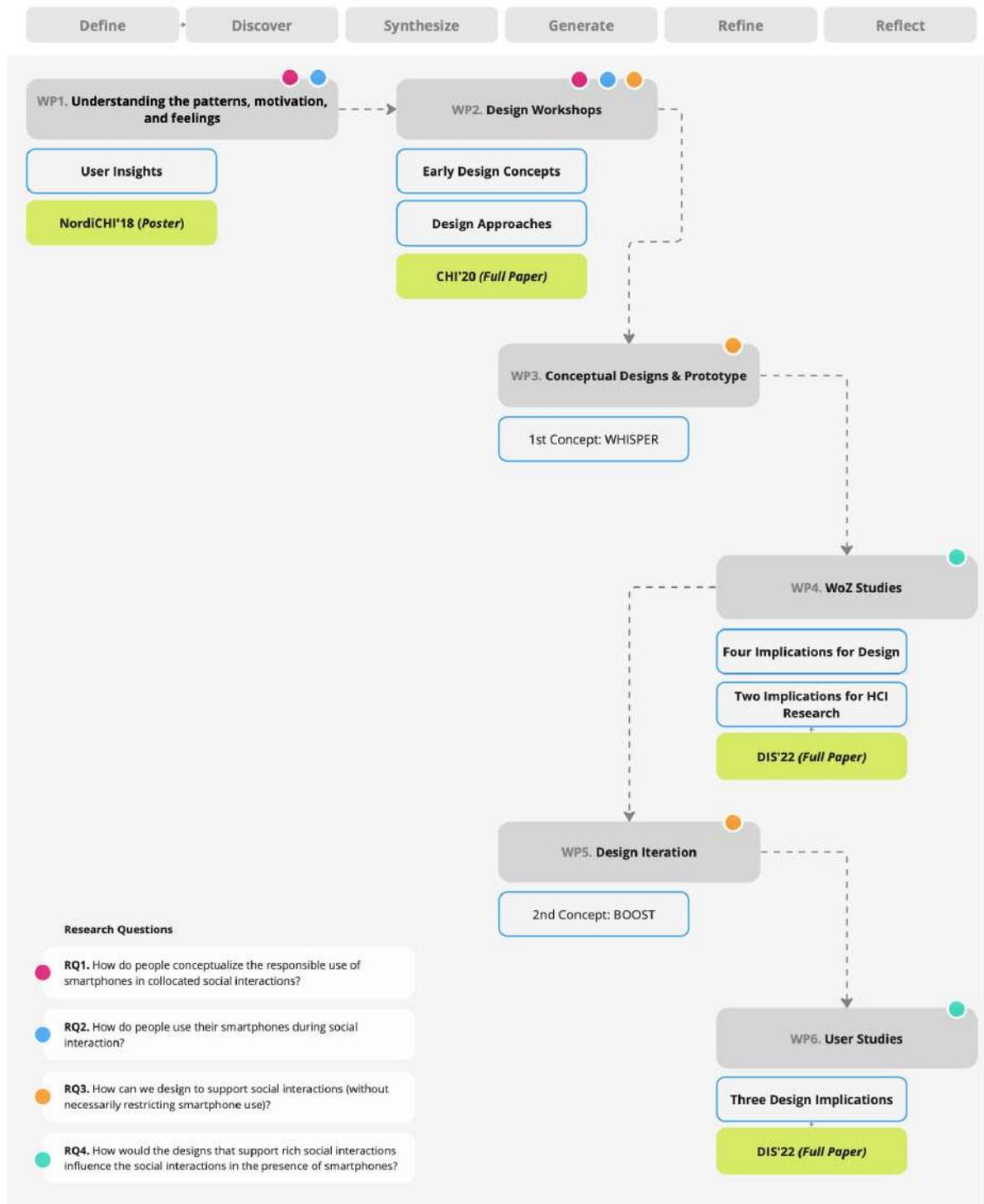

*Figure 1.1 Ph.D. Journey*

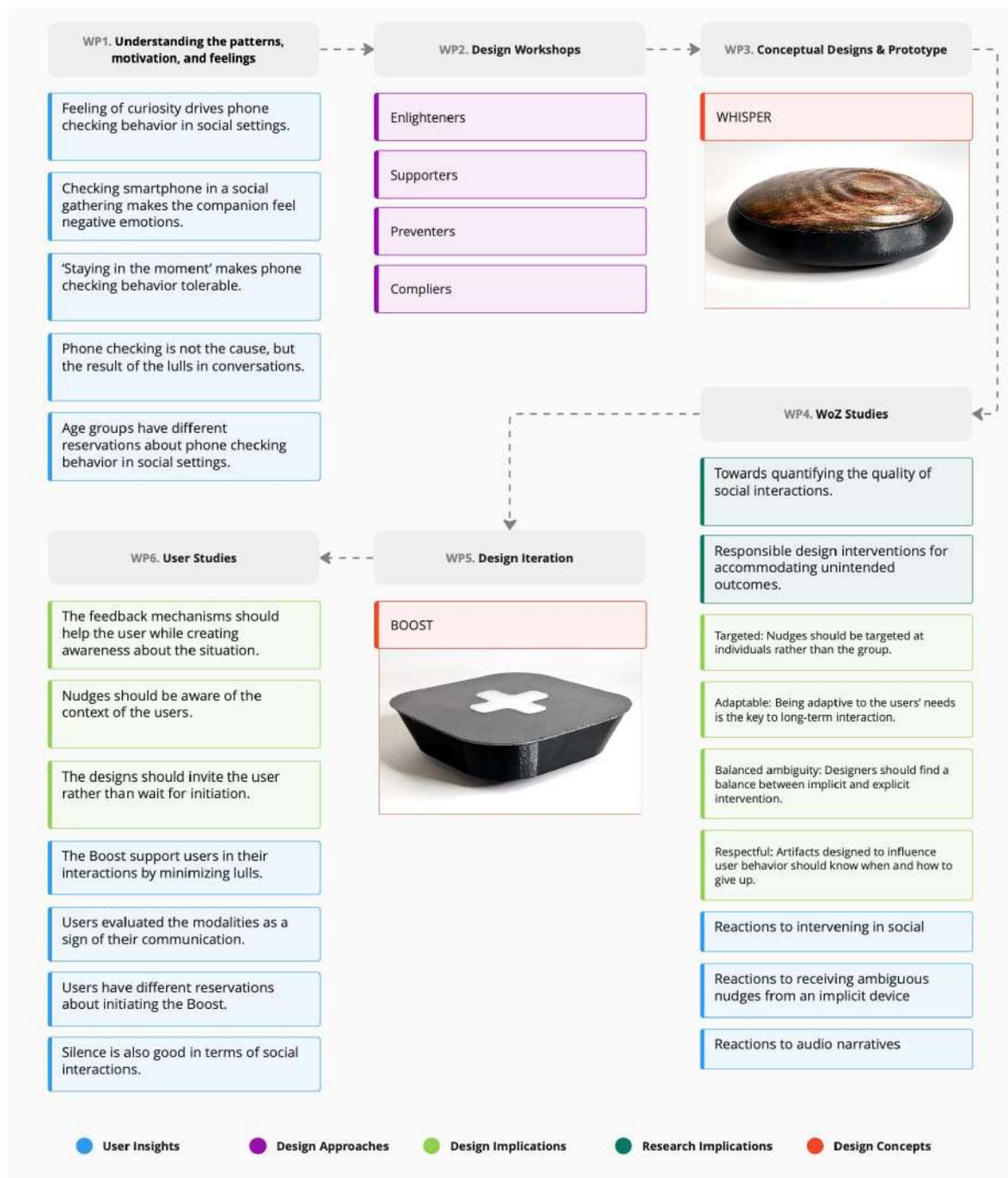

*Figure 1.2 The List of Contributions from Workpackages*

### 1.4.1. Knowledge Contribution

*User insights* help designers and researchers understand smartphone use behavior in public settings and its effects on social interactions. *Design Approaches* for supporting richer collocated social interactions from the users' perspective. While approaches can help designers determine a roadmap for supporting rich interactions, *recommendations* *and implications* provide them with more specific and contextual guidance on designing technologies to support these collocated interactions. These recommendations and implications may inspire designers and researchers to develop interactive technologies that help users manage their excessive smartphone use during social interactions and sustain their digital well-being for social interactions.

### *1.4.2.    Artifact Contributions*

In addition to the Knowledge contribution, we also provide Design Concepts and Design Artefacts to support richer collocated social interactions without necessarily restricting smartphone use. These research artifacts helped us conduct user studies and gain rich user insights. While designing these artifacts, we considered the insights and design implications we revealed from user studies.

## *1.5.    Dissertation Overview*

This thesis is composed of nine chapters. In Chapter 1, we introduce a summary of the thesis and information provided in the following chapters. Chapter 2 presents the related work that frames this thesis's relevance. Chapters 3-9 cover four main research steps of the thesis. In Chapter 3, we provided the study of marginal observations and three focus group sessions with different age groups to uncover users' motivations for using smartphones and feelings towards smartphone use during social interactions, as well as techniques they use to deal with this usage behavior (RQ1, RQ2, RQ3). In Chapter 4, we extend our participant pool by conducting another set of focus group sessions with similar demographics to advance this understanding. This exploration helped us understand smartphone use in social interactions and strategies to mitigate its use during these interactions by making our data crystallized. As a continuation of this chapter, we conducted Design Workshops in which we provided the user insights that we gained in previous studies to the designers and asked them to ideate solutions to support rich social interactions. We explored the design space for the context and identified four design approaches to mitigate smartphone use during social interactions. Based on this exploration, in Chapter 5, we present our design process for an intervention (i.e., WHISPER) to enrich social interactions and support digital well-being for social interactions. We created a semi-working prototype of this concept and used it in the user studies in Chapter 6. In these user studies, we assessed the impact of WHISPER on users' social interaction and developed strategies and implications for designing similar solutions (RQ4). With the implications gathered from these 'Wizard of Oz' sessions, we iterated our design and developed our second intervention, BOOST. We provide our design process for a fully working prototype in Chapter 7 and the results of user experiments in Chapter 8. Lastly, in Chapter 9, we present the main takeaways of the thesis along with limitations and directions for future research

### *1.6. List of Publications*

#### *1.6.1. Publications within the thesis scope*

- Paper 1: Hüseyin Uğur Genç, Fatoş Gökşen, and Aykut Coşkun. 2018. **Are we 'really' connected? Understanding smartphone use during social interaction in public settings.** the 10th Nordic Conference on Human-Computer Interaction.

  https://doi.org/10.1145/3240167.3240235

- Paper 2: Hüseyin Uğur Genç and Aykut Coskun. 2020. **Designing for Social Interaction in the Age of Excessive Smartphone Use.** In Proceedings of the 2020 CHI Conference on Human Factors in Computing Systems. https://doi.org/10.1145/3313831.3376492

- Paper 3: Hüseyin Uğur Genç, Duru Erdem, Çağla Yıldırım, Aykut Coskun. 2022. **Mind the WHISPER: Enriching Collocated Social Interactions in Public Places through Audio Narratives.** In Designing Interactive Systems Conference.

  https://doi.org/10.1145/3532106.3533507

- Paper 4: (Will be submitted) Hüseyin Uğur Genç and Aykut Coşkun. **Boosting the Collocated Social Interactions.**

#### *1.6.2. Publications outside the thesis scope as a part of research training*

- Gül Kaner, Hüseyin Uğur Genç, Salih Berk Dinçer, Deniz Erdoğan, and Aykut Coşkun. 2018. **GROW: A Smart Bottle that Uses its Surface as an Ambient Display to Motivate Daily Water Intake.** Extended Abstracts of the 2018 CHI Conference on Human Factors in Computing Systems (CHI EA '18).

# Chapter 2

# BACKGROUND

## *2.1. Smartphones in Users' Lives & Why Do They Use These Devices*

The importance of smartphones in individuals' lives is skyrocketing day by day. Application ecosystems have reached an incredible diversity with increasing processing power and enriching features. Thus, people started to handle all their work on smartphones, where one can find a mobile application for each different need. Now, users can call a cab or remotely control their homes with a touch. Students can do their homework or research by using this medium. During the pandemic days, as people separated from their beloved ones, they cared for each other via smartphones. As of 2019, 35% of the world's population owns a smartphone, and this rate goes up to 76% in developed countries (Bahia & Suardi, 2019). However, the age of having a smartphone has dropped to 12 (Rideout & Robb, 2019), and 98% of generation Z are smartphone owners (Mander & McGrath, 2017). In terms of use differences, teens prefer visual apps related to social media, messaging, video, and music more than the older generations (e.g., millennials prefer apps for online shopping and email.) (Pallini & Pallini, 2018). Their use time average reaches nine hours (Rideout & Robb, 2019), while 59% of millennials spend at least four hours per day (Dolliver, 2019).

Researchers have attributed various characteristics to smartphones beyond their functional value. Smartphones extend the self beyond the human body (Belk, 2013; Clark & Chalmers, 1998). Now, smartphones can be considered as an extension of the users (Katz, 2017; Walsh & White, 2007), part of their identities (Vykoukalová, 2007), or a part of their everyday lives (Ling, 2004). While people are on the way to work, eating with friends, lying in bed, or even crossing the street, smartphones are in their hands and have become an attachment object for individuals (Trub & Barbot, 2016). As this study explores excessive smartphone use behavior during social interactions and solutions to mitigate this behavior, the view of smartphones as attachment objects is highly relevant. Attachment Theory (Bowlyb, 1958) can help understand why individuals use smartphones and why they cannot quickly abandon them. Even though this theory initially defines the relationship between children and their primary caregivers in developmental psychology, it is also constructive in explaining the relationship between adult-adult and adult-object relationships. Since smartphones have similar characteristics to attachment objects, individuals' motivation to use smartphones can be described with the following features of attachment objects:

1) **Associations of positive outcomes**: Since smartphones have numerous features such as constant connectedness with the world, the availability of an unlimited amount of knowledge, communication with social circles, productivity tools, and various sources of entertainment, they produce positive outcomes for users' lives. They make users more efficient by learning new things, growing their careers, and developing businesses (Jewell, 2011). Smartphones provide an easier way to communicate with people, making them feel connected to their communities, families, and friends (Jesse, 2016).

2) **Increasing sense of comfort and stress relief:** With positive outcomes, smartphones satisfy psychological needs such as socializing, autonomy, personal safety, and personalization. Since smartphones serve as attachment objects, the feelings of comfort derived from engaging with the device could also logically result in heightened feelings of relaxation (Fullwood et al., 2017).

3) **Distress when the object is absent:** Another characteristic of attachment objects is when the subject is separated from the object, s/he gets stressed. Several studies (Clayton et al., 2015; Fullwood et al., 2017; Hoffner et al., 2016; Konok et al., 2016; Trub & Barbot, 2016) show that participants' separation from their smartphones makes them anxious.

In summary, smartphones are now considered *adult pacifiers* (Diefenbach & Borrmann, 2019) because they make users less stressed, less bored, and more secure in various situations.

## 2.2. *Smartphone Use & Its Influence on Users' Well-being*

Smartphones have many advantages considering their contribution to people's lives. They make their lives easier in many ways with their applications and technologies. In addition to these functional advantages, smartphones have both positive and negative effects on individuals' physiological, psychological, and social well-being. We have discussed these effects in this section.

### 2.2.1. *Positive effects of smartphone use*

Smartphones offer a wide variety of health benefits to their users. Most smartphones now come equipped with several embedded sensors that detect various conditions, e.g., motion, location, image, and sound, improving health monitoring and diagnostic accuracy (Majumder & Deen, 2019). These sensors can measure vital activities such as heart rate, respiratory rate, and health conditions such as skin diseases and eye diseases, thus turning these devices into health monitoring systems (Majumder & Deen, 2019). Using these sensors, users can track their sportive activities, sleep qualities, or vitals. They can share these data with their medical advisors. Also, some researchers developed interventions for fall detection (Y. Lee et al., 2018) and posture monitoring. Besides, smartphones can be used for lung rehabilitation exercises for chronic pulmonary diseases, e.g., coughing and asthma (Stafford et al., 2016). With the ever-developing technology, researchers can now detect cancerous skin cells by using deep-neural networks on images taken by smartphone cameras (Kuzmina et al., 2015).

In addition to these health benefits, smartphones are also seen as adult pacifiers in the literature and form people's comfort areas (Primack et al., 2017). Smartphones provide users with a range of social and emotional benefits due to the increased convenience and accessibility they enable (Subrahmanyam & Smahel, 2011). For example, when individuals enter an unfamiliar environment, they relieve their stress by spending time with their smartphones (Diefenbach & Borrmann, 2019; Trub & Barbot, 2016). Smartphones enhance romantic feelings (Jin & Peña, 2010; Schade et al., 2013), more significant interaction, and collaboration in learning environments (Gikas & Grant, 2013). Another study shows that online media's high self-disclosure ratio (i.e., revealing information about the self to another person) reduces users' loneliness (Karsay et al., 2019).

With the advantages above, social interaction in an online medium (e.g., via social media apps on smartphones) can lead to many positive outcomes. For instance, online social interaction helps young adults maintain connections to a community, leading to social capital and well-being (Ellison et al., 2007). These interactions also increase social connectedness and improve mental health (Grieve et al., 2013). One study examined a broad range of effective and social outcomes among young adults (Oh et al., 2014) and identified online social interaction as leading to perceived support, positive affect, meaningful interactions, increased perceived community, and greater life satisfaction. Also, among young adults, online social interaction allows individuals to stay connected to previous communities and build networking in their present location (Ellison et al., 2007). Moreover, in 2020, users changed their lifestyles (e.g., social isolation, curfew), and they are separated from each other because of the pandemic affecting the world. However, they try to close this distance by using technology (i.e., video calls and conferencing tools on devices) (Wanga et al., 2020) and diminishing the negative aspects of social isolation through these technologies.

### *2.2.2. Negative effects of smartphone use*

Contrary to the positive effects of smartphone use, there is a lot bigger awareness of the constant connectivity' negative aspects. Thus, this behavior harms users' physical, mental, and social health.

For the physical adverse effects, in a study on excessive smartphone use among university students, 35.9% of the participants suffered from fatigue during the day, and 38.1% experienced a decline in sleep quality (Matar Boumosleh & Jaalouk, 2017). Another study shows that because of the users' posture while they use their devices, they have joint and neck pain (S. Lee et al., 2017). Also, due to excessive texting, smartphones intensify thumb arthritis (Cherry, 2020). As another critical hazard, a study highlights that cell phones emit radiation, affect brain cells, thus mutating existing cells, and cause cancerous cells like brain cancer (Miller, n.d.).

For the effects on individuals' mental health, a study surveying 467 young people about their social media usage habits found that social media use leads to lower sleep quality, lower self-esteem, and higher anxiety levels (Woods & Scott, 2016). The time spent on social media networks may also affect individuals psychologically. Many clinicians observed various mental disorders symptoms, such as anxiety and depression, in people who spend more time on social media (Woods & Scott, 2016). Furthermore, researchers working on social media use found that students who spend more than two hours a day on social networking sites like Facebook or Instagram are more likely to suffer from distress, poor mental health, and even suicidal thoughts (Sampasa-Kanyinga & Lewis, 2015).

A qualitative study with 16 university students revealed that participants described a fear of social exclusion when they separated from their devices (James & Drennan, 2005). One of the studies found (Cheever et al., 2014) that participants separated from their smartphones reported increased anxiety over time. Similarly, another study (Clayton et al., 2015) found that restricting participants from answering their ringing iPhone while performing a cognitive task resulted in diminished performance on the task, higher reported levels of anxiety, and even physiological effects such as increased heart rate and blood pressure.

In addition to this psychological and physical well-being, smartphones have negatively influenced our social relations and face-to-face conversations. Even though users are aware of the effects, they tend to use their devices during social gatherings. However, many studies indicate that excessive smartphone use negatively influences social relations. For example, people become less engaged with their immediate social environment due to heavy smartphone use during social interaction (Brown et al., 2016; Misra et al., 2016; Rotondi et al., 2016; Vanden Abeele et al., 2016). In an empirical study that addresses the impact of smartphone use during dyadic conversations on 238 participants, participants perceived this behavior as less polite and attentive (Vanden Abeele et al., 2016). Another study shows that when smartphone use behavior occurs in interpersonal interaction, the time spent with friends becomes less valuable, positively and significantly related to users' life satisfaction (Rotondi et al., 2016). They enjoyed a meal with their friends less when their smartphones were present. People have tense arousal and boredom because they feel less socially connected and perceive time slower (Dwyer et al., 2018). Even excessive smartphone use is associated with lower relationship satisfaction with the romantic partner (Roberts & David, 2016). Besides using smartphones during social interaction, studies show that the mere presence of a phone on the table (even a phone turned off) changes what people talk about. If we think we might be interrupted, we keep conversations light on topics of little controversy or consequence (Przybylski & Weinstein, 2013). And conversations with phones on the landscape block empathic connection. If two people are speaking and there is a phone on a nearby desk, each feels less connected to the other than when no phone is present (Misra et al., 2016).

Also, the study of (Misra et al., 2016) involves a naturalistic field experiment with 100 dyads that shows that people who have conversations without mobile devices reported higher levels of connectedness and empathy than those who simultaneously use mobile devices. Spending less time with friends means less time to develop social skills. Another study found that sixth-graders who spent just five days at a camp without using screens became better at reading emotions on others' faces, suggesting that new generations' screen-filled lives might cause their social skills to atrophy (Uhls et al., 2014). This being so, an in-person conversation led to the most emotional connection, and online messaging led to the least (Sherman et al., 2013).

In our previous study (Genç & Coskun, 2020), we conducted six focus group sessions with 46 participants. We found that smartphone-checking behavior in a social gathering makes the companions feel negative emotions (e.g., they feel anger, offended, bored, and even worthless). Affirmingly, several experiments show that texting while socializing with another person hurts the perceived conversation quality. Participants considered a conversation with a person who uses a smartphone during the interaction as lower in quality (Vanden Abeele et al., 2016), especially when we remember the breadth and depth of face-to-face communication (Knop et al., 2016). For example, phubbing - the act of snubbing someone in a social setting by concentrating on one's mobile phone negatively affects the perceived communication quality and relationship satisfaction (Chotpitayasunondh & Douglas, 2018; Roberts & David, 2017).

## 2.3. *Solutions to Mitigate Excessive Smartphone Use*

As we stated before, as well as the benefits, smartphones negatively affect people's physical, mental, and social health. This issue brings the interventions aiming to mitigate these adverse effects into prominence for both academics and practitioners. Most market interventions that deal with these problems are currently confined to digital apps for smartphones, web browsers, or computers. A study that analyzed 367 apps for "digital self-control" (i.e., setting use limits for apps and devices) on Google Play, Chrome Web Store, and Apple App Store (Lyngs et al., 2019) provides four main categories for the interventions. 74% of the apps are the most common feature category, which involves blocking or removing distractions. The second category is "self-tracking" (i.e., tracking the amount of time spent on devices) which was used in 38% of apps. The third most common feature (35%) is "goal advancement," which aims to guide users toward the right tasks when using their smartphones by setting time/task goals and reminders. The final most common feature is "reward/punishment," which involves gamification and representation of 'points' gained through the amount of their device and app use. In addition to this study, these findings are supported by a recent review focused on 42 digital well-being smartphone apps from Google Play Store (Monge Roffarello & De Russis, 2019). Their research shows that most state-of-the-art digital well-being apps are not targeted toward enabling users to form new habits but are designed to break existing unwanted habits.

Similar to market interventions, excessive smartphone use has recently attracted researchers working in this field. For example, AppDetox (Löchtefeld et al., 2013) and The SAMS (H. Lee et al., 2014) are two mobile applications that allow users to set rules for the applications they want to use less. Let's FOCUS (Kim et al., 2017) aims to reduce phone use in classrooms by giving context-aware reminders to students. Unlike these solutions focusing on individuals and their intention to regulate their behavior, Lock n' lol (Ko et al., 2016) and NUGU (Ko, Chung et al., 2015) aim to reduce smartphone use by restricting group members' smartphone use time. They use group-limiting mode to limit the application use and to mute notification alerts. Unlike Lock n' lol, NUGU allows users to share their limiting time schedules and contexts that they are willing to limit their smartphone use (i.e., studying, working, etc.).

Similarly, FamiLync (Ko, Choi, et al., 2015) turns use-limiting action into a family activity. All family members' smartphone use statistics are shown in a dashboard that provides social awareness of smartphone use. These examples commonly use strategies restricting users' smartphone use via either individuals' motivation or social facilitation.

Contrary to these, in recent years, the inefficacy of restrictive approaches has aroused attention from HCI researchers. They started to develop solutions that followed a different path in terms of giving feedback about the excessive use of smartphones. For instance, LockDoll (Choi & Lee, 2016) is a doll that notifies group members according to phone use by ambient light and by waving its arm. SCAN (Park et al., 2017) monitors the interaction between group members through built-in sensors and defers notifications until it detects breakpoints in interactions like a moment of silence. However, the need for an overall understanding of the field still merits further investigation in developing effective interventions to mitigate mobile devices' effects.

Our analysis of the related work showed that banning or restricting smartphone use behavior is not the best option to mitigate this use behavior's negative effects. Due to their entanglement with daily life, it would not be easy to convince users to establish responsible use practices without exploring their smartphone use behavior from a holistic and bottom-up approach. (i.e., responsible use of technology).

## 2.4. *Digital Well-being*

Mobile technologies, especially smartphones, have become ubiquitous, and people have started to handle most of their work through these devices. The amalgamation of the digital and physical experiences means that individuals gradually have difficulty understanding the borders and lose control of their interactions with the digital world. In the past years, with the importance of users' relationship with technology, the companies like Facebook, Apple, and Google to introduce tools that help people mitigate their digital exposure (Lyngs et al., 2019) (e.g., Apple's Screen Time[1] to track the amount of time for smartphone and application use and Google's Digital Well-being Experiments[2] such as covering users' devices with a paper envelope to limit the use habit from a critical perspective). These attempts are named *Digital Well-being* in the literature and have become an increasingly important trend in both research and industry.

Digital well-being is an emerging term. Although there are attempts to define this term

---

[1] https://support.apple.com/en-us/HT208982

[2] https://experiments.withgoogle.com/collection/digitalwell-being

(Cecchinato et al., 2019; Shah, 2019), there isn't a well-established agreement on its definition. In general, digital well-being covers the effects of technologies on people's mental, physical, and emotional health. It is often defined in terms of the capabilities and skills that an individual requires to use digital technologies successfully (Shah, 2019) (e.g., being aware of the effect and self-controlling the use habit). This can include recognizing the impact of being online on individuals' emotions, mental well-being, and even on their physical health and knowing what to do if something goes wrong in their digital experiences. The choices can influence users' online digital well-being, the content they see, the interactions they have with others, and even how long they spend engaging with technology and the internet.

JISC recently revisited its definition of digital well-being to capture some of its complexities (Shah, 2019). They have now broadened the scope to focus on the individual and broader societal perspectives (e.g., situations in communities, families, and friends). Thus, this extended definition of Digital Well-being covers exploring and managing digital technologies' impact on people's social well-being (i.e., supporting dyadic and group interactions, reducing social isolations, maintaining relationships and connections with friends, family, and others).

## 2.5. *Digital Well-being for Social Interactions*

In general, this study aims to explore how smartphones mediate social interactions and identify strategies and solutions to responsible smartphone use behavior during these interactions. In other words, it focuses on the social side of digital well-being.

Social well-being, along with physical and mental well-being, contributes to good health. It has been identified by (World Health Organization [WHO], 2006) as a central component of individuals' overall health. In addition to identification, according to self-determination theory, people need to experience a sense of belonging and attachment to other people (i.e., relatedness) to achieve psychological growth (Deci & Ryan, 2008). We live together in groups, clustering in cities and towns with families or friends. Most people spend very few of their waking hours alone. Wanting to feel connected and be around other people is a natural impulse. Individuals' health can be fundamentally influenced by the quantity and quality of their support networks and social connections.

There are numerous situations (e.g., talking with a beloved one, sharing experiences with friends) in everyday life where social interaction would be beneficial, emotionally pleasing, or otherwise desirable. With this positive effect, at the same time as non-existent or insufficient social interaction would be problematic for individuals. Many empirical studies have analyzed the impact of social interactions, social trust, and community cohesion on individual well-being.

For example, a study showed that people feel happier when interacting with close others (Venaglia & Lemay, 2017). Another study (Mueller et al., 2019) found that people tended to feel happiest after interactions with friends, followed by interactions with family members, others, and colleagues. Another study that obtains both self- and observer-reports of social interactions (Sun et al., 2019) showed that people report feeling happier and more socially connected when they spend more time interacting with others. Also, people who report that their relationships are more satisfying and supportive tend to have greater well-being (Lyubomirsky et al., 2005).

Face-to-face conversations are essential to our relationships, our creativity, and our capacity for empathy. While dashing off an email, text, or social media post might give us instant communication and the illusion of connection, it's a real-life conversation that really connects us and gives us valuable social support (Turkle, 2016). Talking to friends, colleagues, and family members to share information, give or receive advice, or just to get perspective helps build rapport, foster a feeling of belonging, increase resilience, and helps us to process things and avoid overwhelm. Face-to-face conversation leads to greater self-esteem and an improved ability to deal with others (Pea et al., 2012). Even small talk is good for your well-being. A study (Ybarra et al., 2011) by the University of Michigan found that short-term face-to-face conversations about the weather or other pleasantries can actually improve cognitive functions in the same way that brain-teaser exercises do. Another research has shown that humans need to communicate with others because it keeps them healthier. There has been a direct link to mental and physical health. For instance, it has been shown that people who have cancer, depression, and even the common cold, can alleviate their symptoms simply by communicating with others. People who communicate their problems, feelings, and thoughts with others are less likely to hold grudges, anger, and hostility, which in turn causes less stress on their minds and their bodies (Wrench et al., 2020).

We can argue that social well-being, which includes social interactions with other people (e.g., family, friends, colleagues, etc.), is essential for individuals' general well-being. However, as technology evolves and mobile technologies become one of the main components of daily lives, their effects on users' social well-being as we mentioned in the previous sections, also increase drastically. They negatively influence social interactions, such as damaging intimacy and connection between friends, reducing conversation quality (Misra et al., 2016; Sprecher et al., 2016), and making companions feel awkward and excluded in social settings (Humphreys, 2005).

In sum, digital well-being for social interactions is a recent part of the Digital Well-being term and covers the impact of technologies and digital services on users' social health. This being the case, exploring the social side of digital well-being is underexplored. To gain an in-depth understanding of this side, there is a need to explore the dynamics between mobile technologies (e.g., smartphones) and interpersonal interactions.

## *2.6. Solutions to Enrich Collocated Interactions*

Contrary to these use-limiting concepts, the inefficacy of restrictive approaches has aroused attention from HCI researchers in recent years. Previous research investigated how to use technological devices to enhance collocated interaction. These enhancements can be grouped into different categories: "facilitating ongoing social situations, enriching means of social interaction, supporting a sense of community, breaking ice in new encounters, increasing awareness, avoiding cocooning in social silos, revealing common ground, engaging people in collective activity, encouraging, incentivizing or triggering people to interact." (Olsson et al., 2020).

A study (Jarusriboonchai, Memarovic, et al., 2014) mentions that even when people are physically collocated, they can create "cocoons" or bubbles using mobile devices that might reduce their collocated social interactions. The researchers developed PicoTales (Robinson et al., 2012) to overcome this problem, a storytelling device that allows people to co-create stories while collocated. The prototype consists of a projector and a phone to create a shared experience where people can project simple sketches to continue the story. Unlike these examples that trigger users to interact with each other using mobile devices, some studies give feedback about users' social interaction. Conversation Clock (Bergstrom & Karahalios, 2007) is a table that visualizes the auditory input in face-to-face communication. It provides visual feedback to the users about their conversations and allows them to observe their contributions to the conversation.

FishPong (Yoon et al., 2004), for instance, is a collaborative and cooperative interactive game designed to serve as an icebreaker, enhancing people's social interaction. Cuesense (Olsson et al., 2015) is a wearable display that shows some of the user's social media content related to the person encountered. It is designed to increase awareness and be an icebreaker in first encounters. Similarly, BubbleBadge (Falk & Björk, 1999) is a textual display that provides supplementary information to enhance collocated social interactions, which is worn like a brooch. The information displayed by BubbleBadge can break the ice in new encounters and trigger interactions in the later phases. Another study explored ways to enhance social interaction between strangers with Social Devices which have audio-based interfaces (Jarusriboonchai, Olsson, et al., 2014). These devices start to talk to each other and users during social gatherings to improve social interaction. They interact with users by asking questions or giving them random topics (e.g., movies and plans for the rest of the day).

## 2.7. *Summary of the Related Work and the Gap*

In the literature, there are many alternative solutions; those are valuable in terms of expanding the design space for solutions mediating excessive smartphone use. However, most of the interventions in research and industry follow similar approaches. First, their target is the physical and psychological effects of the problem on personal health. Their main features involve restricting, goal setting, reminding, and reward/punishment mechanisms (i.e., interventions which follow top-down approaches). Under these circumstances, as suggested in a workshop conducted in CHI 2019 (Cecchinato et al., 2019), digital well-being interventions should move beyond a focus on restricting and showing screen time approaches. Parallel with this direction, JISC, also recently updated the term Digital Well-being, and they extended the guide to make the term cover the social side of the problem.

In light of all these previous studies and emerging trends, we can say that mobile technologies ease and improve our lives in terms of many aspects despite their negative effects. Recently, during the pandemic, people have used their mobile devices to reduce the impact of social isolation. Thus, we need alternative methods beyond restricting smartphone use. With these motivations that we stated above, this study explores;

- The smartphone-use behavior of individuals during social interactions,

- Designing interventions to generate solutions that can support social interactions without necessarily restricting smartphone use,

- Identifying implications for designing such solutions

# Chapter 3

# UNDERSTANDING THE PATTERNS, MOTIVATIONS, AND FEELINGS OF USERS

**Research Questions**

**RQ1** How do people conceptualize the responsible use of smartphones in collocated social interactions?

**RQ2** How do people use their smartphones during a social interaction?

**Related Paper**

**NordiCHI 2018** - Are we "really" connected? understanding smartphone use during social interaction in public

As smartphones have become ubiquitous devices, there are many contexts in which we can see the negative effects of their excessive use. In this chapter, we focus on examining their negative effects on daily social interactions. In the scope of this understanding, we propose an alternative approach to deal with the problem of excessive smartphone use. We illustrated this approach with a study on smartphone use during social interaction occurring in public settings, including observations and focus groups. From this research step, we revealed an overall understanding of the users' context and two themes related to users' feelings, strategies, and insights about smartphone use.

## 3.1. Observations

We first made unstructured observations in four different coffeehouses to identify the behaviors pertaining to smartphone use during social interactions. In these observations, we noticed that people frequently check their phones during a conversation. This behavior creates silent moments in conversations and causes people to give superficial and delayed answers to questions.

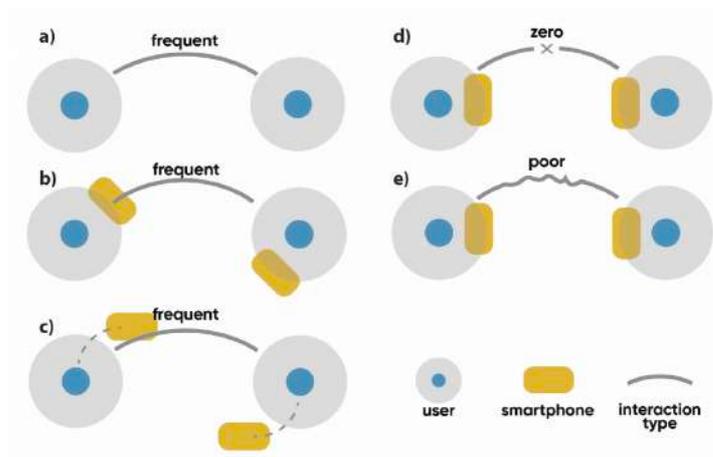

*Figure 3.3 Illustration of the cafe observations*

These observations can be summarized in the same order with the Figure 3.3 as;

a. Users, who are not using their smartphones, maybe more engaged in social interaction because the absence of smartphones creates less distraction.

b. Smartphone use may provide a medium for interaction. For example, we observed that two participants tried to take the best photo of the coffee and scenery by arranging the items on their desks together.

c. Users may use their smartphones to enrich their interactions. Some individuals showed digital content on their smartphones to their companions. (e.g., internet memes and received messages). We observed that the interaction among group members becomes frequent and longer in such cases.

d. Smartphone use may end the users' interactions with each other. We observed that as users started to use their phones, their communication with others was broken down.

e. Smartphone use may make users' responses delayed, superficial, and short. We observed that users' attention to their social interactions might be distracted by their smartphones. For example, notifications and online conversations influence the quality of social interactions.

## *3.2.  Focus Groups*

These general results showed that phone checking might seriously impact the quality of social interaction and, thus, motivated us to gather deep insights about this behavior and its effects. We conducted three focus group sessions with participants (24 in total) representing three different age groups (ages between 18-25, 26-40, and 41-55) to deepen the understanding of smartphone use behavior. We explored the situations that trigger smartphone use during social interaction, people's motivations for using and not using their smartphones, as well as their feelings and reactions toward people who constantly interact with their phones instead of engaging in conversation. We identified two overarching themes based on this exploration.

**Theme 1: Staying in the moment makes phone-checking behavior tolerable**

Although participants who use their smartphones in social settings acknowledged the distractions they lead to, they said that they tend to justify this behavior by giving a valid reason, for example, stating the importance of a message. One interesting finding was that the phrase "I am with you" is many times more than enough justification for the interruptions in interactions. For many participants, being or not being with the person next to them is the key factor in tolerating phone-checking behavior.

**Theme 2: Phone checking is not the cause, but the result of the lulls in conversations**

Another provocative finding is that participants opposed the idea that smartphone use kills conversations. They claimed that the phone is not the cause of the lulls in conversations; rather, it is the result of these lulls. In other words, people tend to start using their phones when there is a lull in a conversation. They further stated that smartphones could even be used to deal with lulls in conversations. Smartphones may fire up a conversation by giving a subject to talk about, such as presenting content from social media networks.

Paper 1:

# Are We 'Really' Connected? Understanding Smartphone Use During Social Interaction in Public

# Are We 'Really' Connected? Understanding Smartphone Use During Social Interaction in Public

**Hüseyin Uğur Genç**
Koç University – Arçelik Research
Center for Creative Industries
34450 İstanbul, Turkey
hgenc17@ku.edu.tr

**Aykut Coşkun**
Koç University – Arçelik Research
Center for Creative Industries
34450 İstanbul, Turkey
aykutcoskun@ku.edu.tr

**Fatoş Gökşen**
Department of Sociology
Koç University
34450 İstanbul, Turkey
fgoksen@ku.edu.tr

**Abstract**
Excessive smartphone use has negative effects on our social relations. Previous work addressed this problem by allowing users to restrict their smartphone use. However, as this strategy requires users to have high levels of self-regulation, it may not be effective for individuals without an explicit intention to change their behavior. We propose an alternative approach to this problem, i.e. identifying ways of reducing smartphone use without restricting its use. We illustrated this approach with a study examining smartphone use during social interaction in public settings. Based on four unstructured observations in different coffeehouses and three exploratory focus groups with different age groups, we identified two themes in relation to smartphone use in public settings and discussed their implications for designing solutions that aim to enrich social interaction without limiting smartphone use.

**Author Keywords**
Smartphone use; social interaction; design for behavioral change; focus group

**ACM Classification Keywords**
H.5.m. Information interfaces and presentation (e.g., HCI): Miscellaneous;

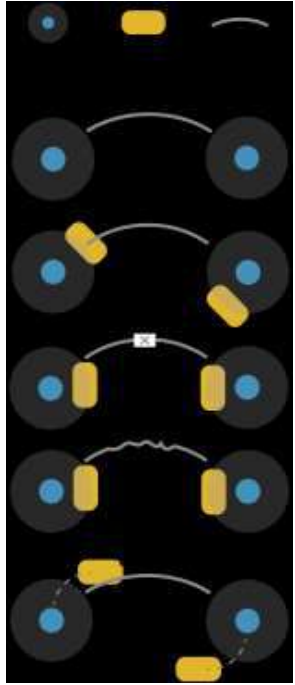

**Figure 1.** Social Interactions in coffeehouse observations

a) Users who are not using smartphones engage in an interaction.

b) Smartphone use provides a medium for an interaction.

c) Smartphone use ends the users' interactions.

d) Smartphone use makes user's responses delayed, superficial and short.

e) Users use smartphones to enrich their interactions.

**Introduction**

Smartphones have become inseparable parts of our daily lives. We are always connected; constantly checking e-mails, shopping, streaming videos, surfing the net and so on. Despite the various benefits it offers, this "being always connected" situation can have negative effects on our mental health [16,20], physical health [20] as well as our social relations [10,18].

HCI literature provides studies aimed at reducing these negative effects. Majority of these studies, however, commonly preferred strategies helping users to restrict their smartphone use. For example, AppDetox [12] and The SAMS [11] are two mobile applications which allow users to set rules for the applications they want to use less. Let's FOCUS [6] aims to reduce phone use in classrooms through giving context-aware reminders to students. Unlike these solutions focusing on individuals, Lock n' lol [8] and NUGU [9] aim to reduce smartphone use by restricting group members' smartphone use time. In a similar vein, FamiLync [7] turns use-limiting action into a family activity so that each family member can participate in this activity. Others followed a subtler approach in terms of giving feedback about excessive use. For instance, LockDoll [3] is a doll that notifies group members according to phone usage by an ambient light and by waving its arm. SCAN [14] monitors the interaction between group members by built-in sensors, and defers notifications until it detects break-points in interactions like a moment of silence.

Although these studies claimed that they decreased excessive smartphone use, the strategies used to influence user behavior has two limitations. First, behavior change strategies allowing users restrict their smartphone use can only work when 1) the users are aware of the negative effects of their use-behavior, 2) have an intention to change this behavior [1] and 3) have high levels of self-regulation to maintain this change [21]. Thus, such strategies may fail to mitigate smartphone use of individuals with low level of self-regulation, who are more likely to be addicted to their smartphones than individuals with high level of self-regulation [4,5]. Second, using these strategies may even create unintended outcomes. For example, the results of a recent study showed that teens, who had to limit their social media use involuntarily, experienced negative feelings and increased the time they spend in social media after the break period was over [2].

We propose an alternative approach to deal with the problem of excessive smartphone use. Inspired by practice-based design [17], our approach aims to investigate the practice of smartphone use and identify ways of reducing its use without restriction. We illustrated this approach with a study on smartphone use during social interaction occurring in public settings. To understand this practice, we first made unstructured observations in four coffeehouses and then conducted three focus group sessions with participants representing three different age groups (Table 1). We explored the situations that trigger smartphone use during social interaction, people's motivations for using and not using their smartphones, as well as their feelings and reactions towards people who constantly interact with their phones instead of engaging in the conversation. We identified two themes based on this exploration, and discuss implications for designing interactive solutions that enrich the quality of social interaction without limiting smartphone use.

| Groups (Age Range) | Number (Sex) | Mean of Age *(SD)* |
|---|---|---|
| Young Adults I (18-25) | 8 (4 Female, 4 Male) | 19.50 (0.75) |
| Young Adults II (26-40) | 9 (5 Female, 4 Male) | 30.77 (4.46) |
| Middle-Aged Adults (41-55) | 7 (2 Female, 5 Male) | 43.7 (4.03) |
| Total | 24 (11 Female, 13 Male) | |

**Table 1.** Age and sex distributions of focus groups

## Observations and focus groups

We made unstructured observations in four different coffeehouses to identify the behaviors pertaining to smartphone use during social interactions. In these observations, we noticed that people frequently check their phones during a conversation. This behavior creates silent moments in conversations and causes people to give superficial and delayed answers to the questions (Figure 1). These observations showed that phone checking may have negative effects on the quality of social interaction, and thus, motivated us to gather deep insights about this behavior and its effects.

We then conducted three focus group sessions to understand motivations behind phone checking behavior during social interaction as well as feelings and reactions towards this behavior. Focus group participants were required to have a smartphone and use at least one mobile app for social media networks (Table 1). We also used age as an inclusion criterion. Our purpose was to discover possible differences and similarities between different generations [19]. Table 2 summarizes the focus group structure. Each focus group was about two hours. We video-recorded each session and transcribed the videos. We analyzed the data through qualitative coding [13].

## Results

Overall, the results indicate that the main reason behind phone checking behavior is the *feeling of curiosity,* i.e. being curious about what's happening in one's social network, which verifies previous work [15]. When their companion constantly checks his/her phone, participants feel negative emotions such as anger, sadness and resentment. They use various tactics to prevent this behavior such as warning verbally, drawing attention through staring, keeping quiet, shaking legs and sighing as well as using physical force (e.g. hitting head with a pillow). Another higher-level observation is that people from different age groups tend to see phone checking behavior differently. While middle-aged participants consider this behavior as a harmful habit and think that it should be changed, young adult participants thought that it has no negative impact on their lives. For them, this was not a big and general problem. They also don't blame smartphones for low-quality social interaction. They mentioned that banning or restricting smartphone use makes the phone more attractive, supporting previous work [2]. Besides these general observations, we identified two themes.

*Theme 1: Phone checking is not the cause, but the result of lulls in conversations*

Participants mentioned that lulls occur occasionally in conversations. Different age groups reacted differently to lulls. Young adults tolerate these silent moments more than the middle-aged adults do. While young adults don't mind sitting silently and interacting with their phones during a meeting, middle-aged participants tend to end a meeting when the number and the duration of lulls increases. Furthermore, even though all the participants regarded phone checking as an unpleasant behavior, they agreed that the phone is not the cause but the result of lulls in conversations.

Although most of the participants opposed to the idea that smartphone use ceases the conversation, they claimed that smartphone use has both negative and positive effects on the conversation quality. Many said that smartphones may fire up a conversation by giving a subject to talk about, e.g. showing a content from social media networks. But, each group agreed that since smartphones create infinite materials to talk about, they also make the subjects superficial.

> *"...Although there are more topics in social media, they seem superficial. For example, the thing we laughed three months ago does not make me laugh now. I forget it. There is nothing left, it feels like nothing has enriched me."* –Female, Young Adults I

Another finding related to this theme was participants' tendency to isolate themselves from the group and use their smartphones for this purpose. This behavior occurs when their degree of intimacy with their companion is not high, or the conversation is not interesting. They also indicated that they tend to use smartphones for isolating themselves when lulls and silent moments happen in conversations.

*Theme 2: Staying in the moment makes phone checking tolerable*

Participants reported that they tend to use their smartphones in social settings despite the distractions it leads to. They admitted that this behavior would be unpleasant for other individuals in a meeting. They told that they often justify this behavior by giving a valid reason such as an urgent call from a parent. One interesting finding across all age groups is using the phrase "*I am with you*". This phrase seems to convey the message that the user is aware of the social convention she/he is violating and compensates the situation by acknowledging or emphasizing his/her presence in that moment. For many, the staying or not staying with the person next to them is the key factor to tolerate phone checking behavior, and this clarification attempt is seen as a sign of respect to this person. The following quote illustrates this,

> *"I got a very important news, right in the moment I met with my high school teacher, whom I had not seen for two years. 'Give me a second, something happened. After I inform my friends, I will be with you.', I said. After several phone calls and messages, I explained: 'I might check the phone sometimes. Not because I'm less interested in you, but because this is important.' When I make these explanations to indicate my 'being with her.' I feel better."* –Female, Young Adults I

**Discussion**

In the remainder of this poster, we revisit the findings and discuss how they can be used to design technologies aimed at reducing the negative effects of smartphone use on social interaction without limiting its use. The results showed that people tend to keep using their smartphones during social interaction although they are aware of its negative impact. The results also indicated that this use behavior changes according to three dimensions; the type of distraction, the quality of communication and the quality of moment. In this poster, we propose three design implications in line with these dimensions.

The most frequently mentioned reasons for using a smartphone in social settings were participants' curiosity about their social media accounts and their need to isolate themselves when they find the topic of a conversation or a person boring. To address this problem without limiting smartphone use, we propose *Augmented Interactions*, i.e. augmenting objects in a social setting with new affordances when each person manage to "stay in the moment". For instance, when the quality of communication among a group of friends sitting in a café reaches a level, the desk might start to keep a tea hot or a beer cold. If the quality is getting higher, the chair might provide even more comfortable experience. This would raise the group's awareness of communication quality as well as motivate them to stay in the moment for a more rewarding experience.

All of the participants accepted that phone checking behavior provokes negative emotions and that they tend to deal with it by using various personal tactics. Plus, all of them stated that this habit is acceptable when it is justified with a reason such as "Mom is calling". Assigning custom ringtones for certain contacts could be a potential solution to this. However, as also indicated in one of the focus group sessions, this would not respond to the complex structure of social interactions. To address this problem, we propose *Smarter Filters*, i.e. deciding whether a distraction is worth looking based on information about the environment, occasion, time of the day, presence of specific person in the setting and so on. In such a situation, how this distraction is presented to the user is also essential for helping users stay in the moment. The feedback should be easily recognizable by the user but not too distractive for others, e.g. a light code or a tag that pop-ups on the phone or even on a table to show the importance of the notifications. Such a solution would not only allow the person, who needs to check his or her phone, get important updates but also prevent others from experiencing negative feelings.

Another significant finding was that lulls in a conversation, which were considered as a major problem by adult participants, trigger smartphone use. A simple solution to this problem could be a device that randomly offers topics to talk about. However, as also our participants stated, such a solution would create an artificial conversation. To address this problem, we propose *Pop-up Memories*, i.e. raising a question based on common interests or memories of a group by using their online data. This concept would lead to a more realistic experience since trying to answer the popped-up question would provide members with more natural subjects to talk about.

## Conclusion

We presented the results of an exploratory study aimed at examining smartphone use in public settings and identifying ways of reducing its use without restriction. We revealed general observations, identified two specific themes and discussed the implications for design. In the future, we plan to develop conceptual prototypes by using these implications and assess their impact on smartphone use during social interaction.

# Chapter 4

# ADVANCING THE UNDERSTANDING & EXPLORING THE DESIGN SPACE

**Research Questions**

**RQ1** How do people conceptualize and manage digital well-being for social interactions, in relation to smartphone use?

**RQ2** How do people use their smartphones during a social interaction?

**RQ3** How can we design interventions that support digital well-being for social interactions (without necessarily restricting smartphone use)?

**Related Paper**

**CHI 2020** – Designing for Social Interaction in the Age of Excessive Smartphone Use

This chapter advances the understanding that we gained from the observations and the focus group sessions. In the scope of this understanding part, we conduct more focus group studies and revisit the previous data while analyzing the last sessions. This helps us to gain a deeper understanding of smartphone use in public settings and extend our themes. In addition to these focus group sessions, we conducted workshops to explore design space by using the knowledge of the previous studies with professional designers and revealed design approaches to help practitioners while designing for the context of smartphone use.

## 4.1. Focus Groups

We conducted three focus group studies with the same participant demographics. Thus, by increasing the number of focus group sessions to six, we received the opinions of 46 participants in total. Our goal was to further our understanding of users' smartphone use and provide a more comprehensive analysis. With this analysis, we added three themes more on top of our two themes in [Chapter 3](Chapter 3) as follows;

**Theme 3: Feeling of curiosity drives phone-checking behavior in social settings**

Participants are aware of the negative effects of smartphone use; however, they use their devices in social settings to share something and to check notifications or recent updates on their social media accounts. They stated that social media helps them connect with their circles without any need for physical presence. Losing this connection, i.e., being unable to follow what their friends are doing, creates a fear of missing out (Przybylski et al., 2013). They reported that they oftentimes get stressed because of this feeling.

**Theme 4: Checking smartphone in a social gathering makes the companion feel negative emotions**

Contrary to the previous theme, participants accepted that when a person checks his or her phone, they feel angry, offended, upset, bored, and even worthless. All participants agreed that smartphones should not be used while talking about important things, such as counseling a friend after a breakup. They revealed that the frequency and duration of phone-checking behavior are crucial in feeling these emotions. If the checking behavior is frequent and long, most of the participants said that they use various tactics to prevent this behavior, such as verbal warning, irony, physical coercion, and ending the meeting.

**Theme 5: Age groups have different reservations about phone-checking behavior in social settings**

We observed that people from different age groups tend to perceive and react to phone-checking behavior differently. While older generations consider this behavior as a harmful habit and think that it should be modified, young adult participants thought that it has no negative impact on their lives. For them, this was not a big and general problem. Also, young adults tolerate silent moments more than middle-aged adults do. While young adults do not mind sitting silently and interacting with their phones during a meeting, middle-aged participants tend to end a meeting when the frequency of pauses increases.

## *4.2.* **Design Workshops**

With the understanding of the smartphone use behavior that we gained from our observation and focus group studies, we conducted three workshop sessions with 15 professional designers to identify design directions that can be used when exploring the solutions for mediating smartphone use in public settings without banning smartphones. By using this knowledge and the no-ban rule, designers developed 67 ideas in total (Appendix 1), and we identified these ideas into four alternative design approaches (Figure 4.4) as follows;

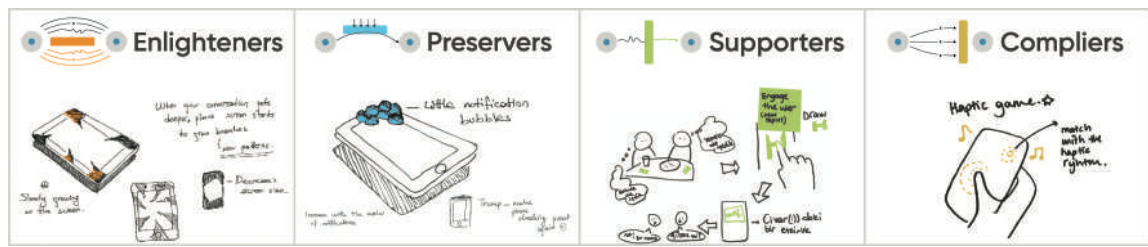

*Figure 4.4 Design approaches and example sketches from Design Workshops*

**Enlighteners:** Focus groups participants stated that they might get lost in their devices once they start to use their smartphones during social interactions. As a result, they become less engaged in the conversation. The ideas under this approach aimed at creating awareness about the quality of social interaction by informing the users. One strategy is to provide physical changes, e.g., a smartphone case that changes its color or heat according to the frequency of smartphone use. The second strategy was making the feedback provided to the user more meaningful, e.g., emphasizing the value of time by informing the users what they can do in real life with the time they spend on their smartphones.

**Preservers**: The ideas under this approach focus on preventing the triggers of phone-checking behavior in social settings, thus preserving the ongoing social interaction between users. Designers, who stated that a large number of notifications and their irrelevancy to the situation were perceived as distracting as they interrupted the conversation, developed smart filters for the management of notifications. The second strategy was making regular notifications as unobtrusive as possible. Designers, who emphasized the importance of eye contact during a meeting observed in the focus group, developed ideas for controlling notifications without interacting with the smartphone. The third strategy was making the behavior of checking notifications more difficult by adding a playful layer to this kind of intervention by using meaningful design frictions (Cox et al., 2016).

Supporters: All designers stated that they consider the amount of speech and the depth of the subject as the variables that determine the quality of the conversation. The ideas developed from this point of view aimed to increase the quality of the conversation among users and help avoid strange lull moments. The first strategy was boosting the conversation. They are based on collecting data about users' interests and presenting the common ones as the proposed topic to initiate a conversation between group members. The second strategy was encouraging group members to find something to talk about themselves rather than suggesting an existing topic. The strained attempts at a conversation would cause artificiality which was also observed in focus groups and tried to overcome this problem by asking questions to the users instead of giving them a direct topic.

Compliers: As identified in the focus groups, users may have the desire to isolate themselves by using their smartphones in social environments. In light of this point, ideas under this approach aimed at providing smooth isolation for users in relation to social interaction. One strategy followed by designers was making the need for isolation highly visible to other group members. Designers thought that this would help users understand one's need for isolation in an explicit way. Another strategy was physically separating the isolated user from the environment to make him or her aware of the consequences of this isolation.

Paper 2:

# Designing for Social Interaction in the Age of Excessive Smartphone Use

# Designing for Social Interaction in the Age of Excessive Smartphone Use

**Hüseyin Uğur Genç**
KUAR, Koç University
İstanbul, Turkey
hgenc17@ku.edu.tr

**Aykut Coşkun**
KUAR, Koç University
İstanbul, Turkey
aykutcoskun@ku.edu.tr

**ABSTRACT**
Excessive smartphone use has negative effects on our social relations as well as on our mental and psychological health. Most of the previous work to avoid these negative effects is based on a top-down approach such as restricting or limiting users' use of smartphones. Diverging from previous work, we followed a bottom-up approach to understand the practice of smartphone use in public settings from the users' perspective. We conducted observations in four coffeehouses, six focus group sessions with 46 participants and three design workshops with 15 designers. We identified five themes that help better understand smartphone use behavior in public settings and four alternative design approaches to mediate this behavior, namely *enlighteners, preventers, supporters,* and *compliers*. We discuss the implications of these themes and approaches for designing future interactive technologies aimed at mediating excessive smartphone use behavior.

**Author Keywords**
Design for behavioral change; Smartphone; Focus group Design workshop

**ACM Classification Keywords**
• Human-centered computing ~ Ubiquitous and mobile computing • Human-centered computing ~ Human computer interaction (HCI)

**INTRODUCTION**
Smartphones accompany us in every aspect of our lives. We have become more dependent on them for almost every daily task, from making a payment to connecting to our homes. While smartphone applications offer promising ways to ease our lives and even to prevent and treat chronic diseases such as diabetes [5] or alcoholism [16], their overuse may lead to physical and mental health problems. These include sight problems, joint pain and neck pain symptoms [30], sleep disturbances and depression [31], and smartphone addiction [29,32]. Furthermore, excessive smartphone use negatively influences our social interactions such as damaging the level of intimacy and connection between friends, reducing conversation quality [36,42], and making companions feel awkward and excluded in social settings [20].

Previously, HCI researchers explored this problem in the scope of behavior change. They offer solutions aimed at reducing the negative effects of excessive smartphone use. However, majority of these solutions commonly benefit from strategies helping users to restrict their smartphone use. For example, providing users with an opportunity to set use-limits for themselves [25,29,33] or setting use-limits to each other via a mobile app [26,27], and reminding excessive use through visual and haptic feedback [8,38]. Diverging from this work, we aim to identify alternative ways of mediating excessive smart phone use without necessarily restricting it. We believe that a wider exploration of this solution space [11] from the users' perspective would make a relevant contribution to the field. It would provide researchers and practitioners working on this problem with more approaches from which they can select according to different behavioral change contexts [2] (e.g., different characteristics of target users, situational factors and so on).

As smartphones have become ubiquitous devices, there are many settings that we can see the negative effects of their excessive use. In this work, we focus on examining the negative effects on daily social interactions. We selected public settings as a case, because it would allow us to understand these interactions easily. In the scope of this paper, we refer a public setting as a physical location where a social group (family, friends, colleagues and so on) meet and spend time together.

Inspired by practice-based design [47], we utilized various data collection techniques to better understand the practice of smart phone use in public settings and identify potential design directions for mediating this behavior. We first made unstructured observations in four coffeehouses. Then, we conducted six focus group sessions with 46 participants representing three different age groups (Table 1). In these observations and focus groups, we explored

- the situations that trigger smartphone use during daily social interactions,
- people's motivations for using and not using their smartphones during these interactions,

- their feelings and reactions towards people who constantly interact with their phones instead of engaging in the conversation.

Based on this exploration, we revealed five themes that provide an account of people's smartphone use behavior in public settings. Then, we conducted three design workshops with 15 designers whose task was to generate solutions for mediating smartphone use in social settings by using these themes as a reference. After analyzing designers' solutions, we identified four alternative design approaches:

- *Enlighteners*: creating awareness about the quality of social interaction by informing the users.
- *Preservers*: aiming to mitigate the triggers of phone checking behavior in social settings.
- *Supporters*: increasing the quality of the conversation among users and helping avoid the strange lull moments.
- *Compliers:* providing smooth isolation for users in relation to social interaction.

We believe that these themes and design approaches can inspire designers and researchers in developing future interactive technologies that help users manage their excessive smartphone use.

### RELATED WORK

#### Excessive smartphone use and its influence on our mental health and social relations

Excessive smart phone use may have adverse effects on our mental health. For example, surveying 467 young people about their social media usage habits during the day and night time, researchers found that social media use leads to lower sleep quality, lower self-esteem and a higher level of anxiety [50]. The time spent on social media networks may also have psychological effects on the individuals. Many clinicians observed the symptoms of various mental disorders such as anxiety and depression in people who spend more time in social media [50]. Furthermore, researchers working on social media use and time spent on these networks found that students, who spend more than two hours of a day in social networking sites like Facebook or Instagram, are more likely to suffer from distress, poor mental health, and even suicidal thoughts [46].

Many studies indicate that excessive smart phone use have negatively influences our social relations. For example, people become less engaged with their immediate social environment, due to heavy smart phone use during social interaction [1,7,36,45]. The time spent with friends becomes less valuable, which is positively and significantly related to life satisfaction [45]. They enjoy a meal with their friends less when their smartphones were present. People have tense arousal and boredom, because they feel less socially connected and perceive time slower [12]. Even, excessive smartphone use is associated with lower relationship satisfaction with the romantic partner [44].

#### Mediating excessive smartphone use via persuasive technology

Persuasive technologies [13], defined as interactive technologies designed to change users' attitudes or behaviors, has gained significant interest from the HCI community within the last two decades. To date, researchers offered design strategies to motivate behavior change [9,14,37], explored the effectiveness of these strategies [24], provided tools and frameworks helping designers to ideate [34] as well as developed prototypes to motivate various behaviors including sustainable (e.g., [23]) and healthy behaviors (e.g., [43]).

Recently excessive smartphone use has also received attraction from researchers working in this field. For example, AppDetox [33] and The SAMS [29] are two mobile applications which allow users to set rules for the applications they want to use less. Let's FOCUS [21] aims to reduce phone use in classrooms through giving context-aware reminders to students. Unlike these solutions focusing on individuals and their intention to regulate their own behavior, Lock n' lol [26] and NUGU [27] aim to reduce smartphone use by restricting group members' smartphone use time. They use group-limiting mode to restrict the application use and to mute notification alerts. Differently from Lock n' lol, NUGU allows users to share their limiting time schedules and, contexts that they are willing to limit their smartphone use (i.e., studying, working etc.). Similarly, FamiLync [25] turns use-limiting action into a family activity. All family member's smartphone use statistics are shown in a dashboard which provides social awareness of smartphone use.

These examples commonly use strategies restricting users' smartphone use via either individuals' motivation or social facilitation. However, these restricting strategies depend on several conditions to be successful. Users should be aware of the negative effects of excessive smartphone use, have tendency to mediate this use behavior [3] and they need to have high level of self-regulation [51] to insist on this behavior change decision. Also, using these strategies may even create unintended outcomes. For example, the results of a recent study showed that teens, who had to limit their social media use involuntarily, experienced negative feelings and increased the time they spend in social media after the break period was over [4].

There are also other solutions that followed a different approach in terms of giving feedback about excessive use. For instance, LockDoll [8] is a doll that notifies group members according to phone usage by an ambient light and by waving its arm. SCAN [39] monitors the interaction between group members by built-in sensors, and defers notifications until it detects break-points in interactions like a moment of silence. Such alternative solutions are valuable in terms of expanding the design space for solutions mediating excessive smart phone use. Having an expanded solution space would allow applying different solutions, comparing their effectiveness in terms of mediating excessive smart phone use, and selecting the most suitable solutions for different behavioral change contexts (e.g. a social gathering or a meeting). From this viewpoint, we conducted three exploratory studies to expand our understanding of the problem of excessive smartphone use and our solution space to deal with this problem.

We started our exploration with conducting unstructured observations in four coffeehouses. Then, we conducted six focus group sessions with 46 participants to better understand the smartphone use as a practice from the users' perspective. After synthesizing the results of these observations and focus groups into five themes, we conducted design workshops in order to identify design directions for mediating smartphone use in public settings. The remainder of the paper first explains the details of each study. As we followed an iterative data collection process, i.e. each stage fed the next one, we present the method and results of each study separately. Then, we will discuss the overall implications of our work for designing interactive technologies aimed at mediating excessive smart phone use in social settings.

**STUDY 1: COFFEEHOUSE OBSERVATIONS**

We made the observations in four different coffeehouses located in a metropolitan city center. Our purpose was to identify the behaviors pertaining to smartphone use during daily social interactions. We made each observation in the afternoon. They approximately took five hours. The observer (the first author) was a non-participant, there was no contact with the observed population. We paid special attention to not to cross eavesdropping borders during observations. Our aim was neither to get an in-depth speech analysis nor listening to private conversations, but it was to gain an overall understanding of people's behaviors via observing gestures, facial expressions, reactions, silent moments, and speech ratios[1]. As a result of this, we were unable to build rapport or ask questions as new information comes up. The observer's overall attitude was to watch the interactions between people and take notes regarding the influence of smartphone use on these interactions.

In these observations, we noticed that people frequently check their phones during a conversation. This behavior creates silent moments in conversations. It causes people to give superficial and delayed answers to the questions of others in the setting. These observations can be summarized as (Figure 1);

a. Users, who are not using their smartphones, may be more engaged in social interaction because absence of smartphones creates less distraction.
b. Smartphone use may provide a medium for interaction. For example, we observed that two participants tried to take the best photo of the coffee and scenery by arranging the items on their desk together.
c. Users may use their smartphones to enrich their interactions. Some individuals showed digital content in their smartphones to their companions. (e.g. internet memes and received messages). We observed that the interaction among group members become frequent and longer in such cases.
d. Smartphone use may end the users' interactions with each other. We observed that as users started to use their phones, their communication with others were broken down.
e. Smartphone use may make user's responses delayed, superficial and short. We observed that users' attention to their social interactions may be distracted by their smartphones. For example, notifications and online conversations influence the quality of social interactions.

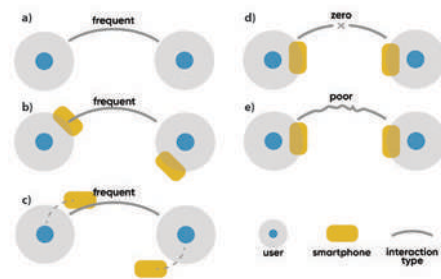

**Figure 1. Identified social interactions during observations**

In summary, these observations showed that phone checking behavior may have negative effects on the quality of social interaction, and thus, motivated us to get deep insights about this behavior and its effects.

**STUDY 2: FOCUS GROUPS**

We conducted six focus group sessions to better understand users' smartphone use patterns in social settings, motivations for using their phones during their social interactions, as well as their feelings and reactions in these situations. 46 participants attended the focus group sessions. All of them have a smartphone and use at least one mobile app for social media networks. We also used age as an inclusion criterion (Table 1). Our purpose was to discover possible differences and similarities between different generations [49].

At the beginning of each session, we asked participants to present themselves by saying their names, ages, occupations, and mobile applications that they frequently use. Then, we asked the following questions;

- *In what kind of situations, you interact with your smartphone and why?*

---

[1] The all three studies did receive an ethical committee approval from Koç University's ethics committee (IRB).

- *In what kind of situations, you don't interact with your smartphone and why?*
- *In these situations, if the person next to you uses his or her smartphone, how do you feel?*
- *How do you react to this behavior?*

For answering these questions, participants were engaged in an individual activity and a group activity. First, they wrote down the answers to post-its separately for five minutes. Second, they grouped all the post-its and extended the initial answers by talking about them for 25 minutes. This structure is inspired from [40]. Our purpose was to gather participants' insights both at the individual and group level.

| Groups (Age Range) | FG1 Number (Sex) | FG1 Mean of Age (SD) | FG2 Number (Sex) | FG2 Mean of Age (SD) |
|---|---|---|---|---|
| Young Adults I (18-25) | 8 (4 Female, 4 Male) | 19.50 (0.75) | 8 (4 Female, 4 Male) | 22.62 (1.79) |
| Young Adults II (26-40) | 9 (5 Female, 4 Male) | 30.77 (4.46) | 7 (3 Female, 4 Male) | 34.3 (4.42) |
| Middle-Aged Adults (41-55) | 7 (2 Female, 5 Male) | 43.7 (4.03) | 7 (4 Female, 3 Male) | 44.6 (3.2) |
| Group Total | 24 (11 Female, 13 Male) | | 22 (11 Female, 11 Male) | |
| In total | 46 Participants (22 Female, 24 Male) | | | |

**Table 1. Age and sex distributions of focus groups**

The duration of each focus group was about two hours. they were video-recorded and transcribed into text, then analyzed by qualitative coding [35]. We first read the transcripts to familiarize ourselves with the data, then we coded them by following a deductive approach, using the questions as categories. Then, we re-coded each category by following an inductive approach, with codes derived from the data. We identified new categories with this approach. After coding the first transcripts separately, we discussed the compatibility of our codes. Then, we continued coding the remaining transcripts with agreed upon codes. This analysis resulted in five themes, which we explain below. It should be noted that, since our purpose is to collect as many different opinions as possible during the focus groups, we did not quantify the number of comments made by each participant.

**Theme 1: Feeling of curiosity drives phone checking behavior in social settings**

Participants are aware of the negative effects of smartphone use. However, they accepted that they use their devices in social settings to share something on social media and to check notifications or recent updates on their social media accounts. The driving force behind this behavior is the feeling of curiosity, which participants mainly associated with their social media accounts. They stated that social media helps them connect with their friends, family and such without any need for physical presence. Losing this connection, i.e. being unable to follow what their friends are doing, creates a *fear of missing out* [41]. They reported that they oftentimes get stressed because of this feeling.

Moreover, they told that the notifications coming from their one-to-one or group conversations in WhatsApp fire up this curiosity. Most of the participants approved that the more the number of notifications (especially successive messages from group conversations) are, the more these notifications stir this feeling. Interestingly, some participants mentioned that they check their smartphones unconsciously. Even though there is no indicator for a notification, they often tend to check whether there is a notification. They highlighted that they perform this behavior more when the phone is on the silent mode.

*"When the phone is on silent mode, I am pressing the home button to open the screen to see if there is a notification to check. 'Is there something to check? Maybe I didn't hear it.' I mean, we have the habit to check the devices unconsciously. This happens very often." – Female, Young Adults II*

**Theme 2: Checking smartphone in a social gathering makes the companion feel negative emotions**

Although participants said that they can use their smartphones during social interaction, they accepted that when a person checks his or her phone, they feel angry, offended, upset, bored, and even worthless. All participants agreed that smartphones should not be used while talking about important things such as counseling a friend after a break-up. They revealed that the frequency and the duration of phone checking behavior are crucial in feeling these emotions. All groups agreed that this behavior can only be justified in situations where using the phone is absolutely necessary such as a call from a family member, school manager or a colleague.

Since each participant agreed that phone checking occurred in a social gathering elicits negative emotions, they use various tactics to prevent this behavior. If the checking behavior is frequent and long, most of the participants said that they usually warn the person performing this behavior verbally. For example, one participant used to warn his friends by ironically saying *"You will exceed your mobile data quota, use mine." (Male, Middle-aged Adults)*. Other participants mentioned that they use "staring", "keeping quiet", "shaking legs" and "sighing" as tactics to draw attention. Also, some said that they shift their focus onto another thing in such situations or they leave the place. The reactions to these breaks in interactions may be more severe. For example, one participant indicated that she used to warn her friends by physical coercion,

*"One of my close friends is addicted to his phone. He uses his phone a lot. Once, I hit his head with a pillow. In another time, I hide his phone. I did this, not because I feel worthless, but he gets lost in there. You know he cares about you but still, he does play with his phone."* – *Female, Young Adults II*

### Theme 3: 'Staying in the moment' makes phone checking behavior tolerable

Several participants reported that they tend to "stay in the moment" by putting their phones away, turning off the mobile data or using silent mode to avoid distractions caused by notifications. They told that the activities such as working, jaunting or playing with a cat can also enable them to forget their phones entirely.

Others reported that they tend to use their smartphones in social settings although they acknowledged the distractions they lead to. However, these participants admitted that this behavior would be unpleasant for the other individuals in a meeting. They told that they tend to justify this behavior by giving a valid reason, for example, stating the importance of a message. One interesting finding across all age groups was that the phrase "*I am with you*" is many times more than enough justification for the interruptions in interactions. For many participants, the being or not being with the person next to them is the key factor to tolerate phone checking behavior. Moreover, this clarification attempt is seen as a sign of respect to the other person.

*"One day, I saw my high school teacher whom I had not seen him for two years. Right at the moment, I got a very important message which I needed to tell other people. I said to him 'Give me a second, something happened. After I write it, I will be with you.'. After several phone calls and messages, I explained: 'I might check the phone sometimes. Not because I'm less interested in you, but because this is important.' I do these explanations to indicate my 'being with the person in front of me.' I feel better."* –*Female, Young Adults I*

The results indicate that participants tend to convey the message of "staying in the moment" without saying its reason. In all of the focus group sessions, the most mentioned indicators for this phrase is one's making an eye contact with his or her companion while checking his or her smartphone.

### Theme 4: Phone checking is not the cause, but the result of the lulls in conversations

All of the participants regarded phone checking in a social setting as an unpleasant behavior. However, they said that smartphone use has both negative and positive effects on the conversation quality, supporting previous work [12,15]. For example, they opposed to the idea that the smartphone use kills the conversations. They claimed that the phone is not the cause of the lulls in conversations, rather it is the result of these lulls. In other words, when there is a lull in a conversation, people tend to start using their phones. Another aspect that lead people to start interacting with their phones in a social setting, their desire to be isolated from the group for a moment. This behavior occurs when their degree of intimacy with people next to them is not high, and the conversation is not interesting. Participants indicated that they tend to use smartphones for isolating themselves when lulls happen in conversations.

They further stated that smartphones can be even used to deal with lulls in the conversations. Smartphones may fire up a conversation by giving a subject to talk about such as presenting a content from social media networks. But all participants stated that because smartphones can create infinite materials to talk about, the suggested topics may seem superficial to the group. For instance, one participant emphasized that memories are more powerful and meaningful than the content in the social media.

*"…Maybe there are more topics in social media, but they seem superficial. For example, the thing we laughed three months ago does not make me laugh now. I forget it. There is nothing left, it feels like nothing has enriched me."* –*Male, Middle-aged Adults*

### Theme 5: Age groups have different reservations about phone checking behavior in social settings

We observed that people from different age groups tend to perceive and react phone checking behavior differently. While middle-aged participants (41-55) consider this behavior as a harmful habit and think that it should be modified, young adult participants (18-25, 26-40) thought that it has no negative impact on their lives. For young adults, this was not a big and general problem. They also don't blame smartphones for reducing the quality of social interaction. They mentioned that banning or restricting smartphone use makes the phone more attractive, which is in line with previous work [4].

Another difference between different age groups was the way they react to lulls in conversations. All of the participants reported that lulls occasionally occur in their daily conversations. However, it seems that young adults tolerate these silent moments more than the middle-aged adults do. While young adults do not mind sitting silently and interacting with their phones during a meeting, middle-aged participants tend to end a meeting when the frequency of pauses increase.

### STUDY 3: DESIGN WORKSHOPS

In the third study, we conducted three design workshops with 15 junior professional designers. The purpose of these workshops was to identify design directions that can be used when exploring the solutions for mediating smartphone use in public settings without banning smartphones.

|  | **Workshop 1** | **Workshop 2** | **Workshop 3** |
|---|---|---|---|
| Number of Participants | 6 | 5 | 4 |
| Mean Age (Standard Deviation) | 24.6 (2.42) | 26 (6.51) | 22.75 (2.18) |
| Professions (Number) | Industrial Designer (3) Interaction Designer (3) | Industrial Designer (2) UX/UI Designer (3) | Industrial Designer (3) UX/UI Designer (1) |
| Number of Ideas | 21 | 26 | 20 |

**Table 2. The characteristics of workshop participants**

We recruited the designers first by using our personal contacts[2] and then by snowballing method. Six designers attended the first workshop, five attended the second workshop, and four attended the third workshop (Table 2).

The workshops were divided into four sections. In the first part (20 minutes), we presented the insights of the observations and focus groups in order to familiarize designers with the topic. After, we introduced themes and describe them one-by-one. Then, we asked them to generate solutions for mediating smartphone use in social settings. We emphasized that the solutions should not prohibit the smartphone. In the second part (30 minutes), designers started ideating individually because we wanted them to generate ideas without being affected by others. In the third part (90 minutes), they worked in groups. During these three stages, we asked them to develop as many ideas as possible on each theme. The participants drew their ideas on an A4 sheet of paper via free hand sketching (Figure 2). In the last 40 minutes of the workshops, we wanted designers to present their ideas to all the participants.

To analyze the workshop outcomes, first we searched for ideas aimed at banning the phone via external stimulus as well as the ideas that are not in line with the workshop brief. We only found one idea, locking the smartphone physically in case of excessive use. We excluded this idea from the analysis. This resulted in a total of 67 ideas (Table 2). Then, we allocated each idea to the theme it belonged to. This allocation showed that the groups generated at least one idea for each theme. The most used theme was *Theme 1: Feeling of curiosity drives phone checking behavior in social settings* (*N*=29). The least used theme was *Theme 5: Age groups have different reservations about phone checking behavior in social settings* (*N*=6). Thus, it seems that we had a good coverage of solutions addressing each theme. Later, we analyzed the ideas based on the approach taken to mediating smart phone use. This analysis resulted in four design approaches: *Enlighteners* (*N*=21), *Preservers* (*N*=16), *Supporters* (*N*=18) and *Compliers* (*N*=12). In the following, we elaborate on each approach by giving examples from generated ideas. Since our aim was to show the breadth of the design space explored by designers, instead of explaining each idea in detail, we tried to present a range of ideas categorized under each approach.

**Enlighteners**

Focus groups participants stated that once they start to use their smartphones during social interactions, they might get lost in their devices (Theme 1). As a result, they become less engaged in the conversation. Addressing this problem, *enlighteners* aim to create awareness about such situations by informing the users (Figure 2a).

In majority of the *enlightener* ideas, designers developed feedback mechanisms that show conversation quality in a social gathering or the frequency and amount of smartphone use. Overall, there were ideas using common strategies to give feedback such as showing a red alert screen to the user or sending a notification like "Return to chat with friends" as a reminder when they focus on their smartphones. As our aim was to identify alternative approaches to mediating use habit, here we do not elaborate on such ideas. Rather, we explain the ideas which explored different ways of giving feedback.

*Physical Changes to Convey Information*

We identified two interesting strategies followed by designers during the workshops. The first was conveying information through physical changes. One example was a smartphone case that changes its color or heat according to the frequency of smartphone use. Another example solution includes a smartphone which grows old and wrapped by an ivy, as the quality of the interaction increases. By doing so, as the designers described, the smartphone will fuse with the environment and draws an analogy about moments that we've forgotten to use smartphones while we are on exciting conversations. There were also ideas that addressed other senses. For example, changing the music or the smell of an environment according to the conversation quality.

*Meaningful Feedback*

The second strategy was making the feedback provided to the user more meaningful. One of the ideas were based on emphasizing the value of time by informing the users what they can do in real life with the time they spend on their smartphones. For example, when users use their smartphones for 15 minutes, this app gives them a suggestion such as "These 15 minutes, you could go for a run with your friend.". In another idea, when a group of people want to use their smartphones, they escape from their owners and hug each other to remind the user how important to have intimate time with your beloved ones.

---
[2] The authors are design researchers who have close contacts with many professional designers working at industry.

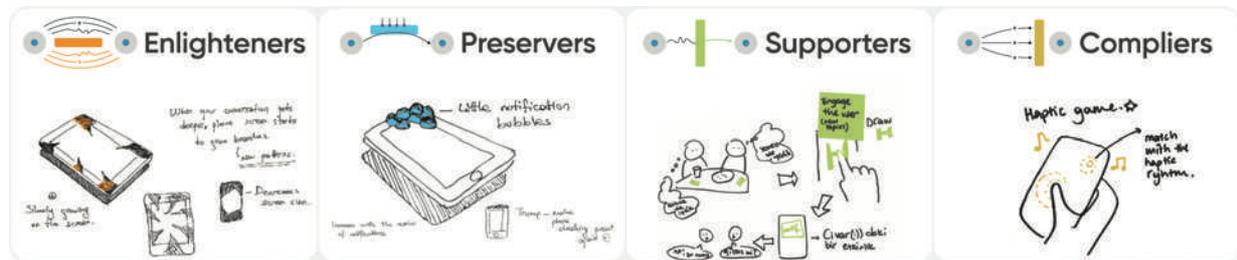

**Figure 2 – Four Design Approaches a) Enlighteners b) Preservers c) Supporters d) Compliers**

**Preservers**

The ideas under this approach focus on preventing the triggers of phone checking behavior in social settings thus preserving the ongoing social interaction between the users (Figure 2b). According to the results of focus groups, these triggers were feeling of curiosity and notifications (Theme 1). We identified three strategies targeting these triggers.

*Smarter Filters*

The first strategy was managing the smartphone notifications in a smarter way. Designers, who stated that a large number of notifications and their irrelevancy to the situation were perceived as distracting as they interrupt the conversation (Theme 2), developed smart filters for the management of notifications. For example, a smart filter app shows the most relevant notifications and mute other notifications by understanding what the user is talking about and whom s/he is with. When the user is talking about business with his or her colleagues at a lunch, s/he will see the notification about financial news. But in a more intimate atmosphere where the user meets with his or her close friends, s/he will see the notifications of concert activity instead of financial news. In another smart filter idea, the designers connected the filtering process to the quality of the conversation and allowed the users receive notifications gradually. This idea was about muting the notifications when the conversation is rich and receiving them as the level of interaction decreases.

*Unobtrusive Notifications*

The second strategy was making regular notifications as unobtrusive as possible. Designers, who emphasized the importance of eye contact during a meeting which was also observed in focus groups (Theme 3), developed ideas for controlling notifications without interacting with the smartphone. For example, in an audible notification scenario that can replace the classical notifications (e.g., vibration, simple sound feedback or displaying in the notification panel etc.), the smartphone says the sender of the notification. (e.g., "Colleague", "Mom"). If the user asks for more detail, it says the broadest keywords that best describe the subject of notification (e.g., "Reports to be completed", "Shopping List"). The user may or may not choose to check the notification after this step. The participants also speculated a wearable apparatus (a very small object such as a wristband or a ring) which provides tactile feedback differentiated according to the notification priority. In addition, there were also ideas that categorize the notifications as unlikely and likely. For example, if receiving a phone call from the user's boss is an unlikely event, the system shows him or her the notification of this event with the iconic images on the back side of the phone and skips the other notifications.

*Invisible Notifications*

The third strategy was making the behavior of checking notifications more difficult via gamification. In [22], the study focused on preventing phone checking via a lockout task intervention. Users are required to type the given digits to check the notifications. However, in our workshops, designers emphasized the playful part of this kind of interventions by using meaningful design frictions [10]. For example, in an idea, designers concealed the source of the notifications, and turned them into colored bricks. They designed games which include color and shape matching interactions through these colored blocks. Another idea was reducing the visibility of the screen and notifications by creating bubbles on the screen for each notification. By doing so, users are not able to see the notifications clearly until they pop the bubbles.

**Supporters**

All designers stated that they consider the amount of speech and the depth of the subject as the variables that determine the quality of the conversation. The ideas developed from this point of view were aimed at increasing the quality of the conversation among users and avoiding the strange lull moments, as mentioned in the focus groups (Theme 4) (Figure 2c). The ideas for the differences between generations (Theme 5) mostly focused on this approach ($N$=5). We identified three strategies used to increase the frequency of interaction between users and to enhance the quality of the conversation.

*Conversation Boosters*

The first strategy was boosting the conversation. Some ideas on this topic are based on collecting data about users' interests and present the common ones as the proposed topic to initiate a conversation between group members. Users can enter their own preferences into these systems manually, or the system can pull data from users' social media accounts automatically. Also, they can collect these data instantly with sensors from the environment.

Designers proposed different alternatives for how these conversation boosters can work. In one of the examples,

when the user draws a question mark or H letter (initial for Help) on the closed phone screen, the phone sends related notification as if the user had received a notification from an ordinary news source (a story about sports news or a cultural event). With this new interaction style, the designers intended that the user does not want his / her call for help to be seen by his / her friends. Another idea was based on reflecting the conversation topics of each table in a restaurant or a coffeeshop to the walls. With this system, it is envisioned that the users that have exhausted their topics will be inspired by the environment.

These boosters were also speculated for the generational differences. Designers who stated that finding common interests between different age groups is hard to reveal and it affects the relationship and interactions between these groups. The ideas aiming at exploring an interesting topic for each individual, when a lull moment occurs, can uncover common conversations between generations and may fill this generational gap.

*Pop-up Memories*
The second strategy was encouraging group members to find something to talk about themselves rather than suggesting an existing topic. The designers, who emphasized the fact that the strained attempts for a conversation would cause artificiality which was also observed in focus groups (Theme 4), tried to overcome this problem by asking questions to the users instead of giving them a direct topic. For example, one group designed a screen saver application that could be installed on smartphones. This application shows the abstract and simple icons floating on the phone, which can initiate conversation in the lull moments. These icons are selected from the intersections of trending topics and interests of the users. With this kind of abstraction and simplicity, the designers aimed to increase the number of topics that can be revealed from an icon. For example, users can think of rain, a film, a song, or an experience that they have when the umbrella icon on the screen begins to float.

*Engaging the "Uninterested" Members in the Conversation*
The third strategy was engaging the group members who want to be out of the conversation as they get bored of the topic. For example, in a situation in which a user is bored, a game that gives him or her the right to change the topic of a conversation. In another scenario, when the user is not knowledgeable with the current topic and bored, an application provides help by explaining the topic or giving some clues about it.

**Compliers**
This approach includes ideas complying with users' intention to be involved in a conversation or not. In particular, they were inspired by users' desire to isolate themselves by using their smartphones in social environments, as identified in the focus groups (Theme 4) (Figure 2d). For this approach, we identified two strategies.

*Prominent Isolation*
One strategy followed by designers was making the need for isolation highly visible for other group members. Designers thought that this would help users understand one's need for isolation in an explicit way. For example, when the user wants to isolate his or herself from the others, as the use of phone increases and as his or her social interaction with the group decreases, an iceberg-like separator rises in front of the user. Similar to this idea, when the user is in the isolation mode, a bubble machine starts to blow bubbles from the user's phone to make the user less visible. In another idea, designers envisioned an AR based application which allows users to wear a virtual animal costume when they need isolation.

*Exaggerated Isolation*
Another strategy was physically separating the isolated user from the environment to make him or her aware of the consequences of this isolation. For example, the chair of the user who wants to be isolated from the group and use his or her smartphone continuously, rises up. Similarly, the desk begins to pull towards the isolated user. With these physical changes, the designers were trying to create physical difficulties to the users, and they wanted to show that isolation should have a cost.

**DISCUSSION**
In this part, we revisit the findings of each study as a whole (observations, focus groups and design workshops), and discuss them as design implications. Our aim in conducting this study was to identify alternative design approaches for mediating excessive smart phone use in daily social interactions, as we argued that restricting smart phone use would not be a sustainable solution for this problem. Overall, the results show that although people are aware of the potential negative impact of excessive smart phone use on their social relations, most tend to use their smartphones during their daily social interactions. It doesn't seem realistic or possible to discard these devices that become more important and more ubiquitous in people's lives. Thus, it appears that the study results provide evidence supporting our initial argument. Furthermore, our work would have several implications for designing solutions aimed at mediating excessive smart phone use in public settings, which we discuss next.

**Informing about the quality of interaction beyond using traditional means**
A big part of the ideas proposed by workshop participants were focused on informing the users about the quality of interaction they have (*enlighteners*). For example, solutions involve informing about the time spend using one's smartphone in a gathering or the frequency of phone checking while your friend are talking. These types of solutions have still potential for mediating smartphone use. However, we think that designers should benefit from non-conventional means of giving feedback when thinking about these solutions. They can redesign the spaces in which users are located. They can give users unusual experiences by augmenting the environment itself and the objects already found in the environment such as using the coffee table to give feedback. At the stage of informing the user about the situation, designers can develop their ideas on different modalities such as sound and smell.

Furthermore, the information given may not be the directly related with the quality of the speech or the amount of smartphone use. In other words, what we call feedback may not be just a status reporter, because user might get used to such feedback after a certain period of time, and the long-term effect may not be sufficient. To avoid this, designers can come up with ideas which contain "meaningful" feedback mechanisms [18] or a feedback which is intended to raise awareness in the user or has different achievements (i.e., rewards, discounts etc.). For example, instead of providing color map to show the level of interaction, a design come up with a suggestion which will have an effect on users' lives such as inspiring them to do cultural activities.

**Triggering nonconventional interactions with the smartphone**
Smartphones are powerful interactive devices that allow us to perform many tasks. The features and abilities of smartphones can be used to trigger new interactions between people, particularly in public settings. For example, a user asking for help during a lull moment in a conversation (Theme 3) can draw a question mark on the closed smartphone screen to get help for the topic to talk, or smartphone case can display icons to initiate a topic.

Besides creating new interactions that help group members engaged in the conversations, smartphones can be also used in situations where such an engagement is not desired. Although the need for isolation (Theme 4) is a negatively perceived concept in social interaction, designers can turn this issue into a non-disturbing phenomenon for the users by creating entertaining solutions (*compliers*) (Theme 2). However, particular attention should be paid to this kind of solution whether entertaining part is valid for users who use the product or not. For example, in the chair idea, which rises up when the user wants to be isolated, mentioned before, is this situation an entertaining one or does it cause a public shaming? Therefore, depending on the context designers may focus on new interactions that invite people into group interaction or create interactions that can change the negative perception of being isolated.

These new interactions could be also used to address users unconscious checking behavior or their tendency to use their smartphones when they are bored. Designers can develop smart phone applications that make this use behavior fun and give users an option to their phone as a fidgeting tool. For example, a game that requires the user to tap the same rhythm given by the vibration of the smartphone without being noticed to the friend or an app that turns the smartphone into a hidden musical instrument.

**Considering the generational differences when designing solutions for mediating smartphone use**
Our rationale for selecting the focus group participants from different age groups was to uncover any differences regarding the reactions towards and motivations for smart phone use during social interaction occurred in public settings. Analyzing the focus group results, we discovered that while young adults are comfortable with using their smartphones during social interactions as well as more tolerant to people who perform this behavior, middle-aged individuals have many reservations regarding this behavior. In general, middle-aged participants (40-55), thought that this behavior is problematic and that it significantly reduces the quality of interaction between a social group. For example, when there is a lull in a conversation, young adults said that they do not mind siting silently and checking their phones. On the contrary, middle-aged adults accused the smartphone use for creating these lulls and stated that they even tend to leave the group if the lulls persist (Theme 5).

However, we realized that none of the solutions proposed in the design workshops addressed such differences. One potential explanation for this may be the fact that since all of the workshop participants were young adults, their solutions might have inspired by their own experiences that they have as a young adult. Another explanation may be the fact that focus groups helped discovering the generational differences in a broad way; that is, it did not allow us to get deeper insights at the individual level. In line with these limitations, in the future, we plan to conduct interviews and co-design workshops with families to understand the generational differences in relation to this excessive behavior better.

**Keeping the attention in the "real world"**
We discovered that, when people are with their companions in a public setting, such as hanging out with friends in a café, there are two major reasons that prevent them from interacting with each other. These are 1) notifications coming from their smartphones that shift their attention from their companions to smartphones (Theme 1), and 2) lulls in the conversation that trigger smartphone use (Theme 4). Design workshop results indicated that this attention shift can be mediated by informing the group members about how it impacts the quality of social interaction (*enlighteners*), preventing or managing the subjects that create the distractions (*preservers*), and supporting the situations which motivate group members to interact with each other (*supporters*).

Many of the ideas categorized under *supporters* was focused on shifting the attention from the virtual world (checking your Facebook page) to the real world (telling a story to your companion). According to the findings, if the users enjoy the topic of a conversation or the moment they are in, they are less distracted and even forget the existence of their smartphones (Theme 3). As designers develop solutions to make a conversation more engaging, they should think about ways that can produce content that is appropriate to each individual's interests and that will not make users feel forced or artificial about the topic. Also, users' past experiences can be used as an inspiration source for the content generation.

While trying to manage smartphone notifications which distract the users during social interaction (*preservers*), designers should generate adaptive solutions that can change according to the context [19]. For example, while a user is in a business meeting, notifications related with other kind of things such as nonserious and nonurgent content should be detected and filtered. In addition to this adaptability, solutions should be aware of important and unlikely events of user's life. They should not deprive users of their smartphone while reducing the distractions of users, for example filtering out an important notification about one's job interview outcome, because a solution that deprives the users might be subject to harsh criticism from users.

In addition to filtering solutions, the notifications themselves and the methods of providing them can be re-designed. For example, rather than using a traditional "beep" sound which distracts all of the group members in a gathering, different modalities such as tactile feedback on the user's skin or narrated feedback can be used to maintain eye contact with others [17]. Alternatively, the way we receive notifications can be gamified via turning smartphones into "pleasurable troublemakers" [28] which makes it difficult (not impossible) to check unnecessary notifications.

**Taking into account users' concerns about approaches**

Mediating excessive smart phone use has many benefits such as overcoming the negative effects of smartphone use on our physical, mental and social wellbeing, as we mentioned in previous work. Furthermore, using bottom up design approaches to mediate this behavior have a potential to increase the effectiveness of the solutions as it might increase users' acceptance. However, since such approaches still rely on changing user behavior through using interactive technologies, they carry the similar risks with other persuasive technologies. In the following, we briefly mention two of these, which should be taken into account while designing future solutions.

Even though the interventions of the designers are meaningful and positive for the users, they may not be welcomed by them, or these interventions may create unintended outcomes [48]. For example, in our focus groups sessions, participants told about solutions aimed at boosting a conversation via topic suggestions could create an artificiality. This may harm the natural interaction as a result of the interference of an external factor. Another issue related with such solutions would be privacy [6], as they rely on users' personal data such as (interests, likes, friends etc.) to identify relevant topics to boost conversations.

Furthermore, the ideas presented under the *compliers* should handle with care. Even though the purpose of this approach is to make users' isolation smoother and more tolerable, they may reduce the interaction between people and make the other users feel awkward and excluded in social settings. Thus, it is essential to learn more about users' reactions towards the solutions generated to identify which of these is the most suitable ones for a specific context and a specific user group.

**CONCLUSION**

In this study, we presented the results of an exploratory study aimed at examining smartphone use during daily social interactions and identifying ways of reducing its use without restriction. First, we made unstructured observations in four coffeehouses, then conducted six focus group sessions with 46 participants to gain a deeper understanding of users' smartphone use behavior. After analyzing these explorations and conducted three design workshops with these themes in order to identify design approaches to mediate smartphone use in public settings. The four design approaches are;

- *Enlighteners*: creating awareness about the quality of social interaction by informing the users.
- *Preservers*: aiming to mitigate the triggers of phone checking behavior in social settings.
- *Supporters*: increasing the quality of the conversation among users and helping avoid the strange lull moments.
- *Compliers:* providing smooth isolation for users in relation to social interaction.

These design approaches would inspire other designers and researchers in designing solutions that can mitigate excessive smartphone use without necessarily restricting it. However, we note that this would not be a trivial effort. First of all, smartphones have become inseparable parts of our lives. We use them for communicating, learning, entertaining, shopping and for many other daily tasks. Second, they are purposefully designed to grab our attention and keep it for a long time through, for example, sending constant notifications to check our phone. So, what role design could have in addressing the broader problem of excessive smartphone use, in particular during social interaction? We can exemplify two potential scenarios that touch upon this question by using *preservers* approach. One scenario could be re-designing public places in a way that they mediate the things that can disrupt human-to-human interaction. For example, when a team enters a meeting room, the room can turn all the phones into silent mode. Another scenario could be designing a new wave of smartphones or apps that prioritize social interaction over interaction in social media. For example, Lightphone[3] is a new generation of phone designed to provide some main functionalities like calling, texting, setting an alarm, and listening to music without sending feeds and notifications. We believe that examining scenarios like these in order to better understand design's role in addressing excessive smartphone is a topic worth exploring in the scope of future work.

---

[3] http://thelightphone.com

Chapter 5

# DESIGNING THE CONCEPT 1: WHISPER

**Research Question**

**RQ3** How can we design interventions that support social interactions (without necessarily restricting smartphone use)?

## *5.1. Concept Development*

Considering both the understanding of the phenomena of smartphone use and design approaches that we gathered from field studies and design workshops (Chapter 3 & 4), we started to generate ideas to support rich social interactions in public places (Figure 5.5). In this idea generation step, we created different types of ideas. For example, one of the ideas is a smart table that provides abstract visuals to the users by changing its surface. These visuals aim to inspire users to find a conversation topic. By tangibly presenting the data, we were aiming to provide users a chance to interact with the environment (i.e., the table and the shapes on it). Another idea was to focus on the beverage orders of the users. In this idea, according to users' choices, we want to provide notifications via a mobile app. For example, once a user orders a cup of black tea (which is a cultural thing in Turkey), we would provide some current news since, in Turkey, it is a common tradition to drink tea in cafes and tea houses for brief or long conversations where people talk about political and financial issues. It is something Turkish people associate with socializing. Within these ideas, we selected the idea of WHISPER and developed it. The idea provides users with conversation starters by giving them audio narratives (i.e., short stories) via tracking the users' interactions. By doing so, it focuses on enriching social interactions between users and aims to invite users to the conversation with each intervention.

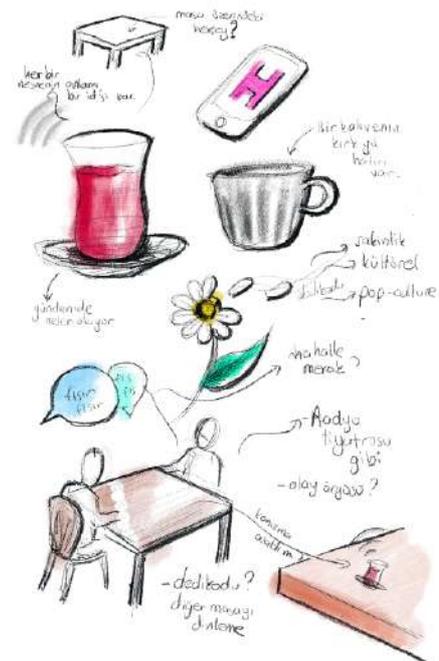

*Figure 5.5 Initial Idea Sketches*

While designing WHISPER, we filtered out ideas that utilized the design approaches of Supporters and Enlighteners. The reason behind this decision is that our participants of the Focus Group sessions were complaining about the lack of conversation topics and lulls in their social interactions. In most of the sessions, individuals said that they needed help to BOOST this interaction, and to do this, they used their smartphones to find content on social media to share with each other. In addition to this behavior, when checking smartphones for notifications, they also stated that sometimes they feel lost and disconnected from their friends next to them. Besides focus groups, in design workshops (Chapter 4), designers frequently stated that the ideas utilizing these approaches (i.e., Supporters and Enlighteners) might be more effective than the ideas utilizing other approaches since they claimed that supporting a richer social interaction is more reasonable than maintaining the current conversation (Preventers) or allowing users to isolate themselves from each other (Compliers).

In addition to that, we revisited the findings from the focus group, which inspired us to design WHISPER (Figure 5.6). The first finding was that users draw a parallel between social media accounts and their neighborhoods, a place where they can listen, follow, and watch lots of details from other lives. For them, the feeling of curiosity is the key factor to engage with something, whether it is the notification on their smartphones or the topic of a conversation such as a change in a friend's life, a travel picture, or a topic of gossip. This curiosity also reveals itself when they are not using social media. For example, some participants in the focus groups indicated that they sometimes start a conversation by eavesdropping on the conversations of the people sitting at the other tables. The second finding was that changes in the amount of communication between users are the key signals for measuring the satisfaction from that conversation. Participants stated that as they talked more, they felt more quality time during the interaction. On the other hand, there were some participants who sometimes preferred silence rather than having a constant conversation.

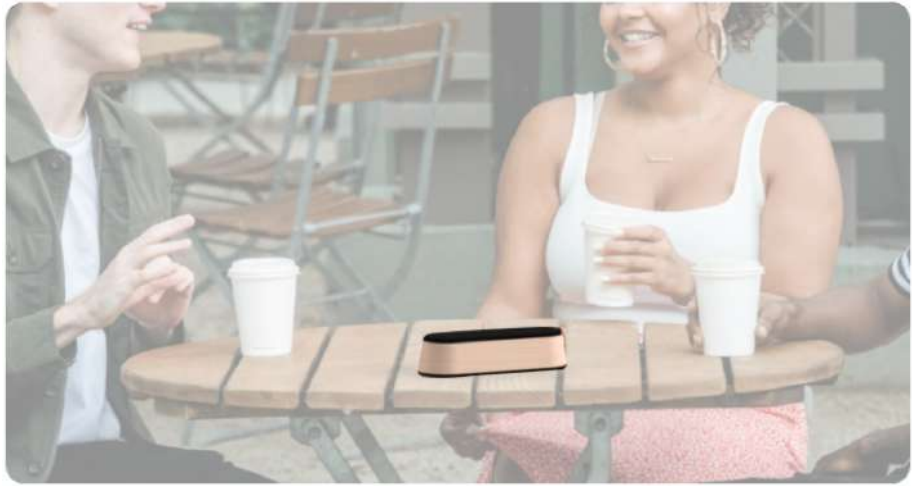

*Figure 5.6 A render for an early version of the WHISPER concept*

Another inspiration point for the WHISPER is audio narratives, which were popular before the screens came into our lives with devices such as televisions, computers, or smartphones. We chose to use audio narratives since we thought that this screenless modality might help us to intervene in interactions between individuals with a less distractive way.

Inspired by the findings about feelings of curiosity and lulls in conversations and audio narratives, we developed our concept, WHISPER (Figure 5.6).

The WHISPER is a smart sound box that tracks lulls in the conversations, and when a lull occurs, it provides short narratives as if other people were talking about a subject (e.g., a detective talking with a witness to solve a crime). By doing such an intervention, WHISPER aims to BOOST a conversation between users by tracking the quality of the social interaction and providing a story that might attract their attention from their phones.

## 5.2. *Concept Refinement*

After early concept generation, we conducted an online concept improvement workshop with 10 professional designers to further develop and refine the concept. This workshop introduced four design approaches (supporters, enlighteners, preventers, and compliers) and the WHISPER concept. We asked the designers to generate solutions that support rich social interactions by tracking lulls in the conversations and by using audio narratives to trigger individuals' curiosity. We formed three groups (groups of 3, 3, and 4), and each group generated ideas by using Miro (i.e., an online collaboration tool), finalizing and presenting their ideas in the Miro Artboard (Figure 5.7). We finished the workshop with a discussion session where all designers gave critics on each project and discussed issues that emerged during ideation. The session lasted six hours, and it was video recorded. As a result of this workshop and previous studies, we identified three design considerations for further developing WHISPER.

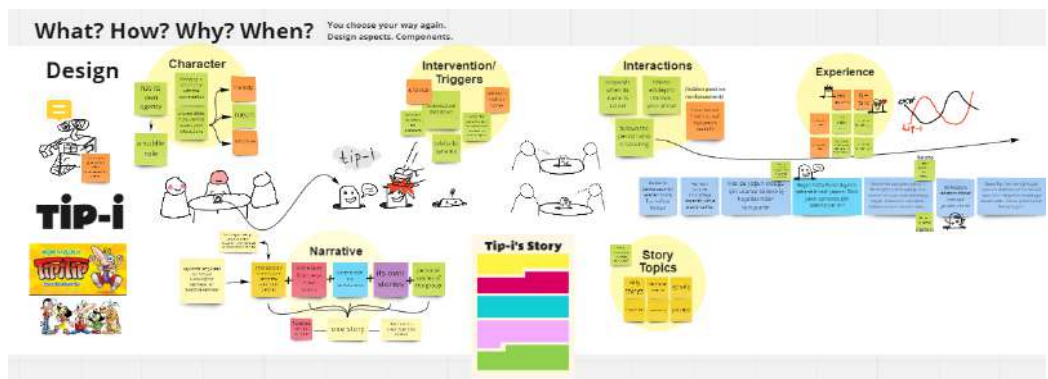

*Figure 5.7 An example Miro Boards from the Idea Refinement Session*

**Consideration 1:** WHISPER should provide unforced interaction: WHISPER was originally designed to be used in public places like cafés and restaurants. Users will be exposed to it without any further intention. Also, in our previous studies (Chapter 3), we found that users sometimes prefer to sit in silence, not interacting with each other verbally though sharing the same space. In the concept development workshop, designers focused on this issue; thus, they decided that any intervention should not force users to interact; rather, it should be triggered by users' intention to interact.

**Consideration 2:** WHISPER should admit failure: In relation to the first point, in the concept improvement workshop, designers also discussed indicated that any intervention in social interaction should not be stubborn. If it is ineffective to generate or BOOST any social interaction between the users, it should not nag users in this regard.

**Consideration 3:** WHISPER should include personally relevant subjects without invading users' privacy: There were many ideas that focused on personalized interventions in the design workshops (Chapter 4), e.g., providing personalized content suggestions for maintaining the conversation. These ideas used users' personal information collected from their social media accounts and smartphones (e.g., events, likes, followings, photos). This issue was a matter of heated debate in the refinement workshops, where designers generated ideas around the WHISPER since collecting and presenting personal data in various ways may be seen as a privacy concern for users.

## 5.3. *Final Concept*

The previous WHISPER concept is a smart sound box that tracks lulls in the conversations, and when a lull occurs, it provides short narratives as if other people were talking about a subject (e.g., a detective talking with a witness to solve a crime) according to the drink coaster that they choose. By doing such an intervention, WHISPER aims to BOOST a conversation between users by tracking the quality of the social interaction and providing a story that might attract their attention from their phones.

We revised our initial concept by taking into account these considerations 1) Interaction Type, 2) Level of Blending with the Environment. Combining the aspects helps us to generate four different versions of the WHISPER (Table 5.1 Aspects of WHISPER Alternatives).

| Concept Name | Interaction Type | Level of Blending with the Environment | Design Decisions | |
|---|---|---|---|---|
| | | | Proximity with User | Form Idea |
| WHISPER LAMP | Implicit interaction | 1 / Highly Concealed | Lamp / Vase etc. | Complying with the environment |
| WHISPER DECO | Implicit interaction | 2 / Concealed | on Table / Wall | Modern shape that resembles ephemerality / no relationship to context / no explicit meanings |
| WHISPER WAVE | Peripheral Interaction *(Narrative Tokens)* | 3 / Noticeable | On Table | Wavy surface / Elegant / Zen |
| WHISPER APP | Focused interaction *(QR & Choice Notifications)* | 4 / Highly Noticeable | Smartphone | Notification and quiz-like app |

*Table 5.1 Aspects of WHISPER Alternatives*

**Interaction Type:** This aspect involves whether a user can interact with the concept in an implicit, peripheral, or focused way in The Interaction-Attention Continuum (Bakker & Niemantsverdriet, 2016).

*Focused Interaction*, the most common type of interaction continuum in today's world, provides users with an explicit medium to interact. In this scenario, users are aware of the WHISPER APP (I.e., The Smartphone App version of the WHISPER Concept). They can start, manipulate and terminate the interaction when they want and if the users are in a lull moment, app notifications invite users to interact.

As the opposite of focused interaction, *Implicit Interaction* is defined as "an action, performed by the user that is not primarily aimed to interact with a computerized system but which such a system understands as input" (Bakker & Niemantsverdriet, 2016). In the WHISPER LAMP and WHISPER DECO scenarios, users don't have any control over the interaction process. The concepts can measure the Level of Social Interaction between users, give the interventions when needed and close themselves when they fail to enhance the social interaction.

Apart from those, *the peripheral interaction* provides information from the computing systems to users in a subtle manner to be perceived by users in their periphery. In this scenario, users are aware of the WHISPER WAVE artifact and the Story tokens on their table.

**Level of Blending with the Environment:** This indicator is related to the level of noticeability of the concept by users. The rank has 4 states:

**1) Highly Concealed:** Users cannot see where the speaker or sound source is.

**2) Concealed**: Users cannot notice the concept unless it intervenes in the social interaction with the stories.

**3) Noticeable**: Users can see the concept and the sound source as soon as they settle down at the tables in the cafe.

**4) Highly Noticeable**: Users can see the concept the same with 'noticeable,' and also, they are invited to interact with the concept with designated indicators such as QR Codes.

To proceed with our initial user studies, we chose our third concept, WHISPER WAVE.

We thought it was the most suitable one since it has Peripheral Interaction and 3rd Level of Noticeability. With its mid-level properties in our aspects, we were planning to explore the other concepts (i.e., WHISPER LAMP, WHISPER DECO, and WHISPER APP) and the users' thoughts on those concepts during studies and interviews.

According to this version (Figure 5.8), WHISPER uses tokens to initialize the device-user interaction. Once a user draws a token from the token holder placed on the device, WHISPER starts to follow the conversation between users. On all the tokens, there are abbreviations for conversation themes (e.g., CR for crime, DR for drama, FA for fantasia). These texts allow users to explore different stories and do not violate their personal information (Consideration 2). After a while, when the amount of interaction between users drops, WHISPER plays the narrative. After three trials, if the narratives are not working to create interaction, the WHISPER reduces the amount of its participation in the context (Consideration 3).

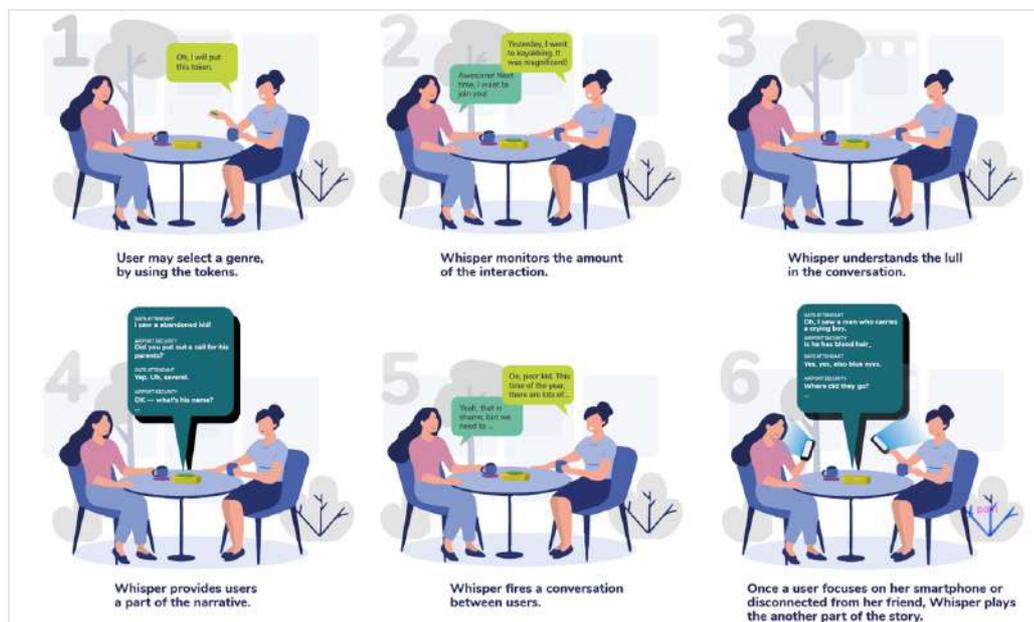

*Figure 5.8 The use scenario of WHISPER*

### 5.4. Prototyping WHISPER

WHISPER is an interactive audial narrative box that has tangible parts (i.e., drink coasters) to provide users the ability to interfere with the audio narrative.

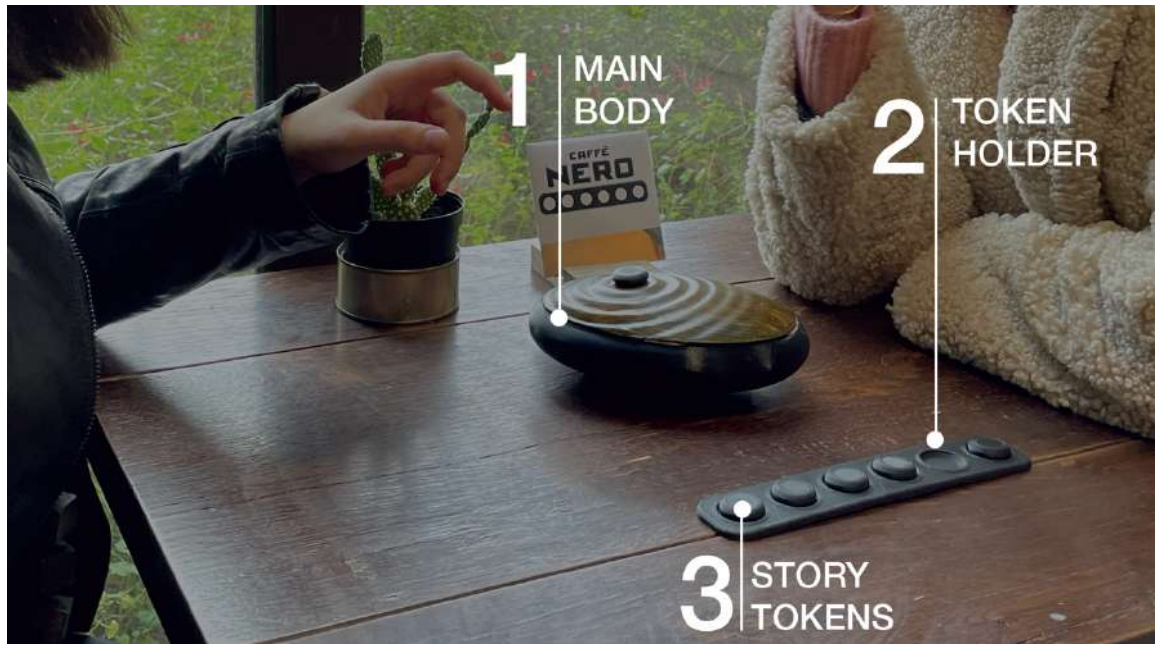

*Figure 5.9 The components of WHISPER*

The basic concept (Figure 5.9) has 1) the main body that includes a Bluetooth speaker to provide the narrative and magnet, 2) a cup that holds the tokens, and 3) story tokens that have genres embossed on and a magnet to stick with the main body.

Lim et al. (Lim et al., 2008) consider prototypes in their role in the evaluation and their generative role in enabling designers to reflect on their design activities in exploring a design space. With this aim, they presented the anatomy of prototypes as a framework for prototype conceptualization. To plan the prototyping stage of WHISPER, we used Filtering and Manifestation dimensions (

Table 5.2) of the Anatomy Framework for prototyping in Interaction Design. With this framework, we managed our process more effectively and efficiently.

| Concept Feature | Dimension |
|---|---|
| High-fidelity Prototype | Appearance |
| Drink coaster made with Cork to initiate the product. | Material, Interactivity |
| A mobile app to change the narrative's plot. | Functionality, Interactivity |
| Just-in-time Personalized Narration | Data |

*Table 5.2 Concept and its classification with Anatomy Framework*

We built a high-fidelity prototype *(Appearance Dimension)* to provide a more realistic experience for the user. We produced all parts of the WHISPER WAVE (i.e., narrative box) using 3D printers (Figure 5.11a). After this stage, we did post-processing for the wavy part of the main body to make it look like made of wood. The wooden look helps us provide more realistic experiences to our participants, and the final product can easily blend with the Café environment. To reach this appearance, we painted the entire surface with white primer spray paint (Figure 5.11b) and sanded it using sandpaper in various thicknesses in order to increase the surface quality (Figure 5.11c). We experimented with various materials to make the surface look like wood (**Error! Reference source not found.**). After this process, we found our method and drew wood age rings on the surface (Figure 5.11d) with a process consisting of spray paint, varnish, aging paste, and alcohol ink to give the appearance of wood.

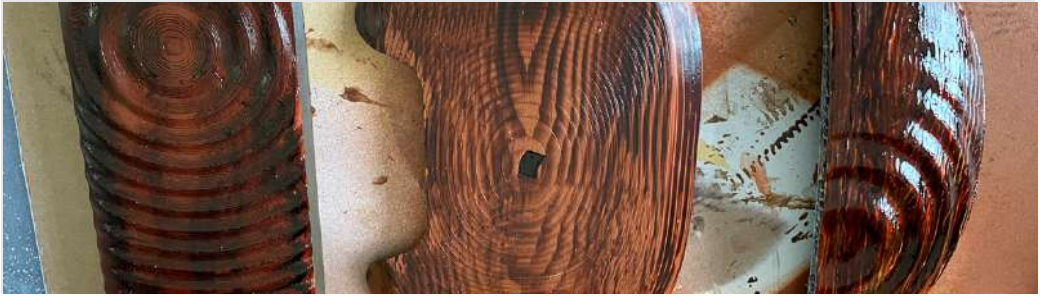

*Figure 5.10 Painting Experiments for Wooden Surface*

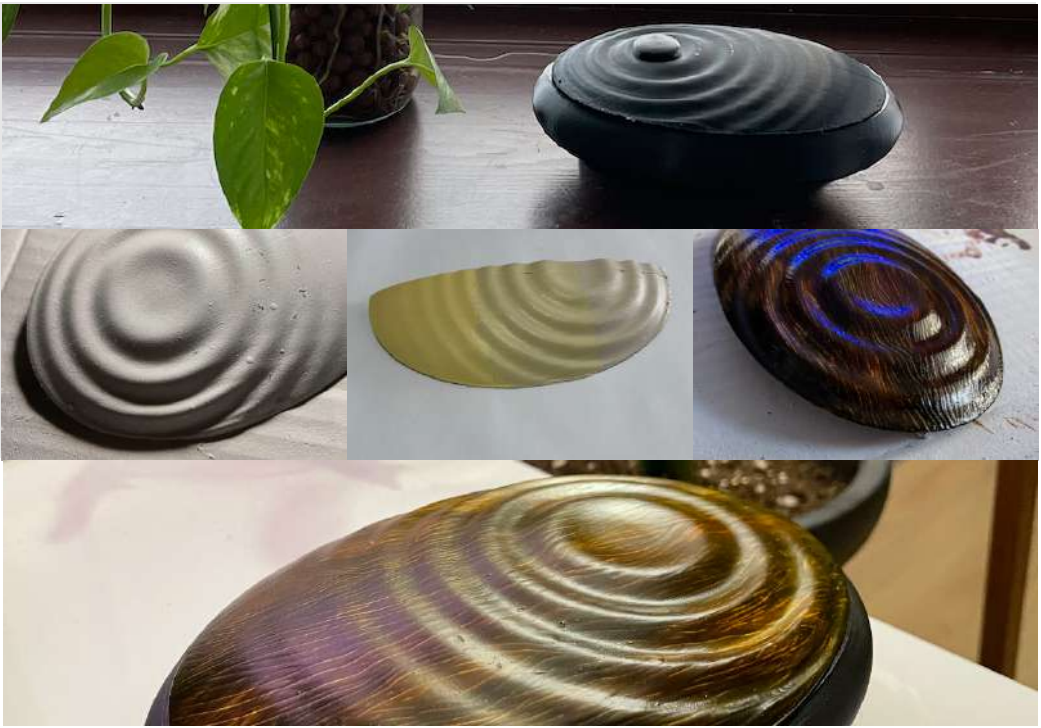

*Figure 5.11 Final Painting Process a) 3D Prints, b) Primed, c) Sanded, d) Aged, e) Final*

In parallel with the prototyping stage, we developed our Audio Drama stories. During this process, we cooperated with an improvisation club in Istanbul (i.e., Çarşamba Pazarı Doğaçlama Kulübü) on a voluntary basis. Three female performance artists from the club agreed to participate in our work. We presented 10 story themes ([Appendix 2](#)) that we prepared beforehand to the participants one by one. While creating all the scenarios, we asked them to imagine themselves sitting at a table in a cafe. Our intention in asking for this was to increase the resemblance of the stories in WHISPER WAVE to the sounds coming from the other tables in the café setting. We made all the sound recordings in the Recording Studio at Koç University, which has professional equipment (Figure 5.12). This collaboration took approximately 5 hours.

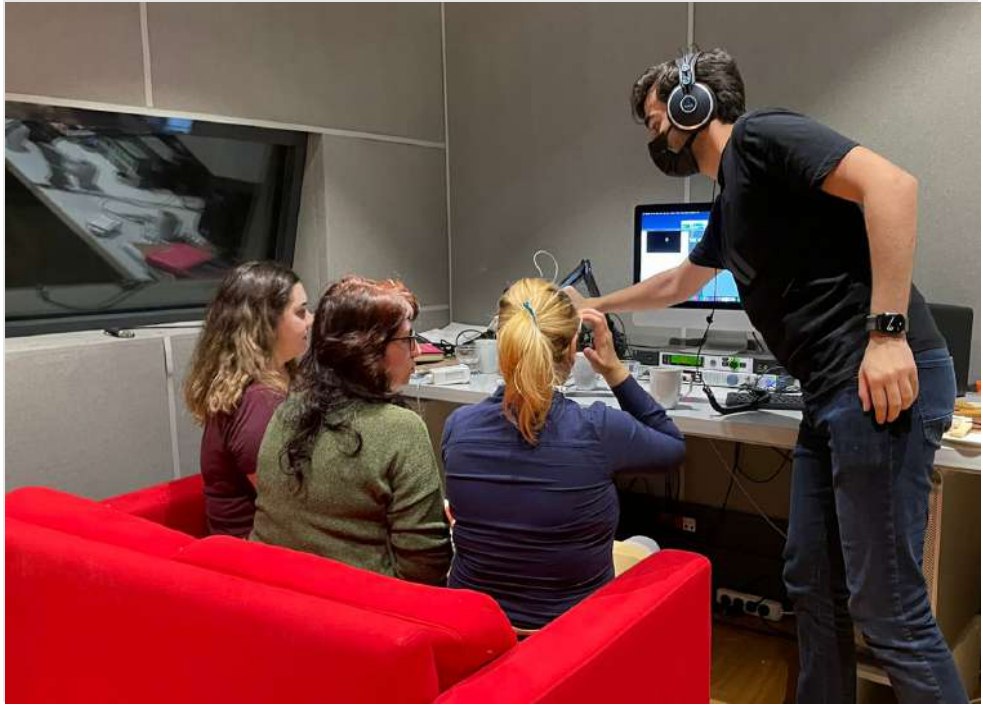

*Figure 5.12 Working with the Improvisation Performance Artists*

After recording the raw sounds, we did post-processing to enhance the quality of the audio (e.g., dialog editing, cropping, and cleaning) and added sound effects and foley to increase the sense of reality (e.g., crying, walking, typing message, ringing) by using audio editing software, Adobe Audition. With the final edits of the performances, we had six stories of various genres (e.g., crime, tragedy, fantasy, sci-fi so on.) in total ([Appendix 2](#)).

# Chapter 6

# ASSESSING THE WHISPER: THE WIZARD OF OZ STUDIES

**Research Question**

**RQ4 -** How would the design interventions aimed at supporting rich social interactions influence the interactions between individuals in the presence of smart phones?

**Related Paper**

**DIS 2022** – Mind the WHISPER: Enriching Collocated Social Interactions in Public Places through Audio Narratives

After building the semi-working prototype, we started to conduct Wizard of Oz user studies in order to understand its effects on the context and users' insights into the concept.

In this section, we present detailed information about our user studies, experiment procedure, and the insights we received from users after the experiment under three themes.

In addition to user insight, we discuss our results in terms of recommendations and implications to provide designers with more specific and contextual guidance on how to design technologies to support these collocated interactions. These recommendations and implications may inspire designers and researchers to develop interactive technologies that help users manage their excessive smartphone use during social interactions and sustain their digital well-being for social interactions.

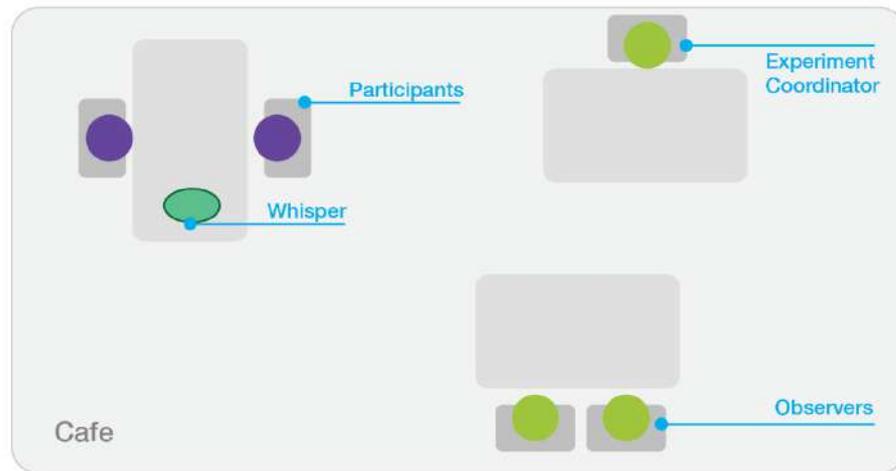

*Figure 6.13 Experiment Layout*

## *6.1. Experiment Setting*

We organized a total of 10 sessions and interviewed 21 people in total (12 Females, 9 Males) In this section, we share these findings from our work.

Similar to our aim to build a high-fidelity prototype, as mentioned before, the experiment setting was as real as possible for the users (Figure 6.13). To satisfy these requirements, we placed our setting to a café at Koç University to invite our participants. The table was decorated with a vase with flowers and a plastic ad holder to blend our concept with the environment (Figure 5.9).

## *6.2. Experiment Procedure*

In this stage, we prepared our experiment procedure. During pilot testing, we refine our procedure and prototype according to the experiences that we gain.

Participants were invited to a one-hour interview about "Technology use during the pandemic" without indicating the real aim of the study. The meeting point was a café on Koç University's campus. They went through three stages;

Stage A - Pre-experiment:

**Step 1:** With the participants arrive at the café, one of the researchers accompanies and tells them to sit at the table we have arranged.

**Step 2:** She shared a QR code linked to a short questionnaire, including demographic information, the Smartphone Use Scale ([Appendix 3](#)), and the Big Five Inventory with users ([Appendix 4](#)).

**Step 3:** After completing the questionnaire, the researcher indicates that we make them wait a bit because we wait for one of the researchers.

Stage B – WoZ Experiment:

During the experiment, we used a semi-automized Wizard of Oz (WoZ) set up which means that some functionalities of prototype were triggered automatically, some were controlled by the researchers.

**Step 4:** The researcher leaves the place.

**Step 5:** Other two researchers start to observe the participants in disguise.

**Step 6:** According to the interaction level between the users (e.g., amount of conversation, dealing with the smartphone, and body expressions), researchers play the tracks from the stories with their devices remotely. During each silence, they play another track. This part of the experiment took 25 minutes.

Stage C – Post-interviews:

**Step 7:** After 25 minutes, the first researcher comes back to the experiment area.

**Step 8:** She explains all the details of the experiment and the video recordings. If a participant does not sign the consent form and does not allow us to have the records, we delete them immediately in front of the participants.

**Step 9:** Once the participants complete the survey, one researcher explains the real aim of the study.

**Step 10:** He conducts the semi-structured interviews. In this part, we aimed to get feedback about the concept and participants' feelings, thoughts, and concerns ([Appendix 5](#)).

## *6.3. Findings and Outcomes*

Although in most of the sessions, participants had good reservations about WHISPER,

with the analysis of our sessions, we discovered several issues that contradict these positive responses. We believe that these contradictory findings will be useful in designing such technologies. We extract the implications of our findings for both design research and practice to demonstrate this usefulness.

We discovered 9 themes under three main topics.

**Reactions to intervening in social interactions**

- Interventions are helpful, but they can be annoying sometimes.

- Silence is a good indicator of a low level of interaction, but not always.

- WHISPER can be a conversation starter both in one-to-one and crowded group settings.

- WHISPER's interventions are perceived as warnings about the low level of social interaction.

**Reactions to receiving ambiguous nudges from an implicit device**

- Implicitness creates surprise and encourages users to explore the product.

- Ambiguity drives users' curiosity and creates a sense of wonder.

**Reactions to audio narratives**

- Audial feedback is good at attracting attention from smartphones.

- The content of the stories is less important than what it triggers.

- Users would like to have control over the stories.

On the side of design implications, we concluded that design interventions (DIs) aimed to enhance co-located social interactions should be *respectful, adaptable, targeted*, and *have a balanced ambiguity*. Furthermore, we provide two broad implications for HCI research (RIs): *the quantifying a subjective notion from users' lives (i.e., social interaction)* and *the responsibility aspect of the persuasive technologies to intervene in social interactions*.

Paper 3:

# Mind the Whisper: Enriching Collocated Social Interactions in Public Places through Audio Narratives

# Mind the Whisper: Enriching Collocated Social Interactions in Public Places through Audio Narratives

Hüseyin Uğur Genç
KUAR, Koç University, Istanbul, Turkey
hgenc17@ku.edu.tr

Duru Erdem
Koç University, Istanbul, Turkey
derdem18@ku.edu.tr

Çağla Yıldırım
Koç University, Istanbul, Turkey
caglayildirim18@ku.edu.tr

Aykut Coskun
KUAR, Koç University, Istanbul, Turkey
aykutcoskun@ku.edu.tr

## ABSTRACT

The quality of social interaction has great importance for psychological and physiological health. Previous research indicates that smartphones have adverse effects on collocated social interactions. Most HCI works addressed this issue by restricting smartphone use during social interactions. Diverging from previous work, we designed WHISPER, an audio narrative box that aims to enrich collocated social interactions without restricting mobile technology use. We conducted a user study in a café environment with 21 participants to understand how users react to WHISPER and how it would influence their social interactions. In this paper, we present the result of this study and discuss four implications for technologies designed to enhance collocated social interactions (Respectfulness, Balanced Ambiguity, Adaptability, and Being Targeted) and two implications for research touching upon the HCI work on Design for Behavior Change and Collocated Interactions (Designing responsible interventions for accommodating unintended outcomes and Quantifying the quality of social interactions).

## CCS CONCEPTS

• **Human-centered computing** → Ubiquitous and mobile computing; Human computer interaction (HCI).

## KEYWORDS

Collocated Interaction, Social Interaction, Design for Behavior Change, Wizard of Oz

## 1 INTRODUCTION

Social interactions are essential for personal growth and social relations. They contribute to physical and mental wellbeing as a central component of individuals' overall health [63]. According to Self-determination Theory, people need to experience a sense of belonging and attachment to other people (i.e., relatedness) to achieve psychological growth [15]. Wanting to feel connected and be around other people is a natural impulse. Individuals' health can be fundamentally influenced by the quantity and quality of their support networks and social connections [5].

However, the ubiquity of smartphones in our daily interactions increased the number of distractions and elements of attraction that overshadow social interactions. Users have begun to use their mobile devices during their collocated social interactions, which are the synchronous and direct interactions between people nearby [48]. Now, in social meetings with friends, users can record that moment as a video or photo, browse social media, share digital content, or even instantly communicate with people they are not with. Despite these advantages, smartphone use has adverse effects on individuals' physiological and psychological health [33–35] and their relationships. It may damage the level of intimacy and connection, reduce interaction quality [42, 52], and make companions feel awkward and excluded in social settings [23].

It is not easy for users to unplug from their smartphones as these devices have become more and more integrated with their lives. There have been interventions to reduce excessive smartphone use, both from research and industry. Most of these interventions utilize features that involve restricting, goal setting, reminding, and reward/punishment mechanisms, following a top-down approach. However, previous research indicates that digital wellbeing interventions should move beyond a focus on restricting and showing screen time approaches, as suggested in several studies [3, 6].

Addressing this need, we strived to understand how social interaction can be enriched in the presence of smartphones without essentially restricting their use. Within the scope of this study, we refer to social interaction as "the reciprocal interactions among individuals that happen during face-to-face encounters through verbal communication." We examined the following research questions: 1) How do users perceive technologies that monitor the interactions between collocated users and nudge them to maintain a conversation without limiting smartphone use? and 2) How would such technologies influence collocated social interactions? In order to answer these questions, we first designed a research prototype called WHISPER. This prototype is an interactive audio narrative box that gives short, pre-recorded audio stories during lulls in the conversations. We chose to focus on lulls since previous HCI work examining smartphone usage during social interactions [20] showed that smartphone checking behavior results from poor interaction (e.g., lulls in conversations), not the other way around. Second, we conducted a user study with 21 participants, ten sessions in total.

We make the following contributions to literature. We present a novel design, WHISPER, for reducing excessive smartphone use during social interactions without necessarily restricting it. We identified four design implications to guide designers in devising technologies to enhance collocated social interactions (1-*Respectfulness*, 2-*Balanced Ambiguity*, 3-*Adaptability*, and 4-*Being Targeted*). We present the implications of our results for HCI work on Design for Behavior Change and Collocated Interactions by discussing the responsibility of designers in creating interventions to accommodate *unintended outcomes* and the issues that might arise as a result of monitoring and intervening in social interactions.

## 2 RELATED WORK
### 2.1 Collocated social interactions

There are numerous situations in everyday life where social interaction would be beneficial, emotionally pleasing, or otherwise desirable (e.g., talking with a beloved one, sharing experiences with friends). Non-existent or insufficient social interaction would be problematic for individuals. Many empirical studies have analyzed the impact of social interactions, social trust, and the sense of belonging to a community on individual wellbeing. For example, a study [64] showed that people feel happier when interacting with close others. Another study [43] found that people tended to feel more satisfied after interactions with friends, followed by interactions with family members, others, and colleagues. Another study that obtains both self- and observer- reports of social interactions [60] showed that people report feeling happier and more socially connected when they spend more time interacting with others. Also, people who report that their relationships are more satisfying and supportive tend to have greater well-being [38].

Collocated interactions are critical for our relationships, creativity, and empathy. While sending an email, text, or social media post instantly links us with the community and gives individuals a sense of connection. However, it is a real-life interaction that truly connects us and provides us with essential social support [61]. Sharing information, giving, or receiving advice, or getting perspective from friends, coworkers, and family members helps create rapport, foster a sense of belonging, boost resilience, and help us process things and prevent overload. Face-to-face conversation leads to greater self-esteem and an improved ability to deal with others [51]. Even small talk is good for users' well-being. A study [67] found that short-term face-to-face conversations about the weather or other niceties can increase cognitive skills in the same way that brain-teaser activities do. Another research [66] has shown that humans need to communicate with others because it keeps them healthier. There has been a direct link to mental and physical health. People with cancer, depression and even the common cold improved their symptoms by talking with others. People who communicate their issues, feelings, and opinions with others are less likely to harbor grudges, resentment, or hatred, resulting in less mental and physical stress.

We can argue that social interactions with other people (e.g., family, friends, colleagues) are essential for well-being.

### 2.2 Effects of smartphone use on individuals' social interactions

Mobile application ecosystems have reached an incredible diversity with increasing processing power and enriching features. Now, users can call a cab or control their houses remotely just with a touch. Smartphones do not only satisfy such utilitarian needs. Smartphone applications, with these capabilities, promise to make lives more accessible (e.g., navigation, contactless payment) and even to prevent and treat chronic disorders like diabetes [4] and alcoholism [22]. During the COVID-19 pandemic, they helped people communicate and socialize with each other during lockdowns.

Despite these benefits, excessive smartphone use may cause physical and mental health issues (e.g., joint and neck pain, sleep disturbances, depression, smartphone addiction) [33–35]. Furthermore, smartphone use may adversely affect users' social interactions and relations. For example, people become less engaged with their immediate social environment due to heavy smartphone use during social interaction [1, 9, 42, 57]. In an empirical study that addresses the impact of smartphone use during dyadic conversations on 238 participants, participants perceived this behavior as less polite and attentive [1]. Another study shows that when smartphone use occurs in interpersonal interaction, the time spent with friends becomes less valuable, which is adversely and significantly related to users' life satisfaction [57]. They enjoy a meal with their friends less when their smartphones are present. People have tense arousal and boredom because they feel less socially connected and perceive time as slower [16]. Even excessive smartphone use is associated with lower relationship satisfaction with the romantic partner [54]. Besides using smartphones during social interaction, studies show that the mere presence of a phone on the table (even a phone turned off) changes what people talk about. If we think we might be interrupted, we keep conversations light on topics of little controversy or consequence [52]. Also, conversations with smartphones block empathic connection. If two people are speaking and there is a phone on a nearby desk, each feels less connected to the other than when there is no phone present [42].

One study involves a field experiment with 100 dyads [42] that shows that people who have conversations without mobile devices reported higher levels of connectedness and empathy than those who simultaneously use mobile devices. Spending less time with friends means less time to develop social skills. Similarly, a study found that sixth-graders who spent just five days at a camp without using screens became better at reading emotions on others' faces, suggesting that new generations' screen-filled lives might cause their social skills to atrophy [62]. This being so, an in-person conversation led to the most emotional connection, and online messaging led to the least [58].

Another study [20], conducted with 46 participants in six focus group sessions, found that smartphone checking behavior in a social gathering makes the companions feel negative emotions (e.g., anger, offended, bored, and even worthless). Affirmingly, several experiments show that texting while socializing with another person hurts the perceived conversation quality. Participants considered a conversation with a person who uses a smartphone during the interaction as lower in quality [1, 29] (e.g., phubbing - the act of snubbing someone in a social setting by looking at a smartphone instead of paying attention [40], negatively affects the perceived interaction quality and relationship satisfaction [12, 55].)

However, as technology evolves and mobile technologies become one of the main components of daily lives, their effects on users' wellbeing and social interactions also increase drastically.

### 2.3 Solutions to mitigate excessive smartphone use

Most of the market interventions that deal with the negative impacts of these smartphones are currently confined to digital apps for smartphones, web browsers, or computers. A study that analyzed 367 apps for "digital self-control" (i.e., setting use limits for apps and devices) on different application stores [37] provides four main categories for the interventions. The most common feature category involves blocking or removing distractions. The second category is "self-tracking" (i.e., tracking the time spent on devices). The third most common feature is "goal advancement," which aims to guide users towards the right tasks when using their smartphones by setting time/task goals and reminders. The final most common feature is "reward/punishment," which involves gamification and representation of 'points' gained through the amount of their device and app use.

Similar to market interventions, excessive smartphone use has recently attracted researchers working in HCI. For example, AppDetox [36] and The SAMS [32] are two mobile applications that allow users to set rules for the applications they want to use less. Let's Focus [28] aims to reduce phone use in classrooms by giving context-aware reminders to students. Unlike these solutions focusing on individuals and their intention to regulate their behavior, Lock n' lol [30] and NUGU [31] aim to reduce smartphone use by restricting group members' smartphone use time. They use a group-limiting mode to limit the application use and mute notification alerts. Unlike Lock n' lol, NUGU allows users to share their limiting time schedules and contexts in which they are willing to limit their smartphone use (e.g., studying, working). SCAN [50] monitors the interaction between group members by built-in sensors and defers notifications until it detects breakpoints in interactions like a moment of silence.

Similarly, LockDoll [11] is a doll based on smartphone use draws group members' attention by changing its lights or weaving its hands. Social Display [24] is an additional display that informs surrounding people about their mobile activities by showing the app they are currently using. However, most of these studies are solutions that aim to enable people who do not know each other to interact using smartphones or focus on giving feedback on the quality of their interactions [48]. Therefore, the need for an overall understanding of the field still merits further investigation in developing effective interventions to mitigate mobile devices' effects.

There are many alternative solutions valuable in expanding the design space for solutions mediating excessive smartphone use.

However, most of the interventions, both in research and industry, follow similar approaches (e.g., restricting, goal setting, reminding, and reward/punishment mechanisms (i.e., interventions which follow top-down approaches). These strategies depend on several conditions to be successful. Users should be aware of the adverse effects of excessive smartphone use, tend to mediate this use behavior [2], and have a high level of self-regulation [69] to insist on this behavior change decision. Also, using these strategies may even create unintended outcomes. For example, the results of a recent study showed that teens, who had to limit their social media use involuntarily, experienced negative feelings and increased the time they spent on social media after the break period was over [3]. Under these circumstances, as suggested in several studies [10, 20], the digital wellbeing interventions should move beyond a focus on restricting and showing screen time approaches.

### 2.4 Solutions to enrich collocated interactions

Contrary to these use-limiting concepts, the inefficacy of restrictive approaches has aroused attention from HCI researchers in recent years. Previous research investigated how to use technological devices to enhance collocated interaction. These enhancements can be grouped into different categories: "facilitating ongoing social situations, enriching means of social interaction, supporting a sense of community, breaking ice in new encounters, increasing awareness, avoiding cocooning in social silos, revealing common ground, engaging people in collective activity, encouraging, incentivizing or triggering people to interact." [48].

A study [25] mentions that even when people are physically collocated, they can create "cocoons" or bubbles using mobile devices that might reduce their collocated social interactions. The researchers developed PicoTales [56] to overcome this problem, a storytelling device that allows people to co-create stories while collocated. The prototype consists of a projector and a phone to create a shared experience, where people can project simple sketches to continue the story. Unlike these examples that trigger users to interact with each other using mobile devices, some studies give feedback about users' social interaction. Conversation Clock [8] is a table that visualizes the auditory input in face-to-face communication. It provides visual feedback to the users about their conversations and allows them to observe their contributions to the conversation.

FishPong [68], for instance, is a collaborative and cooperative interactive game designed to serve as an icebreaker, which enhances the social interaction of people. Cuesense [47] is a wearable display that shows some of the user's social media content related to the person encountered. It is designed to increase awareness and be an icebreaker in first encounters. Similarly, BubbleBadge [17] is also a textual display that provides supplementary information to enhance collocated social interactions, which is made to be worn like a brooch. The information displayed by BubbleBadge can break the ice in new encounters and trigger interactions in the later phases. Another study explored the ways to enhance social interaction between strangers with Social Devices which have audio-based interfaces [26]. These devices start to talk to each other and users during a social gathering to improve social interaction. They interact with users by asking questions or giving them random topics (e.g., movies, plans for the rest of the day).

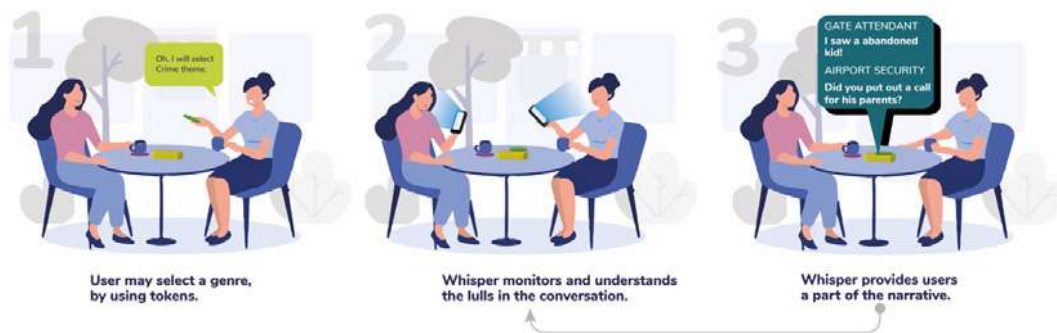

Figure 1: Interaction Scenario of WHISPER.

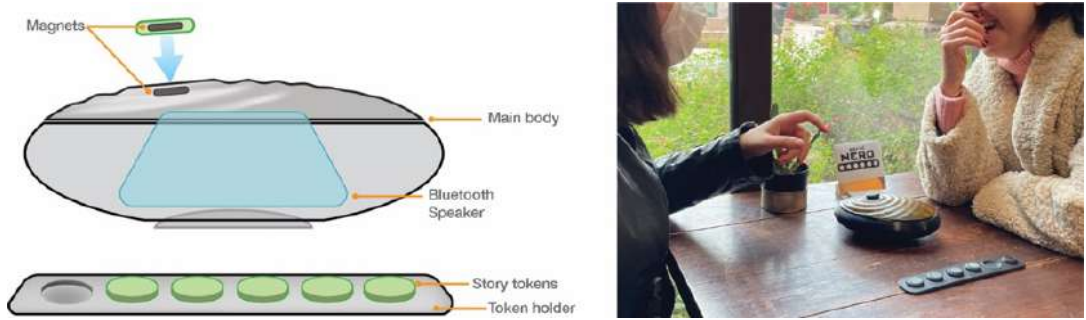

Figure 2: (Left) The cross-section of WHISPER, (Right) The Prototype in the Experiment Area

## 3 DESIGNING AND PROTOTYPING WHISPER

We aim to design a device that enriches social interactions in the presence of smartphones. Looking at the literature, we found a study exploring ways of enriching users' collocated interactions without limiting their usage of smartphones [20]. In this study, the researchers offered four different design approaches. *Enlighteners* raise awareness about the quality of social interaction. *Preservers* reduce the triggers that cause people to use their phones. *Supporters* improve the quality of conversations and help users avoid unexpected lulls. *Compliers* provide users with smooth isolation from social interaction when they do not want to interact. While designing WHISPER, we were inspired by the *Enlighteners* and the *Supporters'* because they are directly related to enriching the conversations and highlighted as the most promising approaches to intervene in social interactions [20].

WHISPER is an interactive audio narrative box. It follows lulls in the conversations to decide when to intervene. When there is a long lull or when individuals start using their smartphones instead of interacting with each other, it starts a pre-recorded audio story (20-30 seconds) to create a moment of curiosity, to gain their attention away from their phones, and to open up a new conversation. To create stories that attract attention, we were inspired by the notion of "keeping an ear out," i.e., actively trying to hear something in the environment. Hence, the stories included conversations from other tables (e.g., friends discussing their romantic relationships). With the story tokens, users can change the genre of the narrative (e.g., comedy, mystery). If no token is attached to the WHISPER, it plays an audio narrative from a random genre. WHISPER acts as a supporter and provides a potential topic to initiate a conversation. It acts as an enlightener, as the initiation of a story indicates low interaction between collocated people (Figure 1).

### 3.1 Creating the prototype of WHISPER

WHISPER consists of three parts: 1) the main body, which has a Bluetooth speaker to give the narrative and a magnet, 2) a holder that contains the tokens, and 3) story tokens with genres imprinted on them and a magnet to attach to the main body (Figure 2).

While designing WHISPER, the wavy form was chosen to create a resemblance to the sound waves (Figure 3e). The center point of the ripple pattern, where the magnetic tokens are placed, was designed to create intuitive interactions. We created a high-fidelity prototype to provide the participants with a more realistic experience. We used 3D printers to create all parts of the WHISPER (i.e., the narrative box) (Figure 3a). Following this step, we used post-processing to make the wavy area of the main body seem like wood (Figure 3). The wooden appearance helped us achieve a more finalized feel with its natural and cozy qualities, which gave our participants a more realistic experience. We applied these treatments because we wanted WHISPER to be an implicit and ambiguous product which can blend with the café environment in appearance and as an object which does not demand too much attention from the users when there is an active conversation. Thus, it provides users with implicit interaction [6], only when there are silent moments and when people check their phones. In addition, WHISPER's operating logic is ambiguous (i.e., why and when it intervenes in social interaction) and is unknown to the first-time users. WHISPER does not have a display explaining its purpose, or it does not have a button allowing participants to initiate interaction. It is only triggered when there are silences occur[1].

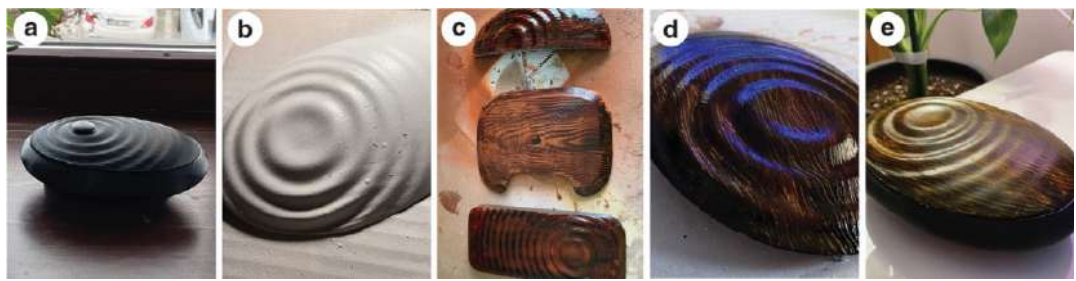

Figure 3: The Prototyping Steps of WHISPER.

Table 1: Final Narratives and Plots

| No | Genre | Plot |
| --- | --- | --- |
| 1 | Fantasy | While the two sisters play in their garden, they discover a mysterious gate and pass into another dimension. In the dimension they pass through, the world is different. |
| 2 | Tragedy | A woman has lost her 5-year-old child. She went to the police station, made her First Notice of Loss application, and waited in a cafe. Her sister came to see the woman. |
| 3 | Romance | A group of friends meets at the cafe. One of them fell in love with the boyfriend of her friend, who was not with them, and the guy said that he liked that girl. They discuss how to explain this to their close friends. |
| 4 | Sci-fi | Oxygen begins to be sold for money, and a large tax is taken from it. Since the two people sitting in the cafe are aware of this, they start arguing and discussing how to get oxygen cheaply. |
| 5 | Crime | One day, the wife of a wealthy businessman living in a mansion is found dead at home with her cat. One of the suspects and the journalist woman investigating the murder meet at a cafe. She claims she knows who the killer is. |
| 6 | Comedy | Three women are sitting in a cafe. They talk about their husbands. One is disturbed by how reckless her husband is; the other suffers from her husband's dependence on her, and the third is single. |

### 3.2 Creating the audio narratives

We collaborated with an improvisation club in Istanbul to create the stories during this procedure. Three female performers from the club volunteered to participate. We gave ten narrative starters (e.g., A group of friends learns that their best friends were cheated. They debate whether to tell this to him.) that we had prepared in advance for the participants one by one. We instructed them to imagine themselves seated at a café table while developing the stories. Our goal was to make the stories in WHISPER resemble more like the voices coming from the other tables in the café atmosphere. All the recordings were done with professional recording equipment in the recording studio. We post-processed the recordings to improve the audio quality (e.g., dialog editing, clipping, and cleaning) and added sound effects and foley to strengthen the sense of reality (e.g., walking, laughing, typing message ringing). We ended up with six stories of diverse genres (e.g., crime, tragedy, fantasy, sci-fi) (Table 1) [see supplementary material for example, narratives]. We divided each narrative into approximately 30-second segments. With these segmented narratives, we were able to play the next part of the same story during each lull moment.

---

[1] During the experiments, we did not inform participants about WHISPER's purpose to maintain its ambiguity for the participants.

## 4 USER STUDY

Our aim was not to design a commercial product and test its feasibility and acceptance by the users. Instead, we wanted to create a research prototype [45] to examine the following research questions: 1) How do users perceive technologies that monitor the interactions between collocated users and nudge them to maintain a conversation without limiting smartphone use? and 2) How would such technologies influence collocated social interactions?, in order to answer these questions. Thus, we built a semi-working prototype and used Wizard of Oz (WoZ) [14] as a method. We organized a total of 10 sessions. Nine of these sessions consisted of dyads, and one included a triad. Twenty-one people (12 females, nine males) participated in our field studies. All the participants were university students, as excessive smartphone use was widespread among the young population (e.g., Generation Z) [53]. We recruited them via sending recruitment emails to university channels and posting on social media channels. We asked volunteers to bring a friend to the experiment. On top of collecting demographics, we asked participants to state their level of intimacy with their friends out of 10, which indicates the perceived qualities of their relationships, and we asked them about their social media usage by using [27]. Participant characteristics are shown in Table 2.

Table 2: Demographics of the Participants

| No | Session | Gender | Age | Social Media Use (5) | Level of Intimacy (10) | Friendship Duration (Years) | No | Session | Gender | Age | Social Media Use (5) | Level of Intimacy (10) | Friendship Duration (Years) |
|---|---|---|---|---|---|---|---|---|---|---|---|---|---|
| 1 | 1 | Female | 19 | 3.87 | 8 | 2 | 12 | 6 | Male | 24 | 3.75 | 5 | 4 |
| 2 |   | Male | 21 | 3.37 |   |   | 13 |   | Female | 22 | 3.00 |   |   |
| 3 | 2 | Female | 21 | 2.75 | 8 | 1 | 14 | 7 | Male | 21 | 3.75 | 7 | 12 |
| 4 |   | Male | 20 | 2.12 |   |   | 15 |   | Female | 20 | 2.5 |   |   |
| 5 | 3 | Female | 22 | 5.0 | 9 | 2 | 16 | 8 | Female | 19 | 1.9 | 6 | 2 |
| 6 |   | Female | 21 | 3.5 |   |   | 17 |   | Female | 20 | 2.5 |   |   |
| 7 | 4 | Male | 19 | 3.5 | 5 | 3 | 18 | 9 | Male | 22 | 3.75 | 7 | 1 |
| 8 |   | Female | 23 | 3.12 |   |   | 19 |   | Female | 22 | 2.75 |   |   |
| 9 |   | Male | 24 | 3.75 |   |   | 20 | 10 | Male | 22 | 3.25 | 9 | 8 |
| 10 | 5 | Female | 21 | 3.37 | 6 | 3 | 21 |   | Male | 22 | 3.25 |   |   |
| 11 |   | Female | 23 | 3.12 |   |   |   |   |   |   |   |   |   |

Table 3: Number of audio narratives played for each session during lulls

| Session Number | 1 | 2 | 3 | 4 | 5 | 6 | 7 | 8 | 9 | 10 |
|---|---|---|---|---|---|---|---|---|---|---|
| Number of audio narratives played | 3 | 2 | 5 | 3 | 2 | 4 | 2 | 2 | 4 | 2 |

Without indicating the real aim of the study (i.e., enriching collocated social interactions), participants were invited to a one-hour interview about "Technology use during the pandemic." The reason behind concealing our study aim was to observe users' interactions as naturally as possible without priming them. The meeting point was a café at the authors' university campus. Participants went through three stages. For each step, we have the approval of the university's ethics committee. In the first stage, one of the researchers accompanied and told the participants to sit at the table. We shared a short online questionnaire including demographic information. After completing the survey, a researcher informed the participants that the interview would start in a few minutes. While the participants were waiting, the other two researchers observed them in disguise. Researchers remotely played the story tracks during each silence according to the interaction level between the users (e.g., two minutes of complete silence without any interaction). We played at least two narratives each session, at most five (Table 3). This part of the experiment took 25 minutes. In the last stage, we explained the real aim of the study and introduced WHISPER and its story tokens. Then, we conducted semi-structured interviews asking about their perceptions of the quality of social interaction, positive and negative impressions about WHISPER, their concerns towards the concept, and their expectations. All stages video-audio recorded and took approximately one hour.

We have two data sources for analysis: observation notes and interview transcripts. While interviews (i.e., participants' insights into WHISPER) serve as the main source, we used observation notes for 1) guiding interviews (e.g., researchers asked about the reason for a particular behavior during the interview) and 2) supporting interview results (e.g., describing a situation mentioned by a participant during the interview). Qualitative coding was used for the analysis of the sessions [41]. After familiarizing ourselves with the data by reading the transcripts, we selected two different transcripts to code them separately. The first three authors coded the transcripts individually, which will be later compared in a collective meeting to find and resolve inconsistencies. During coding, we employed both a deductive and an inductive approach. For the former, we used the interview questions as categories (e.g., Positive impressions regarding WHISPER or participants' concerns). For the latter, we retrieved codes from the interview and observation data, and we re-coded each category (e.g., Curiosity makes the users tamper with the device). With this method, we were able to discover new categories. We reviewed the compatibility of our codes after coding the first transcripts individually, which standardized them. We then proceeded to code all transcripts using the agreed-upon codes. Each time a new code emerged, the coders came together and updated the codebook after a short discussion. After the coding was finished, we categorized users' insights by affinity diagramming by collaboratively identifying thematically related codes, e.g., the code "intervention annoys the user and creates stress."

## 5 RESULTS

The design of WHISPER helped us explore three questions. 1) How would users react to the idea of intervening in their social interactions during lulls? 2) How would they react to receiving nudges from an implicit device blended with the environment? 3) What do they think about hearing an audio narrative during lulls in the conversations? Hence, we categorized the users' insights into these three questions. Our analysis revealed contrasted but rich insights into each subject.

### 5.1 Reactions to intervening in social interactions

We identified three themes regarding participants' reactions to the idea of intervening in social interaction.

*5.1.1 Interventions are helpful, but they can be annoying sometimes.* During the experiments, there were times of uninterrupted silence between participants (see Table 2). Parallel with the literature [20]; most users stated that these silences disturbed their interaction, creating an urge to end the gathering. One participant explained this by saying, *"Full silence stresses me out, and the need to constantly try to bring up a topic is exhausting" (P12)*. Therefore, participants found receiving interventions during the silences appropriate (N=11).

On the other hand, 13 participants indicated their concerns about the intervention, while three reported mixed opinions. They evaluate these interventions as crossing the limits by forcing them to talk all the time. One participant stated that she could get up from her seat and sit at another table to get rid of the WHISPER's nudges. Another participant took WHISPER during the experiment and handed it to the café staff. When we asked the reason for this behavior in the interviews, she stated that it discomforted her friend. She further explained by saying:

> *"Like someone else's child, he keeps coming and pulling my arm. It would bother me if it forced me into this situation. Always desire to talk. Always talk; it wants us to talk. It can turn into a dystopia. It turns into an authority controlling our interactions. If it were real life, I would get up and move to another table." (P6)*

*5.1.2 Silence is a good indicator of a low level of interaction, but not always.* One of the frequently discussed issues during the interviews was the time of interventions, which is directly related to the quality of the social interaction. By design, WHISPER intervenes in social interaction only during silent moments. Overall, participants understood this behavior in a way that WHISPER measures the level of their social interaction. Five stated that silence is an accurate indicator of a low level of social interaction. However, eight brought a different perspective by stating that social interaction should not be measured by the amount of speech and the silences in between. They said they do not always need to talk to people they feel close to and that the silences during social interactions are also pleasurable moments. One participant illustrated this with the following:

> *"I don't think it makes sense. Should we always talk about something? I am not sure. Maybe when you're on a first date, yes, it may help. But I have known her for a long time, and we can spend some time together without talking." (P14)*

Participants from four sessions were critical of the issue of talking constantly. They said how much fun they had from that moment (e.g., how much they laughed or how happy they felt) and how close they felt to the person in front of them determines the quality of social interaction. In addition, they stated that the feelings elicited after interaction are also an essential factor in determining its quality. In this case, they indicated whether they felt relieved after an interaction or whether the conversation added value to them could be used as a measurement. One participant said, *"If you feel bad... Getting up from that table feeling relieved from these negative emotions and feeling close to the person in front of you are the things that make that interaction quality higher." (P12)*

Three users suggested that different bodily metrics (e.g., blood pressure, anxiety level) can be tracked to measure the quality of interaction accurately. Still, six participants approached this suggestion cautiously as devices that collect and process such detailed data create a feeling of insecurity and make them feel dystopian. One participant explained this by saying:

> *"[Regarding his comment on other measurement options]... Dystopian! It's dystopian! Seems like a very dystopian product to me. I couldn't even focus elsewhere. My blood pressure? I don't know to share this information with a person but not a device. I don't know; it bothered me a lot." (P4)*

*5.1.3 WHISPER can be a conversation starter both in one-to-one and crowded group settings.* Participants stated that WHISPER could be useful in two social contexts, including one-to-one and crowded group meetings. Regarding the former, two of them said they needed a third person because they were worried about silence in one-to-one or small group meetings. One of them explained it as follows;

> *"The silence that occurs in a meeting makes me feel very uncomfortable. I need the presence of a third person in one-on-one meetings. Therefore, the presence of this product creates confidence in me, I believe. Besides being a good conversation starter, the concept can arouse good feelings." (P2).*

Seven of the participants also indicated that their need for WHISPER should be evaluated according to the level of intimacy of the relationship. There is no need for such a third-party agent to be among close friends and relatives. On the other hand, four participants emphasized WHISPER's potential as an icebreaker with newly met or not very close people. Users stated that the added value of WHISPER could be more significant in these contexts due to reasons such as the lack of common topics and interests to talk about or the existence of the introverts in the group.

Regarding the latter, seven participants stated that the probability of the lulls in large groups is lower than in smaller groups. They said that there are topics to talk about in crowded groups and that a new topic could be constantly sparked. In this type of group dynamics, participants stated that WHISPER could be used as a group activity:

> *I think it can be even more functional in the social interactions of large groups. Everyone's level of intimacy is different, and everyone has a different relationship* *with each other. On that level, opening such an exciting topic with WHISPER is better than someone trying to do it. Because sometimes, if people are not so outgoing and avoid socialization, the environment can be dull. I think this is a product that can fill the gap of this type of people. (P20)*

*5.1.4 WHISPER's interventions are perceived as warnings about the low level of social interaction.* During the interviews, we told participants that stories were played when their conversation quality was low. In response, eight participants considered listening to a story as feedback on their engagement quality. Although nine participants responded positively to this situation, others stated that listening to a story and receiving feedback in this sense can make them uncomfortable. One participant said he would try to be more talkative because of this warning mechanism and expressed these thoughts as;

*"It's worrisome to think that our engagement quality is so bad now that WHISPER is speaking. I would try to open a conversation when this device played a story. I would try to have better-quality interaction. I would feel bad. That isn't good. It's like a warning. I don't think I would want such a thing. We talk about what we want, but this is not its business to correct us? It cannot interfere with us." (P7)*

## 5.2 Reactions to receiving ambiguous nudges from an implicit device

Our analysis revealed two themes regarding their reactions to WHISPER's ambiguous and implicit nature.

*5.2.1 Implicitness creates surprise and encourages users to explore the product.* The implicitness of WHISPER, which aims to distract users as little as possible, has received general appreciation from the participants. They said that WHISPER fits well with the environment, the view of the café, and the objects in the environment. On the other hand, in two sessions, participants did not realize the existence of WHISPER till it played a story for the first time. The first intervention prompted most users to stare at each other and then interact with the device. Some users brought WHISPER close to their ears and tried to listen (Figure 4). Others played with the wavy surface to see if they could interact by assuming it may have a touchable surface. When the interaction between the participants faded after a while, users who received the second intervention paid more attention to understanding what was being said. Two participants stated that they intentionally waited for the sound since they did not know when it would come.

In addition to this 'standby' behavior and interaction with the main body, users mostly interacted with tokens. When they felt the magnetic pull between the tokens and the center point of WHISPER's ripples, they tried to explore how the tokens could be used. Most users tried each token one by one (N=9), while others put the tokens on top of each other (N=3) to see whether they could get any reaction from WHISPER (Figure 4). In two sessions, participants also noticed that the tokens changed the genre of audio narratives and stated that they were trying to get more information from the tokens using the NFC (i.e., near field data transfer) feature on their smartphones (Figure 4).

*5.2.2 Ambiguity drives users' curiosity and creates a sense of wonder.* The ambiguity about the product (i.e., being unaware of its aim and features) aroused both positive and negative feelings in users. Similar to listening to a conversation from other tables, WHISPER fires curiosity with its audio narratives. At the first moment of the audio intervention, all users showed signs of surprise, as they did not understand what was happening. They tried to figure out the reason behind sounds. A user thought someone had forgotten a Bluetooth speaker and tried to find that person.

*"I thought someone had connected this device by mistake. Something is coming from that person's phone, something I know too - there's a popular video on Twitter; for example, I expected to hear something familiar like this. That's why I was like, 'What is this?' I wondered whose it was in the café and tried to guess its owner (P10).*

WHISPER's ambiguous design aroused a sense of wonder in all participants. Participants did not know how to use the product or when to interact with WHISPER throughout the study. In eight sessions, participants stated that the concept is too vague in terms of its features and logic and that this curiosity and fidgeting behavior is effective in this short time (i.e., the duration of the experiment). They highlighted that not knowing answers to why and when WHISPER intervene might negatively affect the use of WHISPER in the long run.

## 5.3 Reactions to audio narratives

WHISPER relies on audial feedback to implicitly inform users that their level of interaction is low and utilizes stories to nudge them to engage in a conversation with the other. Analyzing participants' reactions to these features, we identified four themes.

*5.3.1 Audial feedback is good at attracting attention from smartphones.* As intended, all participants found the idea of using sound for interventions very effective. We observed that, once an intervention had been given through sound, while three participants shifted their attention from their smartphones to WHISPER, four put their phones aside and talked to their friends. One participant compared the audial feedback with visual feedback and stated that the former is more effective in shifting her attention from her phone as it dominates visual perception in the café setting. In three sessions, the participants stated that they could not hear the stories clearly due to the noise in the environment (e.g., high volume of background music or a crowded environment). Although this was mentioned as a disadvantage of sound modality, users said that missing the story is unimportant and that even words can suffice to fire a conversation. Affirmatively, one participant stated that he could only hear the word "gate" from the story (Table 1 – No 1). This created a conversation topic with his friend because it reminded him of an incident that happened to his girlfriend.

*5.3.2 The content of the stories is less important than what it triggers.* All the participants knew the stories were fictive, although we tried to make them genuine by using drama. While three participants would resemble the experience of hearing the stories to eavesdropping and expect to hear "real" conversations from other tables, five participants see the product as the third member of the group, who is nudging them to speak by providing some keywords. In addition to this, although most of the participants liked the stories and stated that they could start a conversation among themselves (N=18), they seemed to be less interested in the story's content. They perceived the stories as a trigger for starting a conversation. One participant explained this by;

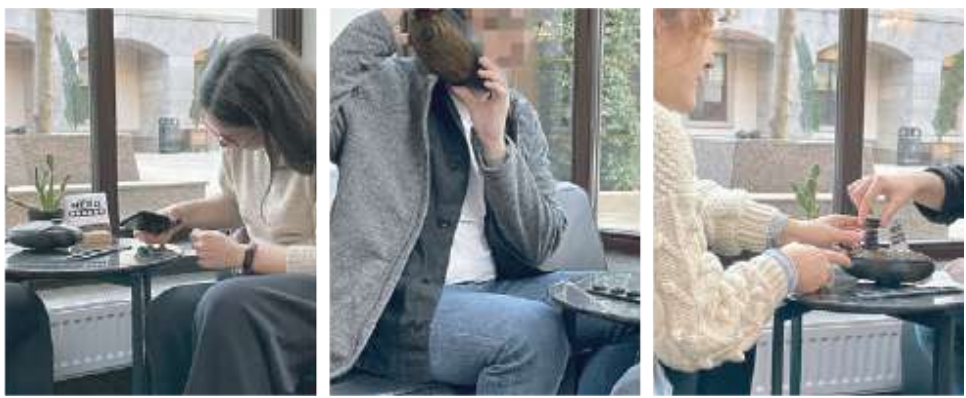

**Figure 4: (Left) Exploring tokens via NFC, (Middle) Listening to Whisper up to his ear, (Right) Stacking tokens to explore interactions**

> "One story caught my attention. I told her about it, and we talked for 5-10 seconds, but it did not attract her attention. Talking about the story is somehow productive, but we need to discuss what it implies. We can talk about the concept of time in the movie, not the movie itself. Talking about the story can be tedious, but what it evokes is more engaging." (P9).

In addition, a participant explains his experience of WHISPER creating a bond with her friend as follows;

> "The toxic lovers in the story. This is a topic that we talked about and made fun of. We looked at each other, laughed, and felt that we had agreed there." (P5).

Five participants stated that the idea of listening to the conversation of the other table in the café was a valid behavior for them. However, they also drew attention to the fact that the effect of this behavior could wear off in time. They offered some ideas for maintaining this effect (e.g., stories can be personalized according to what people have in common, or WHISPER can ask questions instead of telling stories). In the quote that follows, one participant states that one way to avoid this situation might be to play content that contributes to the conversations.

> "It is very attractive to hear something from the other tables. Only when this happens in real life, do we talk about it. Since this was a fictionalized thing, it wouldn't interest me after a while. Maybe something could enrich our conversation, like a paragraph from a book that we both like. I would love to use it. It must have contributed to my speech." (P19)

*5.3.3 Users would like to have control over the stories.* All the participants liked the idea of choosing stories or directing the stories by using the tokens. However, four participants preferred not being able to select the stories as they believed it was not in parallel with the idea of listening to what was happening at the next table. A participant explains the attractiveness of the uncertainty of events in a café as follows;

> "I liked its spontaneity and being something separate from me and my phone. Directing the story destroys this spontaneity. I don't know 'who' is talking 'what' at another table. I don't know about WHISPER's stories either. Not being able to select genres resembles it. That's why I didn't want to choose a genre." (P6).

## 6 DISCUSSION

Our aim in this study was to explore how users perceive WHISPER as a product that subtly measures people's social interaction quality and intervenes in their interactions and how it would influence their social interactions. Overall, the participants indicated that the concept is a valuable tool to enhance the quality of their social interactions. After WHISPER's every intervention, we observed an increase in participants' interactions. We also identified several issues that conflict with these positive reactions. These include whether silence can be used for measuring the quality of social interactions, whether WHISPER is more effective in small or crowded groups, and whether WHISPER's ambiguity affects positively or negatively. Since our goal with the study is not to validate our concept but to examine users' behavior and reactions to interactive technologies designed to boost social interaction, we believe that these conflicting insights would be beneficial for designing such technologies. To illustrate this usefulness, we discuss the implications of our findings for both design research and practice.

### 6.1 Implications for Design

Based on insights we gathered from the sessions, we conclude that design interventions to enhance collocated interactions should be *respectful*, *adaptable*, *targeted*, and *have a balanced ambiguity*.

*6.1.1 Respectful: Artifacts designed to influence user behavior should know when and how to give up.* As Fogg highlights [18], computers can be proactively persistent and implement their persuasive strategies repeatedly since they do not get tired. Similarly, WHISPER applies its persuasive strategy (i.e., providing audio narratives) at every lull moment without getting tired. Although we designed WHISPER as subtle as possible (i.e., it blends with the environment as it resembles a decorative object and only functions when there is a lull in conversation) and based on users' everyday practices (e.g., listening to a conversation at another table), some participants stated that they were uncomfortable with the WHISPER's constant nudges during lull moments. This implies that even a concept that subtly influences user behavior can backfire [59] and create an unintended effect, as seen in the case of a participant taking the device and handing it over to café personnel. Thus, artifacts designed to prevent lulls in face-to-face conversations should know how and when to give up if they cannot achieve the intended behavior of users. For example, when WHISPER invites users to interact for the first time, it might try to give interventions a couple of times by increasing the time intervals, e.g., it can wait for 10 minutes after the first intervention, and it can wait 20 minutes after the second intervention. If the users' social interactions are not affected positively despite these attempts, it might turn off and become an ordinary accessory in the environment. In this way, WHISPER might accept its ineffectiveness and not disturb users further.

*6.1.2 Balanced ambiguity: Designers should find a balance between implicit and explicit intervention.* Ambiguity can be seen as a design opportunity, as Gaver et al. remarked [19]. Parallel with our study; users may find it *frustrating*, sometimes *intriguing*, *mysterious*, and *delightful*. People are compelled to interpret these ambiguous situations, which leads them to begin struggling conceptually with systems and their settings. Thus, this struggle is resulted in more profound and more intimate relationships with the meanings supplied by those systems [19]. Inspired by this, we designed WHISPER in a way that does not reveal the meaning, reason, and timing of the interventions (i.e., narratives). This was preferred to ensure that users were affected by WHISPER's presence as little as possible during their social interactions (e.g., focusing on intervention rather than interaction). However, the implicitness of the concept, including its design and intervention logic, was one of the points participants contradicted themselves. Although some stated that ambiguity and implicitness are not a significant problem for them, we realized that they could not position this concept in their interactions as they cannot understand the purpose of WHISPER due to this over-implicit nature. As participants mentioned, there is also a surprise effect since WHISPER is an unusual product that offers completely different interactions. From this point of view, we can assume that WHISPER may disappear or be ignored once the surprise effect wears off. WHISPER should provide a balanced ambiguity in terms of its interventions and presence to prevent this. Hence, users can be more involved in the backstage of the concept. By revealing the interaction details and rules to the users, inviting the user to interact with the product should be considered in such a design to make the product more meaningful and overcome the short-term surprise effect. To provide this, before the audio narrative intervention, visual clues on WHISPER (e.g., light) can be used as tickets to interactions for users. By increasing or decreasing the blink rate of this feedback (note that it should give up after a certain time according to the 6.1 – Respectfulness approach), the user's attention can be drawn. Thus, if the users want to interact, they can volunteer to initiate this interaction by putting one of the tokens they choose.

*6.1.3 Adaptable: Being adaptive to the users' needs is the key to long-term interaction.* We observed that WHISPER is seen as a conversation starter by the participants. While this feature of WHISPER is appreciated by users, as they pointed out, the desired topics for conversations vary between individuals and the people around them. In the light of these differences, most users offered their suggestions, such as choosing the genre of the story or interfering with its narrative. The primary motivation for these is that people have a wide variety of things they like, their past experiences, common interests with their friends, and the topics for that moment. Therefore, they expect more topic-specific or personalized content, especially for longer, deep, ongoing conversations. All of these show that it is essential to design adaptive systems that can instantly react to users' emotions, thoughts, and needs and is shaped for a specific social interaction.

Unfortunately, the development of customized products raises many privacy problems along with advantages. Providing a meaningful, personalized experience to the user and keeping this experience consistent (which can be of interest to the user and most precise in terms of timing) often requires collecting more user data. Although users admit that their personal devices (e.g., smartphones, smartwatches) collect many data, as we have seen in our interviews, they do not prefer the use of this data by third-party technologies, e.g., WHISPER. Some participants even evaluated data collection for a personalized experience and measuring moments of lulls as "dystopic." Considering this, processing and presenting the user's data in WHISPER might inconvenience them. We have seen that the token system also relieves this discomfort in users. This system can be further developed and used during interaction may be advantageous for such a system. For example, we can allow users to make combinations by matching different tokens. By doing so, WHISPER can provide an enriched interaction suitable for them but does not violate their privacy by collecting personal data.

*6.1.4 Targeted: Nudges should be targeted at individuals rather than the group.* WHISPER aims to arouse users' curiosity by playing a pre-recorded narrative as if they were hearing real conversations between other people in a café environment. Although we designed WHISPER to trigger conversation subtly using this feature, some of the participants who understood the purpose of the prototype perceived it as a warning about their conversation quality. They felt bad each time WHISPER began a narrative, thinking their interaction quality was low. This implies that even though informing the user about an undesired behavior (e.g., not interacting with a collocated person) is a common strategy used in behavior change technologies [65], it may create unintended effects [59]. One way to address this could be making the nudges subtler and targeting them towards individuals rather than the group. For example, for WHISPER, using hidden directional speakers (e.g., ultrasonic sound systems) for giving audio narratives that only one person can hear instead of everyone at the table might be a solution to this concern of the users. By not knowing the source of the sound, they might open a conversation without knowing that their conversation quality is low and focus on the content that they hear instead of the product itself.

In addition to this, the individualized targeting feature of directional speakers can be utilized for other advantages. During our experiments, the audio narratives sometimes did not trigger anything in people; in some sessions, participants did not engage with the device despite the story. However, we found that certain words in stories prompt users to recall their memories and share them with others. Users can also be provided with asymmetrical information to trigger this sharing behavior further. For example, different parts of the same narrative can be communicated to different users by using directional speakers. In such a situation, users can be expected to share the missing pieces in the story with each other by creating curiosity about the targeted narratives. Complementing this asymmetric information by mutual sharing can provide a practice that can enrich users' social interactions by arousing the feeling of curiosity, as stated in previous studies [49].

## 6.2 Implications for Research

In addition to the implications that we gathered for designs to enhance collocated social interactions, we highlight and discuss our results in terms of quantifying a subjective phenomenon and designing responsible persuasive technologies.

*6.2.1 Towards quantifying the quality of social interactions.* In our user studies, we tracked the quality of social interaction between users by looking at the amount of the conversation. Though our aim was not to precisely measure the quality of social interaction, we observed participants' interactions to regulate WHISPER's behavior during the user study; we see a potential for future studies to explore methods of measuring the quality of social interactions. Looking at the previous work measuring the quality of interactions, one type of study utilizes self- or observer- reports. However, such measurements have disadvantages in subjectivity, credibility, and self-deception [70]. As a more objective measurement form, various devices have been developed in the last 20 years to track variables within the social context. Initial studies aimed to measure the collocated interactions with wearable sensor packages (an infrared (IR) transceiver, a microphone, and two accelerometers) called Sociometers [13]. Later, the solution space was extended with several studies which utilized different sensors or smartphones to measure dimensions of social interactions (e.g., distance, gaze, body position) [21, 39, 46]. In most studies, the level of social interaction was measured by metrics such as the amount of conversation between people (i.e., sound level) and, in some studies, by users' gaze directions.

Although these studies are valuable in measuring the quality of social interactions objectively, we need more research before designing interventions according to these measurements. As far as we have seen in our user study, participants had different reservations about measuring the quality of social interaction by monitoring the lull moments. Rather than being something that can be quantified and measured with scales, e.g., the volume of sound, the quality of social interactions is considered as high or low according to the pleasure taken from the moment or the satisfaction felt after the interaction. Participants described the silent moments as sometimes more valuable than those with conversations. Hence, investigating how the quality of social interaction as a phenomenon with such subjective elements could be observed, measured, and made sense of is worth exploring in the scope of future research.

*6.2.2 Responsible design interventions for accommodating unintended outcomes.* Ethical concerns exist wherever there is persuasion and technology, and examining these concerns has always been one of the cornerstones of persuasive technologies. Topics such as telling others how to live to be healthy, what to eat or how much exercise they should do have been studied extensively in persuasive technology. If the purpose of the methods applied for persuasion is to aim at harming 'good values,' this situation can be directly considered unethical. On the other hand, some strategies aiming at the wellbeing of society and the individual may have both intended and unintended outputs.

We find it worth discussing some ethical issues in WHISPER's use case in this context. WHISPER automatically monitors the interaction quality of users sitting at the table in a café environment. With the decrease in interaction between them, users receive the first intervention by WHISPER. We found that, after this stage, regardless of whether WHISPER increases the interaction of users, these interventions can disturb them and create frustration. In addition, we revealed that these interventions might also cause severe reactions in users, as mentioned in previous studies [20].

Due to its ambiguous and implicit nature, WHISPER's intervention and its reason are hidden from participants so that they can focus on interaction rather than intervention. This means that features of WHISPER, such as 'concealment of the aim' and 'intervening without permission,' exceed two of the main elements to consider when designing persuasive technologies [44] (i.e., autonomy/free choice and surveillance/privacy). In support of this, as we have seen in our interviews, these processes, which are beyond the control of the users, may cause the system to have adverse effects on them or lose its effect. As Berdichevsky and Neuenschwander have pointed out, a projected unintended output is a matter of debate in ethics [7], and the creators of this persuasive technology should pay attention to this issue. Because "Computers Cannot Shoulder Responsibility" [18] for unintended unethical outcomes that can reasonably be foreseen. Designers of this technology are the ones who should assume this responsibility, and such a reasonable forecast necessitates extensive user testing and creators' comprehensive forward-thinking [7].

## 7 LIMITATIONS

Our aim in designing WHISPER was not to create a commercial product that could effectively boost collocated social interactions. Instead, our purpose was to explore the implications of having such a product in a social setting for social interactions. Looking at the results, we gather rich insights, even conflicting, into this issue which might trigger future research. We tried to mimic a genuine experience with the experiment as much as possible. Nevertheless, the participants were volunteers who knew they participated in a study about 'technology use during a pandemic.' Though we canceled the real aim of the study, knowing that they participated in a research study might have influenced their behavior. Hence, in the future, we plan to assess the concept in a more naturalistic setting where we place WHISPER on tables in a café and do a participant observation. Also, we note that the study results are not generalizable to other populations and contexts, as the participants represent students from the author's university. Although this was purposeful as excessive smartphone use was widespread among young people (e.g., Generation Z) [53], in the future, we plan to conduct in-the-wild experiments with a sample consisting of participants having various characteristics (e.g., age, occupation).

## 8 CONCLUSION

Social interactions are essential for people's well-being. These interactions have changed with the proliferation of technological devices in our everyday life. In this paper, we presented WHISPER, an artifact designed to enhance people's collocated social interactions in public places without restricting smartphone use, and a user study assessing users' reactions to it and its potential influence on social interactions. With this study, we have seen that while behavior change technologies can be used to enhance collocated social interactions by encouraging individuals to interact with others in a social setting rather than looking at their smartphones, they might create unintended outcomes such as feeling frustrated when reminded of low social interaction. Hence, we synthesized four design implications to guide designers in mitigating these outcomes. Accordingly, technologies aimed at enhancing social interactions should be respectful to users' desire to have control over their interactions, adaptable to diverse user needs and preferences, targeted towards individuals rather than the group, and provide a balanced ambiguity to create surprise and curiosity without creating confusion. Furthermore, we provide two broad implications for HCI research: *the quantifying a subjective notion from users' lives (i.e., social interaction)* and *the responsibility aspect of the persuasive technologies* to intervene in the social interaction. Considering these implications, we plan to revise the WHISPER concept and conduct new in-the-wild studies exploring how a device that maintains the balance between implicit and explicit intervention does not disturb users with frequent interventions and gives up when it fails would impact social interactions.

Chapter 7

# DESIGNING THE CONCEPT 2: THE BOOST

**Research Question**

> **RQ3** How can we design interventions that support social interactions (without necessarily restricting smartphone use)?

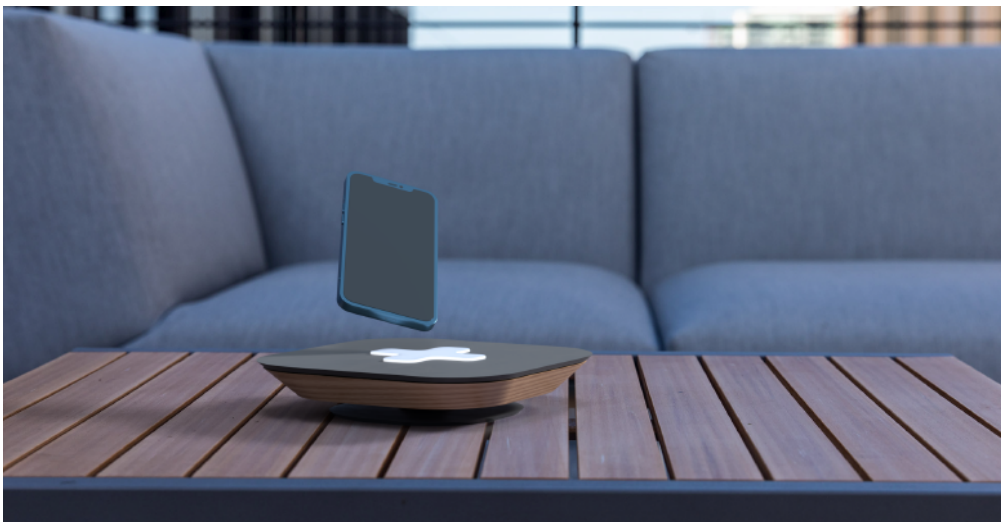

*Figure 7.14 The render of BOOST*

We made another design iteration, BOOST (Figure 7.14), with the information we obtained from the assessment sessions of the WHISPER concept in the previous chapter. In this chapter, we present our second concept's design and prototyping process.

## *7.1. Concept Development*

The BOOST has only a main body and a plus-shaped, translucent rubber surface on the top (Figure 7.14). Different from WHISPER, when the lulls occur, it provides short sentences related to a movie genre (e.g., Mystery). In the BOOST concept, while keeping the audio intervention modality of the previous concept, we have shaped our design criteria based on our three design implications (i.e., *respectfulness*, *adaptability*, and *having a balanced a*mbiguity) and two research implications as follows;

BOOST measures the impact rates of its nudges every time. If it is not successful in increasing the amount of the conversation, the time interval of the nudges increases. In this way, after unsuccessful attempts, it gives up intervening with audio and becomes a subtle feedback provider only with the help of light inside the body. Thus, BOOST concept is respectful to users' choices (Design Implication 1). Also, it knows the users' preferences before social interactions start (Design Implication 3). In the case of our experiment sessions, the device has the movie-genre preferences of the participants and gives the nudge from the pre-selected categories. In addition. To these features, it has visual feedback modality (i.e., LEDs indicating the level of conversation) in addition to audio narratives (Design Implication 4). The study concluded that sound is a dominant modality that cannot be ignored and mentioned that users in a similar setting would like to receive feedback about their conversation without taking any audio narrative nudge.

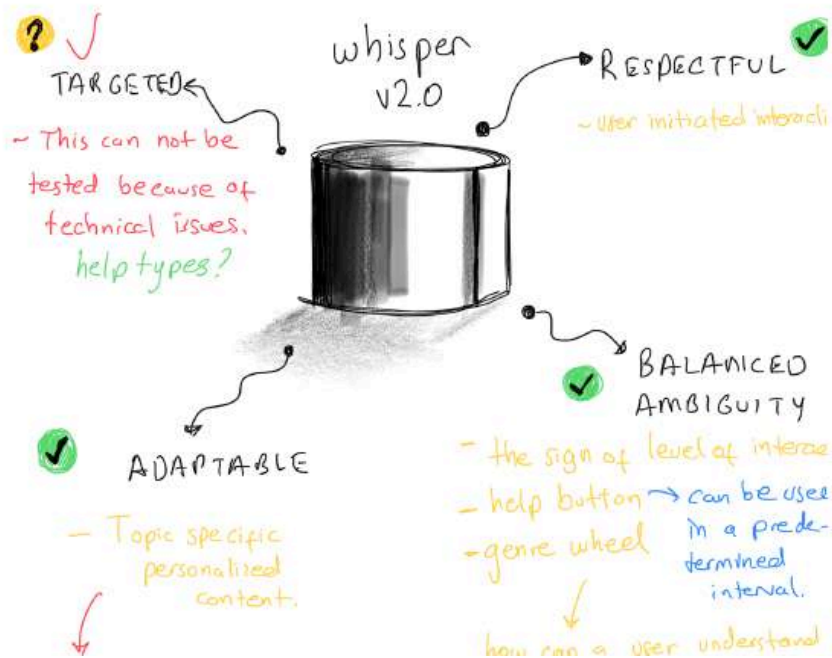

*Figure 7.15 The ideation sketch for the Design Implications*

## 7.2. *Prototyping of BOOST*

In this phase, we increased the prototype's fidelity and made a fully working one. As mentioned in the previous section, the features of this concept that we decided on according to design implications created constraints for us to do WoZ as we did in WHISPER. For example, BOOST automatically decides whether the intervention time intervals should be prolonged by comparing the amount of speech of the users before and after the intervention.

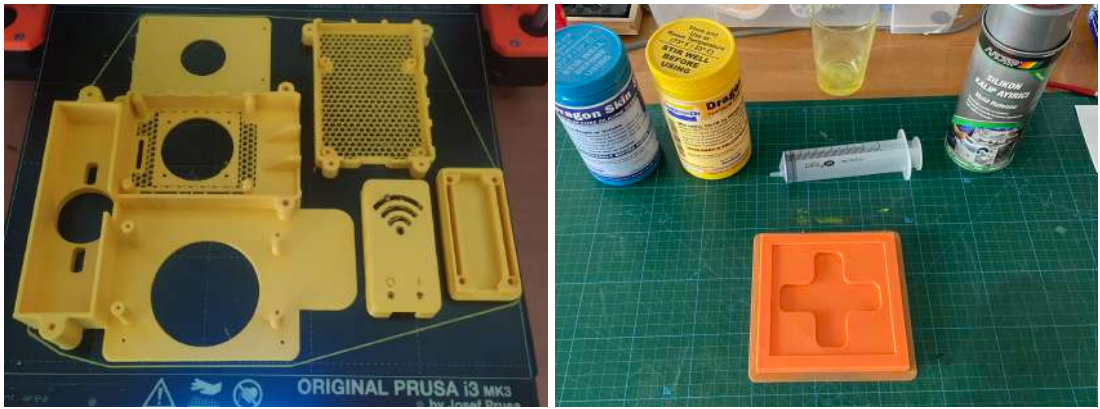

*Figure 7.16 a) 3D Printed casing for the electronics, b) The mold for plus-shaped surface*

Similar to Whisper, we printed all the outer parts of the product and the casing, where the internal electronics will be placed, with PETG material using a 3D printer (Figure 7.16a).

We also printed the negative mold of the translucent plus-shaped part using a 3D printer and produced it by casting two-component silicone material (Figure 7.16b).

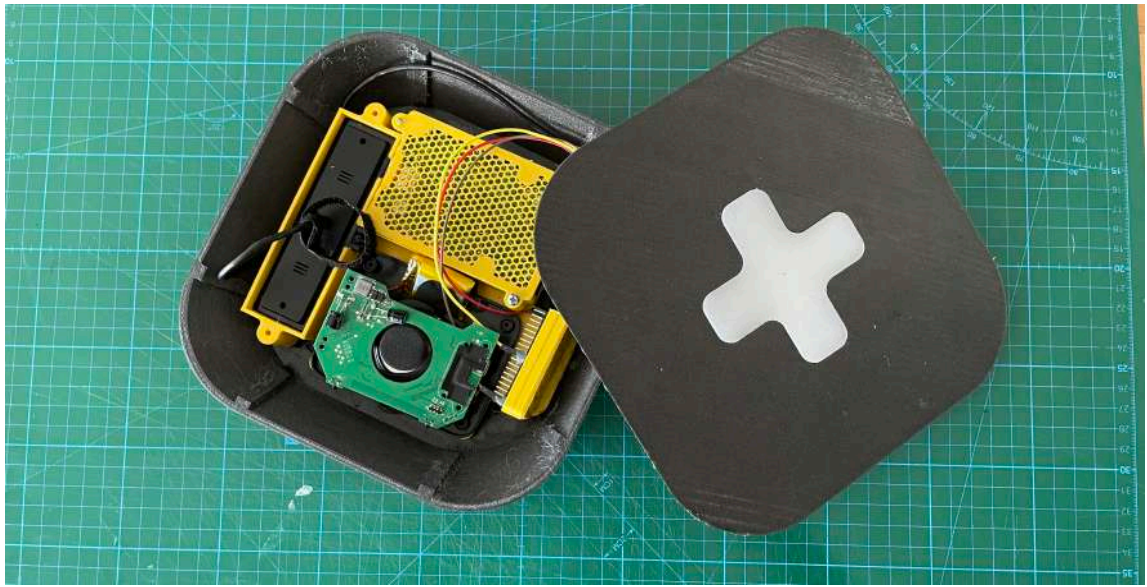

*Figure 7.17 The assembled BOOST*

As for the electronic part of the prototype (Figure 7.17), the product includes 1) a USB microphone that receives users' speech, 2) a Raspberry Pi 4B that classifies these conversations with machine learning, 3) a NodeMCU development board that transfers data to the cloud, and 4) a speaker that delivers voice interventions.

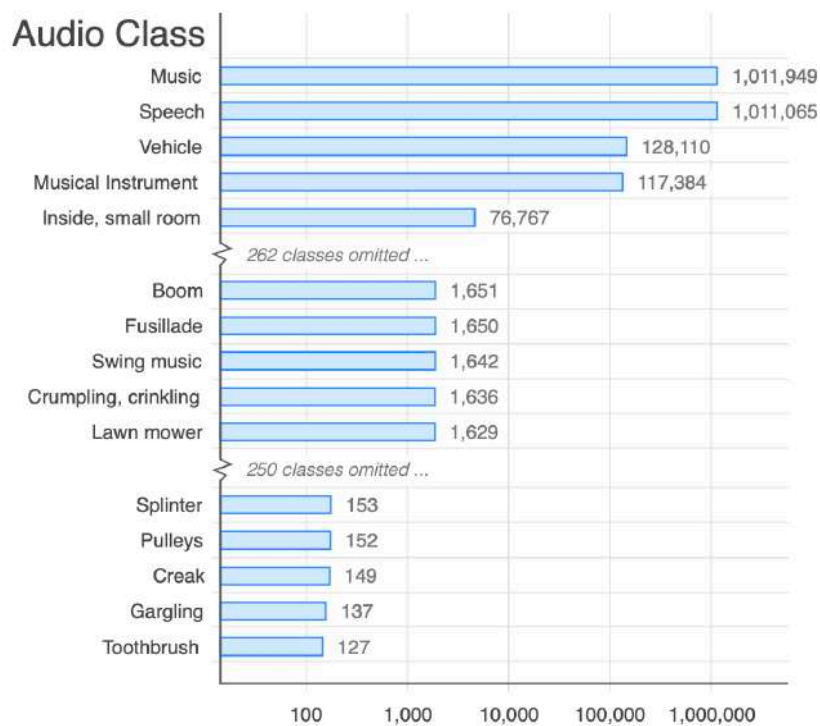

*Figure 7.18 The classification list of YAMNet*

We implemented a pre-trained machine learning model (i.e., YAMNet) with Tensorflow Lite (Abadi et al., 2016) to perform audio classification in real-time inside the Raspberry Pi environment. The model that we used, YAMNet (Google, 2019), is an audio classification model that has been pre-trained on the Google AudioSet dataset to predict 521 different audio events (Figure 7.18), and it helps us determine if there is speech in audio data received from the microphone. In our algorithm, the real-time audio is classified every second, and according to this classification, a score is determined for the amount of conversation users have. If the users' score is below a threshold we set for a certain period, users are given an intervention (Table 7.3). All this data is sent to the Arduino Cloud platform and stored.

| Time | Amount of Conversation | Speech | Intervention |
| --- | --- | --- | --- |
| 2022-11-15 11:54:08 | 100 | TRUE | FALSE |
| 2022-11-15 11:54:09 | 100 | TRUE | FALSE |
| 2022-11-15 11:54:10 | 90 | FALSE | FALSE |
| 2022-11-15 11:54:11 | 50 | FALSE | FALSE |
| 2022-11-15 11:54:12 | 20 | FALSE | FALSE |
| 2022-11-15 11:54:13 | 10 | FALSE | FALSE |
| 2022-11-15 11:54:14 | 0 | - | TRUE |
| 2022-11-15 11:54:15 | 50 | TRUE | FALSE |
| 2022-11-15 11:54:16 | 55 | TRUE | FALSE |

*Table 7.3 An excerpt from a CSV of a session*

As we mentioned before, unlike the previous concept, BOOST has short sentences related to cinema from a certain genre instead of audio stories. In the WoZ studies we conducted in Chapter 6, users stated that the notion of the story or a word told in the story can easily fire a conversation. In addition, they said that they would prefer content with different topics, even if they were from the same genre. For this reason, 90 sentences in six different types from eight genres (Table 7.4) were dubbed by three women and three men.

| Type | Sentence |
|---|---|
| Popularity | Adventure films became popular in Hollywood in the 30s and 40s with the films Robin Hood and Zorro. |
| Example | The Lord of the Rings series is one of the most successful and well-known examples of the adventure field and has greatly increased the recognition of Scandinavian mythology in the world. |
| Actor / Actress | When you think of adventure movies, Daniel Radcliffe and Johnny Depp come to mind with their serial films. |
| Fun fact | Did you know that Johnny Depp's popular Pirates of the Caribbean movie series is one of the most popular examples of this genre and that he was inspired by the Jack Sparrow character, pirate Yusuf Reis? |
| Platform | Uncharted, an adaptation of a video game series about a treasure hunt, can be watched on Netflix. |
| Theme | Some movies in the adventure genre focus on the theme of saving humanity. For example, Interstellar is about astronauts who set out to plug a black hole. |

*Table 7.4 The example sentences for a genre*

# Chapter 8

# ASSESSING THE BOOST: USER STUDIES

**Research Question**

**RQ4** How would the design interventions aimed at supporting rich social interactions influence the interactions between individuals in the presence of smart phones?

This section presents detailed information about our user studies, experiment procedure, and the insights we received from users after the experiment under four themes. With the BOOST prototype, similar to [Chapter 6](#), our aim was i) to validate these implications, ii) to observe the effects of the design concept we created on the collocated social interactions of the users, and iii) to get the users' views on the new concept.

## 8.1. Participants and Experiment Setting

Different from the real-life setting of the WoZ study in the previous chapters, this experiment was held in a lab environment where the participants were invited to a controlled and informant study. In the experiment room, we placed 1) a couch in front of a window with a sight of the forest, 2) our prototype, the BOOST, on a coffee table, and 3) several potted plants (Figure 8.19). With this arrangement, we aimed to provide users with a natural and cozy setting. We invited the participants using the online newsletter platform of the authors' university. We reached 21 individuals, and each participant was asked to bring their friends to the study (42 participants in dyadic groups).

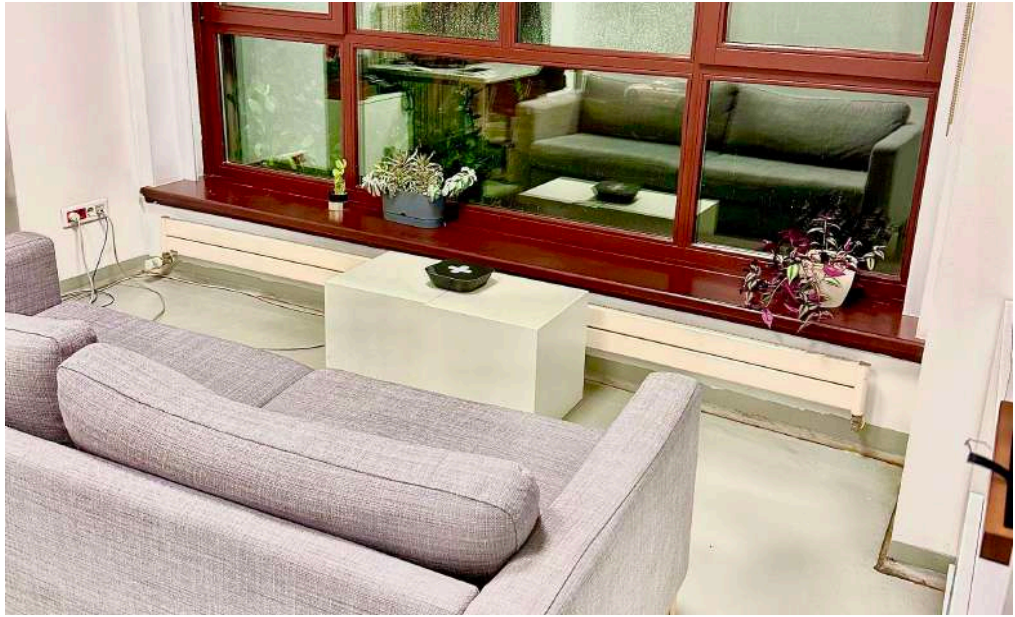

*Figure 8.19 A snapshot from the Experiment Room*

We organized 21 (1 Pilot, 10 Experimental, 10 Control) sessions with dyads. Our first session was a pilot study, and we finalized our experimental setting with this session. Since we changed the arrangement of the couch in the room, we did not include this session. In addition, one participant in the control group sessions did not fill out the post-questionnaire; we excluded this participant's session from the metrics. In total, 38 people (22 Females, 14 Males) participated in the study (Table 8.5). Two participants preferred not to state their genders. The mean age is 21.55 (*SD=2.04*), and the mean duration of their friendship is 3.19 years (*SD=2.29*).

|  | **Group Type** | **Mean** | **Median** | **SD** |
|---|---|---|---|---|
| Age | Control | 21.06 | 21.00 | 1.830 |
|  | Experiment | 22.00 | 22.00 | 2.176 |
| Friendship Duration | Control | 2.67 | 3.00 | 1.372 |
|  | Experiment | 3.67 | 3.00 | 2.848 |
| Gender | Control | 1.67 | 2.00 | 0.485 |
|  | Experiment | 1.70 | 2.00 | 0.657 |

*Table 8.5 Participant Distributions over Groups*

## 8.2. *Experiment Procedure*

Participants were invited to 90 minutes-study and went through three stages;

Stage A - Pre-experiment:

- Participants fill out a short questionnaire including demographic information, the Smartphone Use scale ([Appendix 3](#)), the Big Five Inventory ([Appendix 4](#)), and favorite genres to explore the participants' cinema preferences before the study begins.

Stage B – WoZ Experiment:

- The participants are invited to the experiment room, and researchers serve participants refreshments (i.e., coffee and tea). The researcher explains all the experiment details and the audio-video recorders, and the researcher leaves the room.

- If participants are in the experimental group, the BOOST concept will be turned on and waits for the users' interaction. If they are in the control group, the prototype will not be in sight but still track the interaction between the participants.

Stage C – Post-interviews:

- After 60 minutes, the researcher comes back to the experiment area.

- S/he gives the users a QR code linked to the second questionnaire, including a questionnaire with 15 items for participants' social interaction satisfaction ([Appendix 6](#)), Positive and Negative Affect Schedule (PANAS) (Gençöz, 2000), and the scale for the depth and breadth of the conversation.

- The researcher conducts semi-structured interviews if the participants are in the experiment group. In this part, we aim to get feedback about the concept and participants' feelings, thoughts, and concerns ([Appendix 7](#)).

*Chapter 8: Assessing the BOOST: User Studies*  103## 8.3. Analysis of Data

Throughout the experiments, we collected data from four different sources. First, we collected users' self-reports with pre- and post-questionnaires. Second, the device categorized the interaction between users every second during the sessions and created a separate CSV file for each session (Table 7.3). These files contain information about the lull moments, overall speech ratio, intervention counts, and time stamps. We combined these two sources into one datasheet and ran descriptive (G. Norman, 2010) and inferential statistical analyses, including different t-tests and ANOVAs. We performed these analyses using Jamovi 2.3.19 (The jamovi, 2022). We also placed a GoPro in the experiment room, taking snapshots every 10 seconds. The live stream feature of the GoPro also helped us to take observation notes and monitor whether everything was going as expected. Lastly, we audio-recorded our post-interviews with the users. In total, we collected 285 minutes of interviews, and they were transcribed. The transcripts were then analyzed using MaxQDA 2022, a software enabling easy sorting, structuring, and categorizing large amounts of qualitative data. We used MaxQDA because it speeds up the qualitative evaluation process without suggesting interpretations. In the analysis phase, we used reflexive thematic analysis (RTA) (Braun & Clarke, 2019, 2021). First, three researchers familiarized themselves with the data through transcripts, then discussed and refined initial codes following an inductive approach. We categorized the quotes into seven main groups (e.g., the ways of interacting with the BOOST) and have 268 sub-codes (e.g., Activating means that you need this device). Lastly, we thematized all the codes into 4 groups which we present in the next section.

## 8.4. Findings and Outcomes

The design of BOOST helped us explore three questions. 1) How would users react to intervening in their social interactions during lulls? 2) How would they react to receiving nudges from a smart device? 3) What do they think about receiving audio and visual feedback during lulls in the conversations? While combining our observational notes of sessions with the user interviews, we present four user insights and support these insights with our quantitative findings.

We discovered 11 themes under four main topics.

**The BOOST support users in their interactions by minimizing lulls**

- Nudges are significantly helpful in increasing speech ratio and decreasing negative feelings.
- The BOOST helps users to explore common interests and significantly positively affects their levels of intimacy.

**Users evaluated the modalities as a sign of their communication**

- Users evaluated the concept as a third friend in their social interaction.
- While visual nudges only raise awareness, audio nudges help enrich conversations.
- Visual nudges can be destructive since they are a sign of low-quality interaction.

**Users have different reservations about initiating the BOOST.**

- Activating the device is interpreted as a shameful action.
- Users prefer the device to start automatically rather than having a person start it.
- The design's ambiguity level helps users be aware of the product.

**Silence is also good in terms of social interactions**

- Users can also have quality social interaction in silence.
- Users call people who can stand quietly next to each other as close friends.
- Silence and social interactions escalate each other.

On the side of design implications, we concluded that the design aimed to enhance co-located social interactions should consider the following aspects;

1) Nudges should be aware of the context.
2) The feedback mechanisms should help the user while creating awareness about the situation.
3) The designs should invite the user rather than wait for initiation.

Paper 4:

# Boosting the Collocated Social Interactions

# Boosting the Collocated Social Interactions

**Hüseyin Uğur Genç**

KUAR, Koç University, Istanbul, Turkey, hgenc17@ku.edu.tr

**Duru Erdem**

Koç University, Istanbul, Turkey, derdem18@ku.edu.tr

**Çağla Yıldırım**

Koç University, Istanbul, Turkey, caglayildirim18@ku.edu.tr

**Aykut Coşkun**

KUAR, Koç University, Istanbul, Turkey, aykutcoskun@ku.edu.tr

The quality of social interaction greatly impacts individuals' psychological and physiological health. Previous research indicates that smartphones can have adverse effects on collocated social interactions. Most HCI works addressed this issue by restricting smartphone use during social interactions. Diverging from this previous work, we designed BOOST, an interactive audio narrative box that gives brief facts about a topic (e.g., cinema) during lulls in the conversations with the intention of encouraging face-to-face interaction in the presence of smartphones. We conducted a user study in a lab environment with 38 participants to understand how users react to BOOST and how it would influence their social interactions. In this paper, we present the result of this study and discuss three design implications; (1) Nudges should be aware of the context, (2) The feedback mechanisms should help the user while creating awareness about the situation, and (3) The designs should invite the user rather than wait for initiation, for developing this kind of technology to enhance social interactions without restricting any mobile technology use.

# 1 INTRODUCTION

Social interactions are essential for personal growth and social relations. They contribute to physical and mental well-being as a central component of individuals' overall health [63]. According to Self-determination Theory, people need to experience a sense of belonging and attachment to other people (i.e., relatedness) to achieve psychological growth [13]. Wanting to feel connected and be around other people is a natural impulse. Individuals' health can be fundamentally influenced by the quantity and quality of their support networks and social connections [6].

However, the ubiquity of smartphones in our daily interactions increased the number of distractions and elements of attraction that overshadow social interactions. Users have begun to use their mobile devices during their collocated social interactions, which are the synchronous and direct interactions between people nearby [34]. Now, in social meetings with friends, users can record that moment as a video or photo, browse social media, share digital content, or even instantly communicate with people they are not with. Despite these advantages, smartphone use adversely influences individuals' physiological and psychological health [25–27] and their relationships and damages attention and recognition of nonverbal emotional and social cues.

It is difficult for users to unplug from their smartphones as these devices have become increasingly integrated with their lives. There have been solutions to mitigate excessive smartphone use, both from research and industry. Most of these solutions utilize features that involve assisting, notifying, reminding, tracking, and confronting with excessive smartphone use. However, there is enough evidence that these solutions do not reduce the actual use, as suggested in several studies [3,6].

Addressing this need, we strived to understand how social interaction can be enriched in the presence of smartphones without essentially restricting their use. Within the scope of the study, we refer to social interaction as "the reciprocal interactions among individuals that happen during face-to-face encounters through verbal communication." We examined the following research questions:

1) How do users perceive technologies that monitor the interactions between collocated users and nudge them to maintain a conversation without limiting smartphone use?
2) How would such technologies influence collocated social interactions?

We designed a research prototype [25] called BOOST to answer these questions. This prototype is an interactive audio narrative box that gives short, pre-recorded facts about a topic during lulls in the conversations. We chose to focus on lulls since previous HCI work examining smartphone usage during social interactions [16] showed that smartphone checking behavior does not lead to poor social interactions; rather, it is the poor interactions like lulls in conversations that trigger smartphone use. Then, we conducted a controlled experiment with 38 participants, 19 sessions in total, to understand and measure how this prototype would influence social interactions.

We make the following contributions to literature. We present a novel design, BOOST, to enrich the collocated social interactions in public settings without necessarily restricting mobile technology use. We identified three design implications to guide designers in devising technologies to enhance collocated social interactions; these implications are (1) Nudges should be aware of the context and tailored to the user's needs, (2) The feedback mechanisms should help the user while creating awareness about the situation and (3) The designs should invite the user rather than wait for initiation.

## 2 RELATED WORK

### 2.1 Collocated Social Interactions

There are numerous situations in everyday life where social interaction would be beneficial, emotionally pleasing, or otherwise desirable (e.g., talking with a beloved one or sharing experiences with friends). Non-existent or insufficient social interaction would be problematic for individuals. Many empirical studies have analyzed the impact of social interactions, social trust, and the sense of belonging to a community on individual well-being. For example, a study [43] showed that people feel happier when interacting with close others. Another study [31] found that people tended to feel more satisfied after interactions with friends, followed by interactions with family members, others, and colleagues. Another study that obtains both self- and observer-reports of social interactions [40] showed that people report feeling happier and more socially connected when they spend more time interacting with others. People who report their relationships are more satisfying and supportive tend to have greater well-being [29].

Collocated interactions are critical for our relationships, creativity, and empathy. Sending an email, text, or social media post instantly links us with the community and gives us a sense of connection. However, real-life interaction truly connects and provides us with essential social support [42]. Sharing information, giving/getting advice, or gathering opinions from friends, coworkers, and family members help to create rapport, foster a sense of belonging, boost resilience, and help us process things and prevent overload. Face-to-face conversation leads to greater self-esteem and an improved ability to deal with others [35]. Even small talk is good for well-being. A study [45] found that short-term face-to-face conversations about the weather or other niceties can increase cognitive skills in the same way brain-teaser activities do. Another research [44] has shown that humans need to communicate with others because it keeps them healthier. There has been a direct link to mental and physical health. People with cancer, depression, and even the common cold improved their symptoms by talking with others. People who communicate their issues, feelings, and opinions with others are less likely to harbor grudges, resentment, or hatred, resulting in less mental and physical stress. In short, social interactions with others (e.g., family, friends, colleagues, etc.) are essential for individuals' general well-being.

**2.2 Negative Effects of the Smartphone on Collocated Social Interactions**

Excessive smartphone use may cause physical and mental health issues (e.g., joint and neck pain, sleep disturbances, depression, and smartphone addiction) [25–27]. It may have a negative effect on users' social interactions and relations. For example, people become less engaged with their immediate social environment due to heavy smartphone use during social interaction [2,10,30,39]. In an empirical study that addresses the impact of smartphone use during dyadic conversations on 238 participants, participants perceived this behavior as less polite and attentive [2]. Another study shows that when smartphone use occurs in interpersonal interaction, the time spent with friends becomes less valuable which is an aspect that is adversely and significantly related to users' life satisfaction [39]. They enjoy a meal with their friends less when their smartphones are present. People have tense arousal and boredom because they feel less socially connected and perceive time slower [14]. Excessive smartphone use is associated with lower relationship satisfaction with the romantic partner [37]. Besides using smartphones during social interaction, studies show that the mere presence of a phone on the table (even a phone turned off) changes what people talk about. If we think we might be interrupted, we keep conversations light on topics of little controversy or consequence [36]. Furthermore, conversations occurring in the presence of smartphones block empathic connection. If two people are speaking and there is a phone on a nearby desk, each feels less connected to the other than when no phone is present [30].

**2.3 Solutions addressing excessive smartphone use during social interactions**

There are many alternative solutions valuable in expanding the design space for solutions mediating excessive smartphone use [12,21–24,28]. However, most of the interventions, both in research and industry, follow similar approaches (e.g., restricting, goal setting, reminding, and reward/punishment mechanisms (i.e., interventions which follow top-down approaches). These strategies depend on several conditions to be successful. Users should be aware of the adverse effects of excessive smartphone use, tend to mediate this use behavior [2], and have a high level of self-regulation [47] to insist on this behavior change decision. Also, using these strategies may even create unintended outcomes. For example, a recent study showed that teens, who had to limit their social media use involuntarily, experienced negative feelings and increased the time they spent on social media after the break period was over [5]. Under these circumstances, as suggested in several studies [11,16], digital well-being interventions should move beyond a focus on restricting and showing screen time approaches.

Contrary to these use-limiting concepts, the inefficacy of restrictive approaches has recently aroused the attention of HCI researchers. Several research investigated how to use technological devices to enhance collocated interaction. These enhancements can be grouped into different categories: "facilitating ongoing social situations, enriching means of social interaction, supporting a sense of community, breaking ice in new encounters, increasing awareness, avoiding cocooning in social silos, revealing common ground, engaging people in collective activity, encouraging, incentivizing or triggering people to interact."[34].

A study [19] mentions that even when people are physically collocated, they can create "cocoons" or bubbles using mobile devices that might reduce their collocated social interactions. To overcome this problem, the researchers developed PicoTales [38], a storytelling device that allows people to co-create stories while they are collocated. The prototype consists of a projector and a phone to create a shared experience where people can project simple sketches to continue the story. Unlike these examples that trigger users to interact with each other using mobile devices, some studies give feedback about users' social interaction. Conversation Clock [7] is a table that visualizes the auditory input in face-to-face communication. It provides visual feedback to the users about their conversations and allows them to observe their contributions to the conversation.

FishPong [46], for instance, is a collaborative and cooperative interactive game designed to serve as an icebreaker, enhancing people's social interactions. Cuesense [33] is a wearable display that shows some of the user's social media content related to the person encountered. It is designed to increase awareness and be an icebreaker in first encounters. Similarly, BubbleBadge [15] is also a textual display that provides supplementary information to enhance collocated social interactions, which is made to be worn like a brooch. The information displayed by BubbleBadge can break the ice in new encounters and trigger interactions in the later phases. Another study explored ways to enhance social interaction between strangers with Social Devices with audio-based interfaces [20]. These devices start to talk to each other and users during a social gatherings to improve social interaction. They interact with users by asking questions or giving them random topics (e.g., movies and plans for the rest of the day).

BOOST differs from these previous designs by detecting the silence in the conversations by using a machine learning model which categorizes the sounds (e.g., speech, silence) to decide when to intervene and give nudges in the form of brief audio facts. These nudges consist of short sentences with different genres, and it aims to open new conversations between people to end the silences which can be caused by phubbing behavior. In addition to these audio nudges, BOOST also informed the low interaction by giving visual feedback with its lights. Whenever BOOST detects the silence in the conversation, it turns on its light. After some time, if the low interaction continues, it starts the nudge with a random sentence and aims to increase the users' awareness about the quality of their social interaction. This way, BOOST aims to trigger new and ongoing face-to-face interactions.

## 3 BOOST: AN AUDIO NARRATIVE BOX TO SUPPORT RICH COLLOCATED INTERACTIONS

### 3.1 Concept Development of BOOST

The BOOST has only a main body and a plus-shaped, translucent rubber surface on the top (Figure 1). When the lulls occur, it provides short sentences related to a movie genre (e.g., Mystery). In the BOOST concept, we have shaped our design criteria based on three design implications (i.e., respectfulness, adaptability, and having a balanced ambiguity) and two research implications from our previous study [16]. Accordingly, technologies aimed at enhancing social interactions should be respectful to users' desire to have control over their interactions, adaptable to diverse user needs and preferences, targeted towards an individual rather the group, and provide a balanced ambiguity to create surprise and curiosity without creating confusion. Furthermore, we provide two broad implications for HCI research: *the quantifying a subjective notion from users' lives (i.e., social interaction)* and *the responsibility aspect of the persuasive technologies* to intervene in social interaction.

BOOST measures the impact rates of its nudges every time. If it is not successful in increasing the amount of the conversation, the time interval of the nudges increases. In this way, after unsuccessful attempts, it gives up intervening with audio and becomes a subtle feedback provider only with the help of light inside the body. Thus, BOOST concept is respectful to users' choices (Design Implication 1). Also, it knows the users' preferences before social interactions start (Design Implication 3). In the case of our experiment sessions, the device has the movie-genre preferences of the participants and gives the nudge from the pre-selected categories. In addition. To these features, it has visual feedback modality (i.e., LEDs indicating the level of conversation) in addition to audio narratives (Design Implication 4). The study concluded that sound is a dominant modality that cannot be ignored and mentioned that users who are in a similar setting would like to receive feedback about their conversation without taking any audio narrative nudge.

### 3.2 Prototyping BOOST

In this phase, we increased the prototype's fidelity and made a fully working one and aimed to provide real-like experiences to the user for the concept. As mentioned in the previous section, the features of this concept that we decided on according to design implications created constraints for us to do WoZ as researchers did in previous studies [16]. For example, BOOST automatically decides whether the nudge time intervals should be prolonged by comparing the amount of speech of the users before and after the nudge.

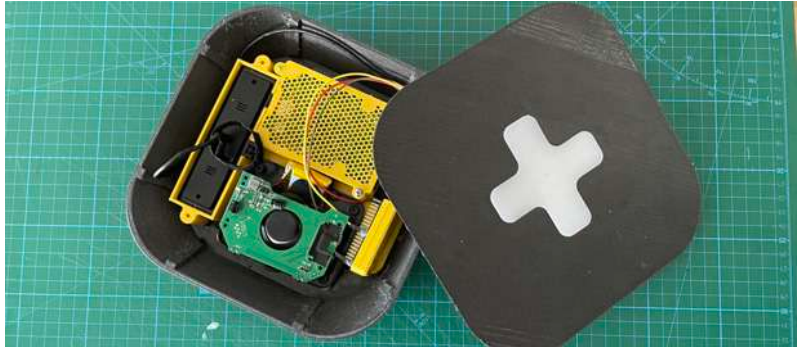

Figure 1 – Left) 3D Printed parts for the electronics, Right) The mold for silicone casting, Bottom) Assembled Parts

As for the electronic part of the prototype, the product includes 1) a USB microphone that receives users' speech, 2) a Raspberry Pi 4B that classifies these conversations with machine learning, 3) a NodeMCU development board that transfers data to the cloud, and 4) a speaker that delivers nudges. We implemented a pre-trained machine learning model (i.e., YAMNet) with Tensorflow Lite [1] to perform audio classification in real-time inside the Raspberry Pi environment. The model that we used, YAMNet [18], is an audio classification model that has been pre-trained on the Google AudioSet dataset to predict 521 different audio events. It helps us determine if there is speech in audio data received from the microphone. In our algorithm, the real-time audio is classified every second, and according to this classification, a score is determined for the amount of conversation users have. If the users' score is below a threshold we set for a certain period, users are given a nudge. All this data is sent to the Arduino Cloud platform and stored. 90 sentences in six different types from eight genres were dubbed by three women and three men (Table 1).

| Type | Sentence |
|---|---|
| Popularity | Adventure films became popular in Hollywood in the 30s and 40s with the films Robin Hood and Zorro. |
| Example | The Lord of the Rings series is one of the most successful and well-known examples of the adventure field. |
| Actor / Actress | When you think of adventure movies, Daniel Radcliffe and Johnny Depp come to mind with their serial films. |
| Fun fact | The Jack Sparrow character was inspired by pirate Yusuf Reis. |
| Platform | Uncharted, an adaptation of a video game series about the treasure hunt, can be watched on Netflix. |
| Theme | Some movies in the adventure genre focus on the theme of saving humanity. For example; Interstellar is about astronauts who set out to plug a black hole. |

Table 1 – Examples Sentences for Adventure genre

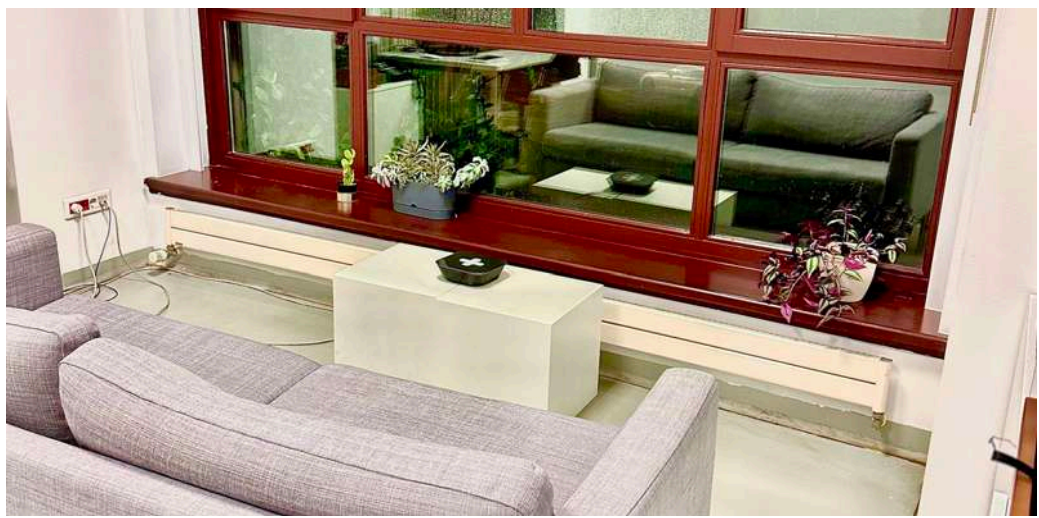

Figure 2 - A snapshot from the Experiment Room

# 4 USER STUDY

## 4.1 Recruitment and Study Setting

This study was held in a lab environment where the participants were invited to a controlled and informant study. In the experiment room, we placed 1) a couch in front of a window with a sight of the forest, 2) our prototype, the Boost, on a coffee table, and 3) several potted plants. With this arrangement, we aimed to provide users with a natural and cozy setting. We invited the participants using the online newsletter platform of the authors' university [4,17]. The reason behind selecting the participants from this generation is that young generations are more likely to use their smartphones excessively. With this announcement, we reached 21 individuals, and each participant was asked to bring their friends to the study.

We organized 21 (1 Pilot, 10 Experimental, 10 Control) sessions with dyads. Our first session was a pilot study, and we finalized our experimental setting with this session. Since we changed the arrangement of the couch in the room, we did not include this session. In addition, one participant in the control group sessions did not fill out the post-questionnaire; we excluded this participant's session from the metrics. In total, 38 people (22 Females, 14 Males) participated in the study (Table 2). Two participants preferred not to state their genders. The mean age is 21.55 (SD=2.04), and the mean duration of their friendship is 3.19 years (SD=2.29).

|  | Group Type | Mean | Median | SD |
| --- | --- | --- | --- | --- |
| Age | Control | 21.06 | 21.00 | 1.830 |
|  | Experiment | 22.00 | 22.00 | 2.176 |
| Friendship Duration | Control | 2.67 | 3.00 | 1.372 |
|  | Experiment | 3.67 | 3.00 | 2.848 |
| Gender | Control | 1.67 | 2.00 | 0.485 |
|  | Experiment | 1.70 | 2.00 | 0.657 |

Table 2 – Participants Demographics

Participants were invited to 90 minutes-study and went through three stages. Before attending the study, participants filled out a short questionnaire including demographic information, smartphone usage habits, and Big Five Inventory with users to investigate the relationship between personality traits and our scenario and favorite genres to explore the participants' preferences before the study begins. During the session, the participants are invited to the experiment room, and researchers serve participants refreshments (i.e., coffee and tea). The researcher explains all the experiment details and the audio-video recorders, and the researcher leaves the room. If participants are in the experimental group, the BOOST concept will be turned on and waits for the users' interaction. If they are in the control group, the prototype will not be in sight but still track the interaction between the participants. After 60 minutes, the researcher returns to the experiment area, and s/he gives the users a QR code linked to the second questionnaire, including the social interaction satisfaction scale, the scale for the depth and breadth of the conversation to explore whether there is an effect of the design concept. The researcher conducts semi-structured interviews if the participants are in the experiment group. In this part, we aim to get feedback about the concept and participants' feelings, thoughts, and concerns.

**4.2 Analysis**

Throughout the experiments, we collected data from four different sources. First, we collected users' self-reports with pre- and post-questionnaires. Second, the device categorized the interaction between users every second during the sessions and created a separate CSV file for each session. These files contain information about the lull moments, overall speech ratio, nudge counts, and time stamps. We combined these two sources into one datasheet and ran descriptive [32] and inferential statistical analyses, including different t-tests and ANOVAs. We performed these analyses using Jamovi 2.3.19 [41].

We also placed a GoPro in the experiment room, taking snapshots every 10 seconds. The live stream feature of the GoPro also helped us to take observation notes and monitor the participants whether everything was going as expected. Lastly, we audio-recorded our post-interviews with the users. In total, we collected 285 minutes of interviews, and they were transcribed. The transcripts were then analyzed using MaxQDA 2022, a software enabling easy sorting, structuring, and categorizing large amounts of qualitative data. We used MaxQDA because it speeds up the qualitative evaluation process without suggesting interpretations. In the analysis phase, we used reflexive thematic analysis (RTA) [8,9]. First, three researchers familiarized themselves with the data through transcripts, then discussed and refined initial codes following an inductive approach. We categorized the quotes into seven main groups (e.g., the ways of interacting with the Boost) and have 268 sub-codes (e.g., Activating means that you need this device). Lastly, we thematized all the codes into four groups which we present in the next section.

# 5 RESULTS

The design of BOOST helped us explore three questions. 1) How would users react to intervening in their social interactions during lulls? 2) How would they react to receiving nudges from a smart device? 3) What do they think about receiving audio and visual feedback during lulls in the conversations? While combining our observational notes of sessions with the user interviews, we present four user insights and, support these insights with our quantitative findings.

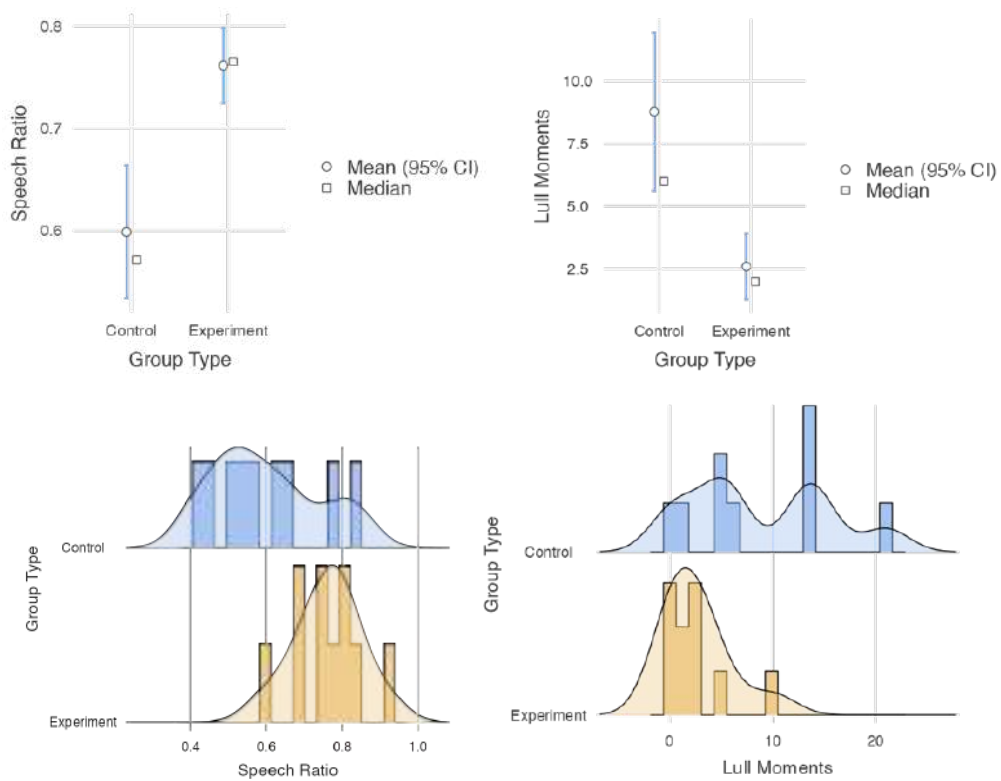

Figure 3 – Upper Left) Mean for Speech Ratio, Upper Right) Mean for Lull Moments, Bottom Left) Distribution of Speech Ratio, Bottom Right) Distribution of Lull Moments

## 5.1 The Boost support users in their interactions by minimizing lulls

During the sessions, there were uninterrupted silences in most groups. These silences did not occur in 4 sessions, three in the experimental group and one in the control group (Figure 3). Most of the experiment participants stated that Boost helped their interactions positively (*N*=15/18).

To support this claim, an independent-samples t-test was conducted to compare Speech Ratios according to Group Types. There was a significant difference in the scores for Speech Ratio in Control Groups (M=0.59, SD=0.14) and Experiment Groups (M=0.76, SD=0.08); t(36)=4.38, p < .001. Also, another independent-samples t-test showed a significance between the lull moments in the group types; t(36)=3.67, p < .001, (Control (M=8.77, SD=6.82), Experiment (M=3.60, SD=3.01)) in Figure 3.

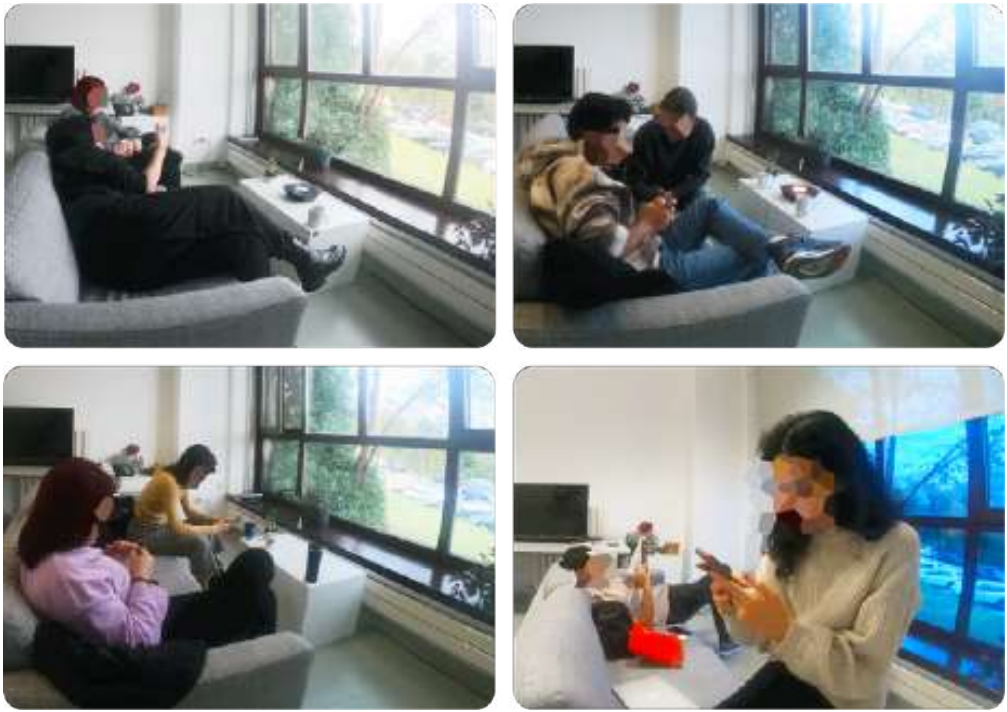

Figure 4 – Snapshots from the sessions, Top) Experiment Groups, Bottom) Control Groups

In our observations (Figure 4), we encountered three different types of Boost's ability to open a conversation; 1) They talked directly about the topic in the nudge (e.g., A sentence about Dune was given, and the participants talked about the movie), 2) They first started to talk about the topic in the nudge, and the conversation evolved to another point (e.g., A sentence about Dune was given, and the participants first talked about it and after that switch to personal tastes.) and 3) they thought the audio nudges were feedback about their silence and they thought that they should talk regardless of this topic and opened another topic (e.g., A sentence about Dune was given, and the participants started to talk about economic changes of the world)

In addition, users said they could not hear every word of the sentences but could only hear some keywords. In general, they state that just hearing these words helps them. Although they started to talk about the topic that Boost brought up, they pointed out that it opened up other topics and touched on each other's personal points. For example, a dyad who had known each other for three months said that after hearing the nudge about the movie, Dune, they discussed the cast and, after, each other's personal tastes about people. Commenting on a similar situation, another participant said it was a good opportunity to discover their common points and get to know each other better.

Parallel to this issue related to common points, it was frequently discussed among the participants for whom such a concept would be better. Some groups noted that there were more awkward silences between people who knew each other less and that they might need a technology like such a product to spark conversation in such a context.

Spearman's rank correlation was computed to assess the relationship between Friendship Duration and Speech Ratio. Contrarily, our results suggest a significant but negative correlation between these two variables, $r(17) = -.44$, $p = .03$. These relationships suggest that while duration of friendship increases, speech ratio decreases as visualized in Figure 5.

Concerning this topic of finding common ground and getting to know each other, when we performed a repeated-measures ANOVA for the intimacy assessment we received from the participants before and after the experiments, the Boost had a statistically significant effect on intimacy levels between dyads ($F(1, 36) = 4.58$, $p = .03$). In addition to that, we performed a one-way ANOVA for the PANAS assessments of the users, and there is a statistically significant difference between the groups ($F(1, 36) = 4.94$, $p = .03$). A Games-Howell post hoc test revealed that negative emotions were statistically significantly lower in the experiment groups ($11.5 \pm 0.37$) compared to the control groups ($16.06 \pm 2.12$, $p = .04$).

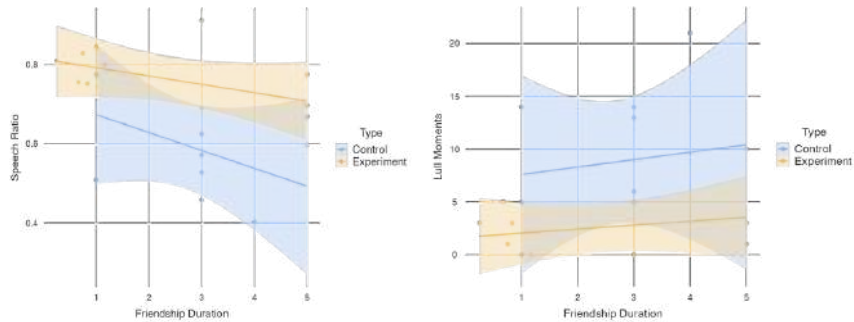

Figure 5 – Left) Speech Ratio vs. Friendship Duration, Right) Lull Moments vs. Friendship Duration

Overall, these results suggest that the design concept, Boost, does have a significant effect on speech ratio, lull moments, negative feelings, and, most importantly, users' intimacy levels. Specifically, our results suggest that when users experience the Boost, their tendency to chat with each other, their feeling of intimacy increases, and negative feelings decrease.

## 5.2 Users evaluated the modalities as a sign of their communication.

In the previous study's results, there are positive and negative comments about a device that monitors users during their social interactions and gives them audio stories to talk about when their interaction quality drops. At the same time, in the previous study, users made analogies for such a product as a scary and dystopic product. Boost, which we designed in accordance with the design implications of the previous study, did not receive negative feedback from users in this direction. Rather, it was seen as a companion that people could carry as a personal device. Some users elevated the status of Boost to a friend who is eager to talk and encouraging others to talk. They emphasized that seeing this product as a companion could relieve people who want to talk or are nervous about speaking.

In addition to Boost's ability to support people's conversations, users stated that they received feedback on their interactions and made both supportive and critical comments about it. Interestingly, most negative comments were on giving feedback with the light as one of the features requested by the participants of our previous study. They evaluated the audio medium as distracting, while a visual medium can be disregarded easily. This appears inconsistent with the previous study's user expectations of seeking visual feedback rather than sound. When asked about the users' opinions on this subject, the most basic concern is that the light is just feedback and a modality that makes them realize that they have failed to communicate with their friends. However, they stated that sound creates a similar awareness but tries to help them differently than light;

> *The sound has an effort. It actually gives you a hand. Right there, the quality of the conversation drops, and it says, "Here's a hint, you can talk about it," but [light] is more like, "there is a problem," you know? (P5)*

Another user commented on the negative effect of the feedback feature as destructive in relationships by saying;

> *Let's say you are sitting with a close friend or girlfriend, communicating or something; then the device turns on the light. It's a very stressful thing. Do we have a problem? Can't we communicate? Should we get therapy? Is our relationship over? Something that can be very devastating. We can't establish healthy communication, you know? The light is on now. (P14)*

In addition to the effect of light on dyadic relations, users said that this situation could also negatively affect them in public settings. According to them, in a cafe environment, the fact that a device like Boost would light up and give feedback when the interaction quality decreased meant that other people would see these "failures."

> *It seems that this light will create a lot of social anxiety. So you look at the device, and there is a perception in a social environment. The perception of others about us. How do we interact with each other? (P6)*

Moreover, some participants interested in their smartphones during the sessions stated that when the light turned on, it caught their attention, and they stopped using the phone. For them, it was a positive invitation to return to social interactions. Against this idea of invitation, three participants stated that when the light on after social interaction drops below a certain threshold, it creates stress for them. As a result, we can say that it is not enough to get feedback on a problem like the weakness of their social interactions, and it will make people feel uncomfortable. Users are looking for a solution that helps them to overcome this problem.

> *I was very nervous [talking about light] about it. That's when I got really nervous. You know, the light is on. I was like, "I need to talk, and I need to talk." I think it would be better if it spoke directly. (P7)*

**5.3 Users have different reservations about initiating the Boost.**

For Boost to start listening to users in the ideal usage scenario, a user had to place their smartphone on the plus-shaped area on the top of the device. However, in our study, we told the participants that from the moment we called the study started, Boost would start tracking their interactions and would give a nudge using light and sound according to the amount of interaction. However, we also wanted to get users' opinions on this issue and got interesting reactions. In a previous study, which conducted six focus group sessions, participants reported that they were uncomfortable with the other person's smartphone use and made warning gestures (e.g., shaking feet, sighing, or taking their friend's phone away). We observed that the interaction scenario to which we refer to this finding creates a different impression when used on a mediator product like Boost. Most participants (N=16) did not find it appropriate to activate a device that will give nudges for low social interactions by showing it to the other person and describing it as a *shame*.

> *It may be a shame, but it depends on the closeness. For example, if you are not interested right now, I do this [putting her smartphone onto the device] instead of saying "let go of your phone" like a friend. I think it's very bad communication, you know? (P9)*

Discussing this scenario, some users stated that activating such a device willingly would mean accepting that social interaction is or will be bad. In addition, many users (N=11) emphasized that Boost should have a structure that can be started with the approval of everyone, not the decision of one side. In addition, they stated that they would prefer it to start automatically as soon as they sit down at the table.

Some users (N=4) wanted the device to have a snooze feature to complement this idea. In addition to making this feature according to the users' wishes, they said that such a product should be aware of the environment and take action by measuring some criteria, e.g., the seriousness of the conversation and the reason for the silences.

In addition to this ambient awareness, all users liked that Boost gradually gave up on the nudge, seeing if it was successful or not. A participant said that such a feature should be implemented in every product interacting with the user. For example, phones working with such an algorithm while giving notifications would remove many distractions in their daily life.

Three users said they couldn't predict what the device would do when they saw Boost outside. Stating that the device has a visible, non-distracting, but interesting design, one emphasized that it should not have a small sneaky design.

> *It's much better this way than something little and sneaky. I felt like it was sitting here*
> *with us. It's a little obvious that there is such a thing here. "There is something on this*
> *table you should know.", its design says. I would sit down while I accepted it,*
> *whatever came from it. But for example, I would feel bad if it wasn't here so obviously*
> *and suddenly a noise started coming from under the table. -P4*

### 5.4 Silence is also good in terms of social interactions.

In one session, users thought that Boost's nudges depended on their own physical movements and that they could activate them using the items they interacted with in the environment. It was one of the sessions where Boost gave the most nudges because they did not talk while looking around. Still, they said they had a lot of fun figuring out Boost's logic.

> *I thought we should have found something. We searched around because the device*
> *said something about Sherlock Holmes. There was a notebook. I opened the notebook.*
> *Maybe there is a message here or something; I was hoping it was not a personal*
> *thing. Well, I opened it, so I read. (P9)*

In support of this example and previous studies in the literature, most participants emphasized that interaction should not be measured by the amount of conversation alone and that they can have a really good time even when people are not talking. Some participants (N=3) stated that they call people with whom they can stand quietly next to each other as a close friends. Participants (N=9) said that their silence with their friends whom they do not feel close to feels more awkward and that these silences with their close friends may have other meanings (e.g., watching the people inside a cafe).

One user interpreted this approach as a cyclical process. She said that she could sit quietly with her friend right now but that this ability was something they gradually gained during their friendship and developed by talking.

> *In other words, we have come to the level of being able to be silent by talking. We seem to have gained the ability to be silent, starting by keeping quiet during our conversations. These are things that feed off and escalate each other. (P9)*

## 6 DISCUSSION & CONCLUSION

### 6.1 Nudges should be aware of the interaction context.

In this study, we monitored the lull moments between conversations to interpret the interaction quality. We provided users with both audio and visual nudges with BOOST in moments of silence that lasted for a certain period. In most sessions, this method gave us accurate results in measuring interactions, but in some sessions, we saw that lull moments could not accurately measure the quality of social interaction. In our user interviews and previous work [16], users stated that silence doesn't mean poor interaction quality.

In addition, these nudges negatively affected some users interacting during the lull moments. For example, in one session, users watched a video on their smartphones together. After the nudge given during their silence, they closed the video and started doing something else. The nudges were given in these moments of silence, where the interaction between the participants was characterized as high, disrupted the users' current interactions, and was seen as coercion.

For these reasons, we should emphasize the importance of creating context-aware designs that better track users' actions at any given moment. For example, knowing the seriousness of the conversation, the current mood of the people, or what they are doing, combined with monitoring the lull moments, can enable more effective designs.

## 6.2 The feedback mechanisms should support the user while creating awareness about the situation.

One of the most striking insights explored in this study was how users compared the effectiveness of the audio nudge with visual feedback on their interactions. Most of the participants expressed their complaints about visual modality, indicating that the light created discomfort because it was perceived as a visual assessment of their low-quality interaction, which was visible to others.

When we asked the participants what the difference was between the two types of nudges, the most basic answer was that visual feedback did not help them in any way and only informed them about their low-quality interaction. According to the participants, although the audio nudges acted as feedback about the situation, they had a structure that would help users to turn a negative situation into a positive one.

This allowed us to reflect on the four design approaches [16] obtained from users and designers through focus group work in our previous work. While the light in BOOST acts as an enlighteners, the audio clips act as supporters. From this perspective, while Enlighteners can inform a user about behavior in a personal setting (e.g., seeing step counts via a smartwatch), they risk eliciting negative feelings like anxiety, stress, and failure when used in a social setting (e.g., seeing the conversation quality via the BOOST). Therefore, it would be beneficial for designers and researchers to consider this situation when developing solutions in this problem area and focus more on solutions that support the interaction. The interventions they developed should not only reveal a problematic situation with feedback but also provide tips to users on how to solve this problem.

## 6.3 The designs should invite the user rather than wait for their initiation.

Another important insight was on activating BOOST. Users did not feel comfortable even with unhelpful visual feedback on their social interactions. Users also said that activating such a device means that they accept that the quality of social interaction is low, or it is interpreted as shameful behavior if one of the parties starts the device. Most participants wanted the device to start automatically and to be notified of this initiation process.

When we evaluate these insights, we see that such devices act as a sign for users during social interactions, a sign for low interaction a sign for one of the parities unhappiness with the current interaction and with the partner. It should be emphasized that this sign can sometimes be perceived negatively. Considering this unintended outcome is one of the ways to develop solutions with high user acceptance. These interventions should invite users to interact rather than wait to be initiated by the users after they need it.

## 7 CONCLUSION

Social interactions are crucial to people's well-being. These interactions have changed with the widespread use of technological devices such as smartphones in our daily lives. In this article, we presented BOOST, a design artifact to improve people's collocated social interactions with a non-restrictive and non-limiting approach for smartphone use, and an experimental user study that evaluates the users' reactions to this concept, its potential, and the effects on the users' social interactions. We recruited 38 participants (19 dyads) and conducted one-hour sessions with the experiment and control groups. In addition to the experimental setup, we conducted semi-structured interviews with the users who experienced the concept and revealed that the design concept significantly and positively affects the dyads' social interactions regarding intimacy, negative feelings, and speech duration. We synthesized the results of these interviews into three design implications to guide designers and researchers in enriching social interactions while mitigating the effects of mobile technologies; 1) *Nudges should be aware of the interaction context*, 2) *The feedback mechanisms should support the user while creating awareness about the situation*, 3) *The designs should invite the user rather than wait for their initiation*.

# Chapter 9

# DISCUSSION AND CONCLUSION

Smartphones have a significant positive and negative impact on social interactions, making their balanced and mindful use an essential component of healthy social relationships. On the plus side, they allow people to keep in touch with friends and family even when they are physically separated. They also allow individuals to exchange knowledge and experiences in real time, which improves social relationships. They can also be used to help with the coordination of social activities. On the downside, these devices can be a distraction during interactions with peers, resulting in decreased face-to-face communication and social skills. Excessive smartphone use can also trigger negative emotions and a detachment from others.

Social interactions and digital well-being play a major role in an individual's life and relationships, as they are often closely intertwined with daily activities and interactions. Maintaining a responsible technology use and high-quality social interactions are crucial for overall happiness, strong and satisfied relationships, and better quality of life. It is essential to be aware of and manage one's digital habits to maintain healthy relationships and avoid negative effects such as social isolation and addiction.

The negative impacts of these devices on our social interactions are undeniable. Most studies deal with this problem by limiting the of these mobile devices. However, smartphones have become an extension of our bodies and are deeply ingrained in our daily lives. Thus, it is neither fruitful nor desirable to restrict smartphone use during social interactions, particularly from the perspective of users. In fact, studies showed that restrictive approaches backfire and create unintended effects on individuals. Against this backdrop, in this thesis, we questioned, "how design could support collocated social interactions without limiting smartphone use." Throughout this journey, we expected to generate design knowledge to support design researchers and practitioners in designing technologies that align with the question. Using a human-centric approach combined with Research Through Design methodology, we understood the challenges and opportunities associated with smartphone use and design solutions tailored to the users' specific needs. By doing this, we produced design implications for solutions supporting healthy and positive social interactions while allowing individuals to take advantage of the many benefits smartphones offer.

We aimed to explore individuals' smartphone use from the standpoint of social interactions and relations (i.e., how smartphone-checking behavior hinders and supports social interaction). To this aim, we conducted in-situ observations and focus group studies on better understanding smartphone use during social interactions. Using the insights derived from these studies, we defined and explored the design space through co-creation workshops with designers and generated diverse ideas for enriching social interactions. We further developed one idea into a research prototype called WHISPER, a box that tells pre-recorded audio narratives when it detects a lull in a conversation. We conducted user studies using the Wizard of Oz method to get users' thoughts and insights into WHISPER. In the second RtD cycle, we designed a second research prototype called BOOST by considering the learnings from this user study. We developed this concept into a fully working prototype and assessed its impact on social interactions in an experimental setting. Thus, we had the chance to deeply observe the usefulness and validity of the information we uncovered and the effects of such a concept on users' social interactions.

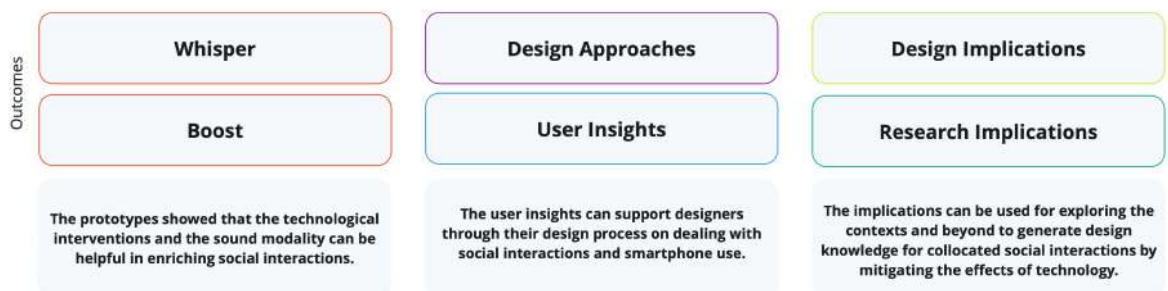

*Figure 9.20 The overall outcomes of the thesis*

The thesis produced three main conclusions. The first, we discovered and showed that technology is double-edged. While it has negative consequences on social interactions, it can be used as a tool to boost these interactions. From the user studies, we gained the point of view that "why and how we use something is more important than the thing itself", and this argument became the supporter of this dissertation. With an iterative design approach, we have shown that designers can create technologies that encourage users to use them more responsibly without banning the use of technology. In other words, we have shown that the negative effect of technology can be mitigated by technology. In this regard, we showed the prototypes, which uses the sound modality to convey information and tailored to different use contexts can successfully and positively affect their social interactions and feelings.

As this thesis focus on a particular direction of a solution, that is, maintaining face-to-face conversations by providing audio nudges, it only covers one portion of the design space for designing technologies to mitigate smartphone use during social interactions without limiting it. We believe that the responsible use of smartphone should and will be an important area for design. We contributed to the literature by providing seven design implications based on series of user studies and these implications revolve around designing the artifact (Balanced Ambiguity), feedback mechanisms (Targeted Nudges, Respectfulness), content (Adaptability, Helping feedback), and also the initiation processes of the interaction (Context-awareness, Invitation Mechanism).

In addition, our research shows that any technology intervening in a user's social interactions might have unintended outcomes and harm the quality of the social interaction. In addition, measuring the quality of social interactions is a wicked and multilayered problem. With the light of these implications, we opened two new research directions for Design & HCI research and there is a need for further investigation if we want to cover the problem area fully.

In the remainder of this section, considering our results, the different perspectives on social interactions and the challenges in targeting them by using technologies, we present the three takeaways of the thesis along with limitations and directions for future research.

## 8.1. *The multidimensional nature of social interactions and the challenges in supporting them*

This work started with the hypothesis that smartphones are problematic for social interactions and relations, and thus they should be restricted. Our initial studies supported this motivation. During our observations in cafes (Chapter 3), we noticed that Phubbing-like behavior patterns have become popular in recent years, and they seemed to affect human relations negatively. However, we also found that people were showing online content to each other with their smartphones or sitting quietly and drinking coffee while having quality time. This variety motivated us to make a deeper exploration of people's social interactions and smartphone use in public settings.

The focus group studies we conducted to achieve this aim revealed that using smartphones during social interactions is perceived as neither bad nor good. This perception changes according to the context and level of intimacy between individuals (Chapter 4). For instance, we found that while some users agreed to remain silent during their social interactions, others said they felt distant from the people they were silent with as if they were not friends. Moreover, while some users complained about the existence of smartphones, others stated that they did not affect their lives negatively and were a part of the natural flow of life. When we consider it as a wicked problem (Buchanan, 1992), these differences draw a direction for us by showing that the source of the problem is not the only one and that this problem is sometimes not considered a problem by the users or the users even are not aware of it. Hence, one of the main takeaways of this thesis is that social interactions are multi-dimensional and can be influenced by many factors, including meeting context, type of relations between individuals, individuals' intention to interact, etc., and the role of smartphones in these interactions is often in flux.

To accommodate the variety in users' perceptions towards and behaviors pertaining to smartphone use during social interactions, we tried to include the user in every stage of the user studies. We followed a user-centered iterative design research process, RtD (J. Zimmerman et al., 2007), by consulting the users in these stages, constantly returning to our previous findings, and iteratively modifying the prototypes. As a result of this iterative and user-centered nature, we managed to develop a final prototype, BOOST, which was found more acceptable by the majority of the users compared to the prototype, WHISPER ([Chapters 6](#) & [8](#)). Therefore, the benefit of RtD as a methodological approach in exploring such a multi-dimensional concept becomes prominent in this thesis. Throughout the journey, we followed an approach to reveal more effective designs with a better understanding of the users. Although the enrichment of collocated social interactions is a problem with different dimensions, we have proven that when users' insights are included, designs and research which affect the majority can be built successfully, and information that covers everyone's needs can be extracted in this process.

### *8.2. Mediating social interactions with technology*

Within the scope of this thesis, we designed research prototypes to enrich social interaction by becoming a part of these interactions. Although we discovered that these prototypes often positively affect users' social interactions, they also raised some concerns. For example, as we have seen in focus groups, even calling smartphone use "excessive" can cause users to get defensive and protective of their phones ([Chapter 4](#)). In addition, although every participant accepted that the enrichment of social interactions was good for the well-being of users, some perceived a product designed for this purpose as a dystopian product.

Another concern was about the privacy and artificiality of the mediated interactions. Almost all respondents said that a product that contributes to their conversations and indirectly contributes to their social interactions needs to know their individual needs. However, the fact that such a device collects digital footprints or listens to users in real-time and interprets the data to nudge them at the right time worried the participants. On the one hand, this data collection raised privacy-related concerns. On the other hand, it raises concerns related to the artificiality of the mediated interaction. In our studies with the second prototype, the participants stated that they could not open up to others and act artificially in an environment where they felt they were being watched ([Chapter 8](#)). Although we clarified that our concepts did not collect or store data, some participants could not trust the device. Thus, a device whose purpose is to enrich people's interactions may sometimes cause poor interaction quality because of the data it collects to be more tailor-made to the user.

Another takeaway of this thesis is that when working on technologies to encourage behavior change designers or researchers should consider the context from multiple perspectives as there can be unintended consequences of a technological intervention despite designers' good intentions (e.g., activating the device is interpreted as a shameful action). In parallel with the first takeaways, we claim that the inclusion of users at every stage of the research process is vital to address this need.

## 8.3. *The disappearance of the research artifacts toward new understandings*

As mentioned in the previous chapters, we followed a design-oriented approach, combining qualitative inquiry with design methods. We conducted observations and focus groups on understanding users. These data collection techniques sometimes create a risk of gathering biased, inconsistent, and very subjective data since the participant might not want to give socially undesirable answers (Opdenakker, 2006). For example, one participant in the focus group sessions said at the beginning of the interview that she did not have any problems using smartphones in social settings and did not worry about this situation. However, towards the end of the study, the same participant admitted that she got angry and smashed her partner's smartphone against the wall because her partner did not listen to her and was busy with his smartphone ([Chapter 4](#)).

Research Artifacts, which we developed with the Research through Design (RtD) approach, offer advantages that can overcome such limitations. In RtD, the design output is not framed as a finished product; instead, it is used to discover information by triggering users' imagination. In the scope of this thesis, the prototypes helped us have a deeper, richer, and more granulated understanding of how people use their smartphones in a social setting and how they would react to nudges trying to get them away from their phones by supporting ongoing conversations. For example, we found that just giving feedback for behavior in a personal context would create discomfort for users, and interventions should be designed to help them instead ([Chapter 8](#)). In addition to that, when the concepts were introduced to users during the interviews, they were able to put themselves in the context while expressing their thoughts, providing a wealth of information about their social interactions. In summary, this study showed us how beneficial research prototypes are for design researchers. Accordingly, we found that such prototypes are useful for

1) Understanding how users interpret the interactions we designed for users and

2) Revealing how and in what ways we can design for subjective phenomena, such as social interactions, are some of the outputs this method brings us.

In every study we conducted, we realized that the concepts we designed were gradually becoming obsolete and that users could have different social interaction practices around these contexts. For example, in the last experimental study with BOOST, we observed that users used the experimental time to socialize, catch up with their friends, and connect with them. The participants in the control group, who did not experience the prototype, reported an increase in positive emotions and a decrease in negative emotions as a result of the experiment. They said it was good to break away from their busy lives and have a one-hour conversation with their friends. This observation led us to conclude that even constructing an interaction scenario instead of the research artifact we developed may positively affect social interactions. Therefore, although the prototypes can nudge people to maintain a conversation during social interaction, they are only one way to enrich social interactions. In this manner, our discussions with users went beyond these artifacts that we made them experience in each study. Thus, while the products we designed to be used in our studies helped us during our research, they left their place to new understandings about the problem space.

## *8.4. Limitations and Future Work*

Our aim throughout the thesis was not to create a commercial product that can effectively boost collocated interactions. Instead, our purpose was to explore the implications of having such a product in a social setting for social interactions. Looking at the findings and the discussion points, it seems that we gather rich insights, even conflicting ones, into this issue which might trigger future research. We designed the BOOST using the design implications we revealed from the user studies with the WHISPER. It received positive feedback from users, and in a sense, we validated the implications. In the continuation of this study, it can be a promising research direction to see how these implications will help designers and researchers design other types of products dealing with social interaction.

In addition, both concepts we designed were tangible artifacts. As our second study revealed, even devising different interaction scenarios without a physical component can be effective and beneficial for enriching social interactions (e.g., an idea for people to find time for each other in a busy work schedule).

Besides these artifact-related directions, we tried to mimic a genuine experience as much as possible with the experiments. However, the user studies for WHISPER and BOOST have a limited number of participants. Although we support our results with quantitative results in Chapter 8, the overall structure of this thesis is a qualitative inquiry. Thus, transferability is one of the important aspects of design studies rather than generalizability (Chow & Ruecker, 2006).

As a continuation of the transferability aspect, we are inclined to include younger generations in this thesis. Even though we had participants from different generations in our focus group studies, we note that the user studies conducted with conceptual prototypes are not generalizable to other populations and contexts, as the participants represent students from the author's university. This was a purposeful decision supported by the idea that excessive smartphone use was widespread among the young population (e.g., Generation Z) (Rideout & Robb, 2019). In the future, we plan to conduct in-the-wild experiments with a sample of participants with various characteristics regarding age, occupation, etc.

# APPENDICES

## *11.1. Appendix 1 – Examples Sketches from Design Workshops*

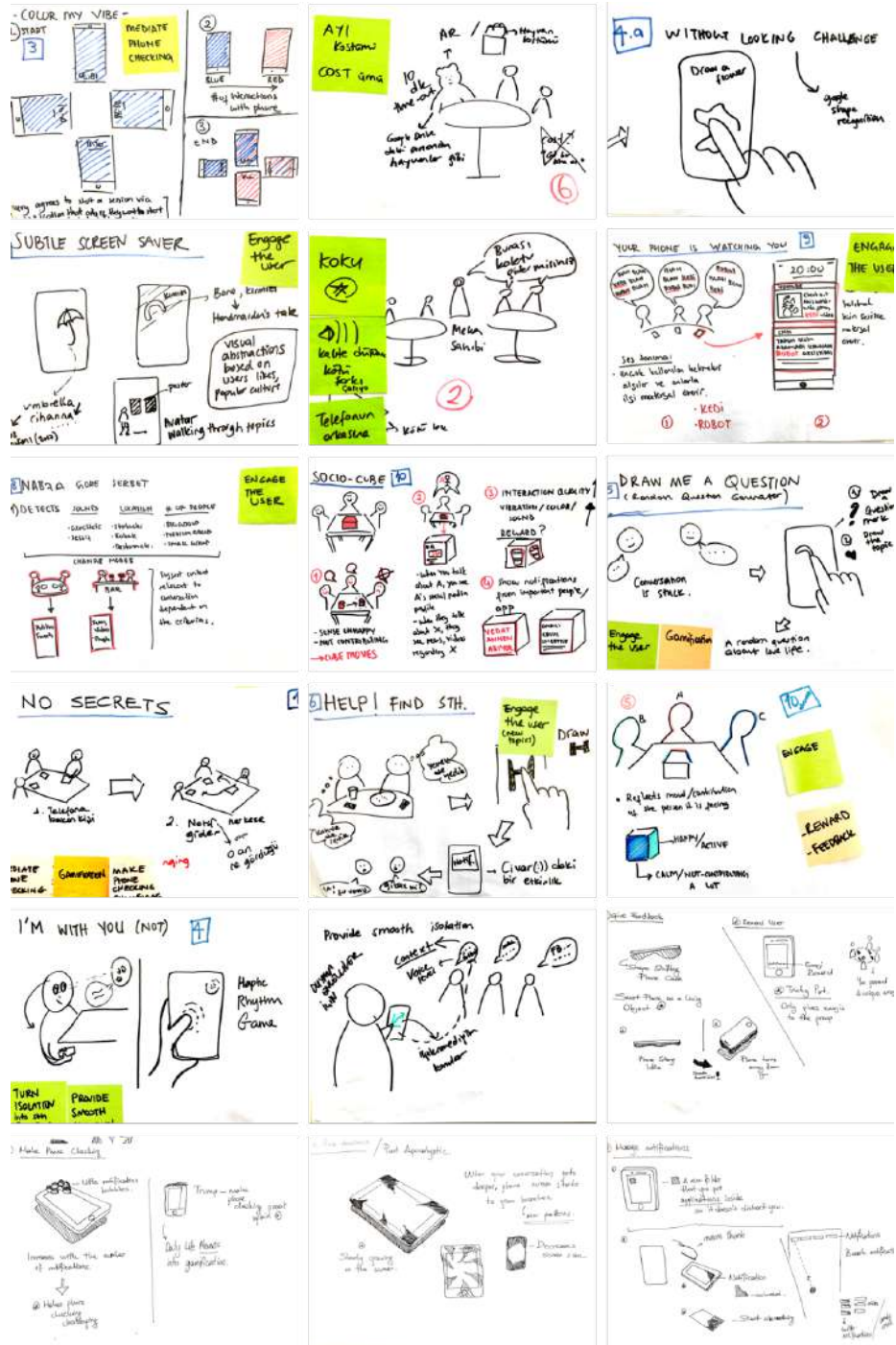

## 11.2. *Appendix 2 - The List of Initial Themes and Selected Genres*

- While the two sisters are playing in their garden, they discover a mysterious wormhole and pass into another dimension. In the dimension they pass through, something is different in their bodies.

- A woman has lost her 5-year-old child. She went to the police station and made her First Notice of Loss application and waiting in a cafe. Her sister come to see the woman. The woman begins to tearfully explain where and when the child disappeared.

- A group of friends meet at the cafe. One of them fell in love with the boyfriend of her friend, who was not with them, and the guy said that he liked that girl. They discuss how to explain this to their close friends.

- One day, the wife of a wealthy businessman living in a mansion is found dead at home with her cat. One of the suspects and the journalist woman investigating the murder meet at a cafe. She claims she knows who the killer is.

- Oxygen begins to be sold for money and a large amount of tax is taken from it. Since the two people sitting in the cafe are aware of this, they start arguing and discussing how to get oxygen cheaply.

- A cigarette addict tries to persuade her friends to sit outside during cold. Her friends do not accept her for a long time, which leads to a cigarette crisis.

- All important conversations take place during the cigarette break at her workplace, who has barely quit smoking. She starts smoking again just to be included in the conversations and they start a fight with her roommate, who is uncomfortable with it.

- This group of friends learned that their friends who were not among them were cheated. They debate whether to tell this to their friends.

- Three women are sitting in a cafe. They talk about their husbands. One is disturbed by how reckless her husband is, the other suffers from her husband's dependence on her, the third is single.

- Two university students are sitting in a cafe and talking about the difficulties they have experienced in online education under the conditions of the pandemic. During a lecture, his family got into a violent fight behind the camera.

You may listen the final stories from this link: https://bit.ly/whisper-stories

### 11.3. Appendix 3 - The Smartphone Use Scale

| Faktörler | Madde | Faktör Yükü | Faktör Öz Değeri | Varyans Yüzdesi |
|---|---|---|---|---|
| Süreklilik | 1. Akıllı telefonumdan uzak kaldığımda kendimi eksik, huzursuz hissederim. | 0,788 | 2,39 | 29,94 |
| | 2. Uyumadan önce ve uyandıktan hemen sonra mutlaka sosyal medya hesaplarımı kontrol ederim. | 0,729 | | |
| | 3. Mobil cihazlarımla (tablet, telefon vs.) devamlı çevrimiçi/aktif bulunurum. | 0,659 | | |
| | 4. Bir şey okuyup çalışırken sosyal medya bağlantımı da kesmem. | 0,599 | | |
| Yetkinlik | 5. Sosyal medya ve internet kullanarak her işimi yapabilirim. | 0,806 | 2,16 | 27,02 |
| | 6. Günlük tüm etkinliklerimi (konuşma, oyun, banka alışveriş vb.) sosyal medya üzerinden yönetebilirim. | 0,751 | | |
| | 7. Yaşamımın her alanında sosyal medyayı aktif kullanırım. | 0,664 | | |
| | 8. Aynı anda hem tablet, akıllı telefon vb. kullanıp hem de diğer işlerimi yapabilirim. | 0,512 | | |
| **Toplam** | | | | 56,96 |

1. I feel incomplete and restless when I am away from my smartphone.

2. I always check my social media accounts before I go to sleep and right after I wake up.

3. I am always online/active with my mobile devices (tablet, phone, etc.).

4. I do not disconnect from social media while I am reading something.

5. I can do everything by using social media and internet.

6. I can manage all my daily activities (talking, gaming, bank shopping, etc.) via social media.

7. I use social media actively in every aspect of my life.

8. You can use both tablet, smartphone, etc. at the same time. I can use it and do my other work.

## 11.4. Appendix 4 - The Short Version of the Big Five Inventory (BFI)

**Measuring personality in one minute or less: A 10-item short version of the Big Five Inventory in English and German**

Beatrice Rammstedt, Oliver P. John

doi:10.1016/j.jrp.2006.02.001

English version.
Instruction: How well do the following statements describe your personality?

| I see myself as someone who … | Disagree strongly | Disagree a little | Neither agree nor disagree | Agree a little | Agree strongly |
|---|---|---|---|---|---|
| … is reserved | (1) | (2) | (3) | (4) | (5) |
| … is generally trusting | (1) | (2) | (3) | (4) | (5) |
| … tends to be lazy | (1) | (2) | (3) | (4) | (5) |
| … is relaxed, handles stress well | (1) | (2) | (3) | (4) | (5) |
| … has few artistic interests | (1) | (2) | (3) | (4) | (5) |
| … is outgoing, sociable | (1) | (2) | (3) | (4) | (5) |
| … tends to find fault with others | (1) | (2) | (3) | (4) | (5) |
| … does a thorough job | (1) | (2) | (3) | (4) | (5) |
| … gets nervous easily | (1) | (2) | (3) | (4) | (5) |
| … has an active imagination | (1) | (2) | (3) | (4) | (5) |

## 11.5. Appendix 5 – Semi-structured Interview Questions for the WHISPER

1. What do you think about your interaction?
2. How did you feel during the interaction?
3. In your opinion, how did Whisper affect your interaction?
4. What aspect of Whisper did you like most?
5. What aspect of Whisper did you hate most?
6. What can be improved for Whisper?

## 11.6. Appendix 6 – Questionnaire for Participant's Social Interaction

**Closeness**

How close do you feel toward the Other after this brief interaction?

**Perceived similarity**

How much do you think you have in common with the other?

How similar do you think you and the other are?

**Awkwardness**

How awkward was the interaction?

**Self-closure**

How much did the Other tell you about himself or herself?

How much did you tell the Other about yourself?

**Liking:**

Overall, how much did you like the other?

Overall, how much do you think the other liked you?

How much would you like to spend time with the other again in the future?

## 11.7. Appendix 7 – Semi-structured Interview Questions for the BOOST

1. What do you think about your interaction?
2. How did you feel during the interaction?
3. In your opinion, how did Boost affect your interaction?
4. What aspect of Boost did you like most?
5. What aspect of Boost did you hate most?
6. What can be improved for Boost?